\documentclass[aps,english,preprint,nofootinbib]{revtex4}
\usepackage{mathrsfs}
\usepackage{graphicx}
\usepackage{amsmath, amssymb}
\usepackage{babel}
\usepackage{color}
\usepackage{slashed}
\usepackage{caption}
\usepackage{subcaption}
\newcommand{\totheleft}{\leftskip=0pt}

\begin{document}

\hfill ACFI-T16-28

\title{Parity- and Time Reversal-Violating Pion Nucleon Couplings: Higher Order Chiral Matching Relations}

\author{Chien-Yeah Seng$^{a,b}$}
\author{Michael Ramsey-Musolf$^{a,c}$}

\affiliation{$^{a}$Amherst Center for Fundamental Interactions,\\
Department of Physics, University of Massachusetts Amherst\\
Amherst, MA 01003 USA}
\affiliation{$^{b}$INPAC, Shanghai Key Laboratory for Particle Physics and Cosmology, \\
	MOE Key Laboratory
	for Particle Physics, Astrophysics and Cosmology,  \\
	School of Physics and Astronomy, Shanghai Jiao-Tong University, Shanghai 200240, China}
\affiliation{$^{c}$Kellogg Radiation Laboratory, California Institute of Technology\\
Pasadena, CA 91125 USA}

\date{\today}

\begin{abstract}
Parity- and time reversal-violating (PVTV) pion-nucleon couplings govern the magnitude of long-range contributions to nucleon and atomic electric dipole moments. When these couplings arise from chiral symmetry-breaking  CP-violating operators, such as the QCD $\theta$-term or quark chromoelectric dipole moments, one may relate hadronic matrix elements entering the PVTV couplings to nucleon and pion mass shifts by exploiting the corresponding chiral transformation properties at leading order (LO) in the chiral expansion. We compute the higher-order contributions to the lowest order relations arising from chiral loops and next-to-next-to leading order (NNLO) operators. We find that for the QCD $\theta$-term the higher order contributions are analytic in the quark masses, while for the quark chromoelectric dipole moments and chiral symmetry-breaking four-quark operators, the matching relations also receive non-analytic corrections. Numerical estimates suggest that for the isoscalar PVTV pion-nucleon coupling, the higher order corrections may be as large as $\sim 20\%$, while for the isovector coupling,  more substantial corrections are possible. 
\end{abstract}

\maketitle

\section{Introduction}

The study of P- and T-violating (PVTV) interactions can be traced back to the
1950s when Purcell and Ramsey proposed searching for the existence of a
permanent electric dipole moment (EDM) of neutron \cite{Purcell:1950zz}. Today, the subject attracts considerable attention as
it is known that CP-violation\footnote{Implying T-violation assuming
that CPT is conserved} is one of the necessary ingredients for explaining
the imbalance between the amount of matter and antimatter of the
current universe \cite{Sakharov:1967dj}. The Standard Model (SM) allows
CP-violating interactions through the complex phase in the
Cabibbo-Kobayashi-Maskawa (CKM) matrix \cite{Kobayashi:1973fv} but
it is insufficient to account for the total observed asymmetry
\cite{Gavela:1993ts,Huet:1994jb,Gavela:1994dt}. Therefore,
alternative sources of CP-violation (CPV) are required.

Assuming that the extra degrees of freedom (DOFs) associated with the beyond
Standard Model (BSM) CPV are heavy, they can be integrated out
of the theory at low energy to obtain effective operators of higher
dimensions that consist solely of SM DOFs. The PVTV components of these
effective operators will in turn generate PVTV low-energy observables, such as EDMs. Current experiments set upper limits on
EDMs, including those of the electron
($8.7\times10^{-29}e\:\mathrm{cm}$, 90\% C.L.) \cite{Baron:2013eja},
mercury atom ($7.4\times10^{-30}e\:\mathrm{cm}$, 95\% C.L.)
\cite{Graner:2016ses} and neutron ($3.0\times10^{-26}e\:\mathrm{cm}$,
90\% C.L.) \cite{Afach:2015sja,Baker:2006ts}. These upper limits imply upper
bounds of the magnitudes of Wilson coefficients of the PVTV effective
operators. When hadrons are involved, translating EDM limits onto these operator bounds is highly non-trivial.  The matching between low-energy
hadronic observables and the Wilson coefficients of operators in
quark-gluon sector involves  various hadronic matrix elements
that are difficult to evaluate from first-principles 
due to the non-perturbative nature of Quantum Chromodynamics (QCD)
at low energy.

In this work we are particularly interested in the PVTV
pion-nucleon coupling constants $\bar{g}_\pi^{(i)}$, where $i=0,1,2$ denotes the isospin.
\cite{deVries:2012ab,Mereghetti:2010tp,Maekawa:2011vs,Bsaisou:2014oka,deVries:2015gea}.
The $\bar{g}_\pi^{(i)}$ govern the strength of 
of long-range (pion-exchange) contributions to atomic EDMs as well as to those of the 
proton and neutron (see, {\em e.g.}, \cite{Engel:2013lsa,deVries:2015gea}). These interactions can be induced by
various PVTV effective operators at the quark-gluon/photon
level such as the $\theta$-term, the quark EDM and chromo-EDM, the
Weinberg three-gluon operator and various four-quark operators. In
particular, if a specific PVTV effective operator breaks chiral
symmetry, then its P and T-conserving (PCTC) counterpart will also generate
corrections to pion and nucleon masses. Consequently,  there
exist matching formulae that relate the induced $\bar{g}_\pi^{(i)}$ and
these mass corrections simply due to chiral symmetry
\cite{deVries:2012ab,Mereghetti:2010tp,Bsaisou:2012rg,Bsaisou:2014oka,deVries:2015gea}. In terms of the SO(4) representation of
Chiral Perturbation Theory (ChPT), the statement above reflects the fact that the PCTC and PVTV components of the
effective operator belong to different components of a single
SO(4) representation; therefore, their hadronic matrix elements are
related through the Wigner-Eckart Theorem. This idea is
practically beneficial as one may then extract the PVTV
hadronic matrix elements for the $\bar{g}_\pi^{(i)}$ from the study of parity- and time reversal conserving (PCTC) hadronic matrix elements that are the pion and nucleon mass
shifts. The latter may be obtained by lattice gauge theory or
other phenomenological approaches. For example, application to the
$\theta$-term with a lattice value of nucleon mass shift yields
$\bar{g}_\pi^{(0)}\approx(0.0155\pm0.0025)\bar{\theta}$
\cite{deVries:2015una}.

It is important to ask how robust these relations are when taking into account possible higher order contributions involving chiral loops and higher order terms in the chiral Lagrangian characterized by addition low-energy constants (LEC's).
As a matter of principle as well as for purposes of numerical precision, one must include these corrections when applying the matching relations. 
For instance, Ref. \cite{deVries:2015una} studied the higher-order
effects to the matching formulae induced by a QCD $\theta$-term in a
three-flavor ChPT. They found that when the matching relation of
$\bar{g}_\pi^{(0)}$ is expressed in terms of the nucleon mass
splitting then its form is preserved by the chiral loop correction.
Consequently, the corrections to the LO matching relations are analytic in the quark masses. 

In this work, we extend the study of Ref. \cite{deVries:2015una} to
cover all effective operators up to dimension 6 that include only
the first generation quarks based on 2-flavor Heavy Baryon
Chiral Perturbation Theory (HBChPT) in the $\mathrm{SU}(2)_\mathrm{L}\times \mathrm{SU}(2)_\mathrm{R}$
representation. First, we perform a general study of how these
operators break chiral symmetry using the spurion
method. This allows us to implement the effects of the chiral
symmetry breaking (CSB) operators to the chiral Lagrangian in a
straightforward manner and obtain the tree-level matching formulae.
Next, we study the chiral one-loop corrections to both
$\bar{g}_\pi^{(i)}$ and the hadron mass shifts. The results are
expressed in the most general form so that they may be straightforwardly
applied to any specific effective operator. Based on the general
formalism above, we study higher-order effects to the matching
formulae induced by the complex quark mass term (induced by the QCD
$\theta$-term), the chromo-EDM and the left-right four quark (LR4Q)
operator that are the only three effective operators in the quark
sector that contain both PCTC and PVTV components
simultaneously and at the same time break chiral symmetry.

Given the length of this paper,  it is useful to summarize here our main
results and point the reader to the respective sections
for details. For convenience, we summarize these features in 
Table \ref{tab:message}, whose content we now proceed to explain.
First, matching
relations exist only for $\bar{g}_\pi^{(i)}$ induced by chiral
non-invariant operators that possess both PCTC and PVTV
components. For these sources, $\bar{g}_\pi^{(i)}$ can be expressed
in terms of mass shifts for nucleon and pion. For chirally invariant
sources, their low-energy PCTC and PVTV effects are not
related by any chiral symmetry and are, therefore, mutually
independent. Hence, there exist no matching relations between
$\bar{g}_\pi^{(i)}$ and hadron mass shifts induced by these sources.
It is also interesting to notice that $\bar{g}_\pi^{(0)}$  depends on the $I=1$ nucleon mass shifts, while $\bar{g}_\pi^{(1)}$
which has $I=1$ depends on the $I=0$ nucleon mass shifts.

Next, we consider higher-order effects, including both one-loop corrections as well as contributions from higher-order LECs. For the loop correction, we find that in many cases a one-loop diagram that corrects
$\bar{g}_\pi^{(i)}$ will have a corresponding diagram with similar
structure that corrects the nucleon mass shift ({\em e.g.}, Figs.
\ref{fig:1a} and \ref{fig:2a}). Furthermore, the CSB vertices in
these diagrams are related by the treel-level matching relations. As a
consequence, the one-loop corrections to $\bar{g}_\pi^{(i)}$ and
nucleon mass shifts induced by these diagrams satisfy the same
tree-level matching relation. There exist  exceptions to this rule, arising from one or more of the following situations: (1) when
the tree-level matching involves $(\Delta m_\pi^2)$,   the
shift of squared pion mass due to the extra operators, the loop
corrections to this term do not have counterparts that correct
$\bar{g}_\pi^{(i)}$; (2) the na\"ive matching between the PVTV
tree-pion coupling $\bar{g}_{\pi\pi\pi}^{(1)}$ and $(\Delta m_\pi^2)$
is spoiled by vacuum alignment, so that  diagrams involving 
insertion of these operators do not satisfy the tree-level matching,
and (3) there are several corrections to the $I=0$ nucleon mass that
do not require extra CSB operators (such as in Fig. \ref{fig:2h},
\ref{fig:2i} and \ref{fig:2k}) so there are no corresponding
diagrams that contribute to $\bar{g}_\pi^{(i)}$. With these
observations in mind, we show that the tree-level matching for
$\bar{g}_\pi^{(0)}$ induced by the $\theta$-term is preserved under
one-loop correction, confirming the result from Ref.
\cite{deVries:2015una}, while matchings induced
by other operators such as dipole operators and four-quark operators
are not respected by loop corrections. On the other hand, contributions from higher-order terms in the chiral Lagrangian do not respect the original matching relations in general, implying a dependence of the matching relations on the associated LECs.

Finally, we estimate the numerical size of higher-order
corrections to the tree-level matching relation using experimental
and lattice-calculated hadron mass parameters as inputs. For each
operator CSB operator $\mathcal{O}$, if the tree-level matching relation has the form $F_\pi\bar{g}_\pi^{(i)}=f^{(i)}$ where $F_\pi\approx 186$MeV is the pion decay constant and $f^{(i)}$ is a function of hadronic mass parameters as well as the PVTV Wilson coefficients, we may characterize the  correction to the LO matching relation as:
\begin{equation}\label{eq:match_deviate}
F_\pi\bar{g}_\pi^{(i)}=f^{(i)}\cdot(1+\delta_{\mathrm{loop}}^{(i)}+\delta_{\mathrm{LEC}}^{(i)})
\end{equation} 
where $\delta_{\mathrm{loop}}^{(i)}$ and $\delta_{\mathrm{LEC}}^{(i)}$ are relative deviations due to one-loop correction and higher-order LECs respectively. We are particularly interested at $\delta^{(i)}_{\mathrm{loop}}$ because $\delta^{(i)}_{\mathrm{LEC}}$ does not involve chiral logs and is therefore suppressed by usual chiral power counting. In principle, one could write down explicit expressions for $\{\delta^{(i)}_{\mathrm{loop}}\}$ as we shall present in the following sections, however their numerical values cannot be determined because they involve the isoscalar and isovector nucleon mass shifts  $\{(\Delta m_N)_\mathcal{O},(\delta m_N)_\mathcal{O}\}$ induced by the operator $\mathcal{O}$ which is not the quark mass operator (except for the case of $\theta$-term). Therefore, in the numerical estimation of $\{\delta^{(i)}_{\mathrm{loop}}\}$ we shall simply set them to zero. The result is summarized in Table \ref{tab:message} where we find that the tree-level
matching formula for $\bar{g}_\pi^{(0)}$ is relatively robust numerically under loop corrections
regardless of choice of the underlying operator while the
status for $\bar{g}_\pi^{(1)}$ in general receive large loop corrections.
On the other hand, the quantity $\delta^{(i)}_{\mathrm{LEC}}$ involves unknown LECs and can only be estimated at present based on rough dimensional arguments. Of course, such estimation may never pretend to be any trustworthy prediction of the actual numerical values of the LECs; in particular, as pointed out in \cite{Wirzba:2016saz}, it makes no prediction to their signs. It therefore only serves to provide a rough estimation to the order of magnitude of the uncertainty brought up by the LECs. We find that the impact of the LECs on the $\bar{g}_\pi^{(0)}$-matching can be as large as $(10-20)$\% while their effect on the $\bar{g}_\pi^{(1)}$-matching is usually not much larger than 1\%. 


Our discussion of this study organized as follows. In Sec. \ref{sec:spurion} we give
introduce a spurion formalism and give a general discussion from the possible forms of the spurion that encode
the explicit CSB effects of the effective operators up to dimension
6. In Sec. \ref{sec:linearOp} we write down the most general form of
PVTV operators as well as PCTC and CSB operators that could
contribute to the loop corrections for $\bar{g}_\pi^{(i)}$ and the
mass shifts of the pion and nucleon. These loop corrections are then
computed in Sec. \ref{sec:loopcorrection} in their most general
form. Based on these results, we perform case-by-case study of the
matching formulae for $\bar{g}_\pi^{(i)}$ induced by different
effective operators, including both loop and LEC contributions, in Sec. \ref{sec:main}. Finally, we shall draw
our conclusions in Sec. \ref{sec:conclusion}.

\begin{table}
\begin{centering}
\begin{tabular}{|c|c|c|c|c|}
\hline Operator & $\bar{g}_{\pi}^{(0)}$ matching &
$\bar{g}_{\pi}^{(1)}$ matching & $\delta^{(0)}_{\mathrm{loop}}$ &  $\delta^{(1)}_{\mathrm{loop}}$ \tabularnewline \hline \hline 
$\theta$-term & LO &
NNLO & 0 & N/A \tabularnewline
\hline chromo-MDM/EDM & LO & LO & 0.021 &
-3.1\tabularnewline \hline 
LR4Q & LO & LO & -0.12
& -3.2 \tabularnewline \hline 
Chiral-invariant
operators & N/A & N/A & N/A & N/A\tabularnewline \hline
\end{tabular}
\par\end{centering}
\caption{\totheleft\label{tab:message}Numerical estimates of the one-loop contribution to the deviation of the tree-level matching formulae. The numerical values of $\delta^{(i)}_{\mathrm{loop}}$ are evaluated at the renormalization scale
$\mu=1$GeV assuming all the nucleon mass shifts induced by non-quark-mass operators are zero. Columns two and three indicate whether the leading matching relation arises at LO, NNLO, or not at all.}

\end{table}

\section{\label{sec:spurion}Chiral Symmetry and The Spurion Method}

It is well known that a massless two-flavor QCD obeys the
$\mathrm{SU}(2)_\mathrm{L}\times \mathrm{SU}(2)_\mathrm{R}$ chiral symmetry defined by the following transformation on the quark
field:
\begin{equation}
Q_R\rightarrow V_RQ_R,\quad Q_L\rightarrow V_LQ_L
\end{equation} where $\{V_R,V_L\}$ are $2\times2$ unitary matrices. Chiral symmetry is explicitly broken in ordinary
QCD only by the quark mass terms. However, when we consider effects
from BSM physics there may be additional higher-dimensional
operators that break the symmetry as well. In general, these
symmetry-breaking terms can always be expressed as products of
$Q_R,Q_L$ with some constant matrices (or products of matrices) in
such as way that if these matrices would transform with a specific
way under the chiral rotation then the corresponding terms would be
chirally invariant. These matrices, known as spurions, are
used to describe the explicit CSB effects in the low-energy
effective theory of QCD because we expect the latter to obey the
same symmetry breaking pattern as QCD itself.

Here we present the most general form of QCD spurion that encodes
the effects from all effective CSB operators up to dimension 6 that
involve only the light quarks and massless gauge bosons. Our choice
of operators are those that obey the SM gauge symmetry at high
energy (see Ref. \cite{Grzadkowski:2010es} for a complete list of
operators). They then undergo electroweak symmetry breaking (EWSB)
where the neutral Higgs is replaced by its vacuum expectation value
(VEV). These operators can be divided into two categories, namely
the quark bilinears and the four-quark operators. Operators in
different categories in general take different form of spurions.

\subsection{Quark bilinears}

At dimension four the only CSB terms are the quark Yukawa coupling
terms that then undergo EWSB to give rise to the quark masses. At
the same time, a non-vanishing QCD $\theta$-term may then be rotated
away using the axial anomaly to be replaced by complex phases in the
quark masses (this procedure will be reviewed in Sec.
\ref{sec:theta}). The resulting Lagrangian will take the general
form 
\begin{equation}
\label{eq:spurion4}
-\bar{Q}_R X Q_L+h.c.\ \ \ ,
\end{equation} 
where $X$ is a complex $2\times2$
diagonal matrix in flavor space and the term would be chirally invariant if $X$ would
transform as $X\rightarrow V_RXV_L^\dagger$ under chiral rotation.

At dimension six the only CSB bilinear operators of quarks are the
$\psi^2H^3$ operators and the dipole-like operators
\footnote{Another operator of the form $i(\tilde{H}^\dagger D_\mu
H)\bar{u}_R\gamma^\mu d_R$, with ${\tilde H}_j\equiv \epsilon_{jk} H^\ast_k$ will be classified as a four-quark
operator after the W-boson is integrated out.}. On the one hand, the
$\psi^2H^3$ operators reduce to complex quark mass terms after EWSB
so we do not need to discuss them separately. On the other hand, the
dipole operators have the general form 
\begin{equation}
{\bar Q}_L \sigma^{\mu\nu} T^A H d_R V^A_{\mu\nu}
\end{equation}
where $T^A$ is a generator of any one of the SM gauge groups and $V^A_{\mu\nu}$ are the corresponding field strength tensor (a similar structure appears for up-type quarks with $d_R\to u_R$ and $H_j\to \epsilon_{jk} H^\ast_k$). After EWSB, the dipole operators reduce to the dimension five forms
\begin{equation}
\bar{q}_L\sigma^{\mu\nu}\frac{\lambda^a}{2}q_RG^a_{\mu\nu},\quad
\bar{q}_L\sigma^{\mu\nu}q_RF_{\mu\nu},\quad
\bar{q}_L\sigma^{\mu\nu}q_RZ_{\mu\nu},\quad
\bar{u}_L\sigma^{\mu\nu}d_RW^+_{\mu\nu}\ \ \ .
\end{equation}
We can neglect the last three operators because their
effects in the generation of pure hadronic operators will be
suppressed with respect to the first either by the electromagnetic coupling strength or inverse powers of the
 heavy gauge boson masses. The remaining operators are 
the flavor-diagonal quark chromo-magnetic dipole
moment(cMDM)/chromo-electric dipole moment(cEDM). 

In terms of the chiral spurion, the cMDM and CEDM operators take the form
\begin{equation}
\label{eq:spuriondipole}
\bar{Q}_R\sigma^{\mu\nu}X \frac{\lambda^a}{2} Q_L G_{\mu\nu}^a\ \ \ ,
\end{equation}
where again $X$ is a
complex $2\times2$ diagonal matrix. We then conclude that the quark bilinears appearing in Eqs.(\ref{eq:spurion4},\ref{eq:spuriondipole})
imply the same form of the spurion, namely:
\begin{equation}
X=a+b\tau_3
\end{equation}
where $\{a,b\}$ are complex numbers. If the spurion would transform
as $X\rightarrow V_RXV_L^\dagger$ under $\mathrm{SU}(2)_\mathrm{L}\times \mathrm{SU}(2)_\mathrm{R}$ then
the Lagrangian would be chirally invariant. Furthermore, any PVTV
effects are contained in the imaginary part of $a$ and $b$.

When the spurion method is applied to the baryon sector of the chiral
Lagrangian, it is convenient to define the following quantities:
\begin{equation}\label{eq:Xtildepm}
\tilde{X}_{\pm}\equiv u^\dagger X u^\dagger\pm u X^\dagger u
\end{equation}
where the subscript ``+" (``-") denotes that the matrix is Hermitian
(anti-Hermitian) and $u$ is a matrix function of pion fields defined in Appendix \ref{sec:blocks}. They ``transform" under chiral rotation as
$\tilde{X}_\pm\rightarrow K\tilde{X}_\pm K^\dagger$. One advantage
of this notation is that it allows us to construct Lagrangian of
which PVTV effects come entirely from the spurion matrix $X$.
For instance, $\tilde{X}_+$ is
parity-even and $\tilde{X}_-$ is parity-odd if $X$ is a real matrix because $u\leftrightarrow u^\dagger$ under P.
Therefore, in LO effective Lagrangian, the
spurion involved should be $\tilde{X}_+$ and not $\tilde{X}_-$
because we require the Lagrangian to be P (and T)-even when the
matrix $X$ is real.

\subsection{Four-quark operators}

Next we study the most general form of spurion fields induced by
dim-6 four quark operators. As explained at the beginning of the section, these operators encode effects of BSM physics at high scale which is assumed to obey the Standard Model $\mathrm{SU}(2)_\mathrm{L}\times \mathrm{U}(1)_\mathrm{Y}$ symmetry, so they are constructed using the $\mathrm{SU}(2)_\mathrm{L}$ doublet field $Q_L$ as well as the singlet fields $\{u_R,d_R\}$. Following the notations in Ref.
\cite{Grzadkowski:2010es}, these operators can be grouped into the
following categories:

\subsubsection{$(\bar{L}L)(\bar{L}L)$:}
The two independent operators could be chosen as
\begin{equation}\bar{Q}_L\gamma^\mu Q_L\bar{Q}_L\gamma_\mu
Q_L,\quad\bar{Q}_L\gamma^\mu\tau^i Q_L\bar{Q}_L\gamma_\mu\tau^i
Q_L.\end{equation} They are both chirally invariant so they do not
give rise to any non-trivial spurion.

\subsubsection{$(\bar{R}R)(\bar{R}R)$:}
There are four independent operators in this category that can be chosen as 
\begin{equation}\bar{u}_R\gamma^\mu
u_R\bar{u}_R\gamma_\mu u_R,\quad  \bar{d}_R\gamma^\mu
d_R\bar{d}_R\gamma_\mu d_R,\quad \bar{u}_R^i\gamma^\mu
u_R\bar{d}_R\gamma_\mu
d_R,\quad\bar{u}_R^i\gamma^\mu\frac{\lambda^a}{2}u_R\bar{d}_R\gamma_\mu\frac{\lambda^a}{2}d_R.\label{eq:RRRR}
\end{equation}
These operators break chiral symmetry as $X_R$ or $X_R\otimes X_R$ where the spurion matrix $X_R=\tau_3$. Chiral symmetry would be preserved if the spurion matrix would transform as
$X_R\rightarrow V_RX_R V_R^\dagger$ under chiral rotation. Here, the notation $A\otimes B$ means that the matrices $A$ and $B$  appear simultaneously in a quark bilinear or a four-quark operator, {\em e.g.} $\bar{Q}ABQ$ or $\bar{Q}AQ\bar{Q}BQ$.

\subsubsection{$(\bar{L}L)(\bar{R}R)$:}
There are four independent operators in this category that can be
chosen as
\begin{equation}
\bar{Q}_L\gamma^\mu Q_L\bar{u}_R\gamma_\mu u_R,\quad
\bar{Q}_L\gamma^\mu\frac{\lambda^a}{2}Q_L\bar{u}_R\gamma_\mu
\frac{\lambda^a}{2}u_R,\quad \bar{Q}_L\gamma^\mu Q_L\bar{d}_R\gamma_\mu d_R,\quad
\bar{Q}_L\gamma^\mu\frac{\lambda^a}{2}Q_L\bar{d}_R\gamma_\mu
\frac{\lambda^a}{2}d_R.\end{equation} These operators break chiral symmetry through a single spurion matrix
$X_R$.

\subsubsection{\label{sec:LRLR}$(\bar{L}R)(\bar{L}R)$:}
There are two operators in this category, namely 
\begin{equation}
\varepsilon^{ij}\bar{Q}_L^iu_R\bar{Q}_L^{j}d_R,\quad\varepsilon^{ij}\bar{Q}_L^i\frac{\lambda^a}{2}u_R\bar{Q}_L^{j}\frac{\lambda^a}{2}d_R.
\end{equation}
Both operators are chirally invariant: for instance, the first operator can be rewritten as $\varepsilon^{ij}\varepsilon^{i'j'}\bar{Q}_L^{i}Q_R^{i'}\bar{Q}_L^jQ_R^{j'}/2$ so its $\mathrm{SU}(2)_\mathrm{L}$ and $\mathrm{SU}(2)_\mathrm{R}$-invariance are explicit. Meanwhile, they allow
complex Wilson coefficients that give rise to PVTV physics. We
can then define their ``spurion" simply as a complex number.

\subsubsection{Induced Left-Right Four Quark (LR4Q) Operator}\label{sec:LR4Qop}
Finally there is another four quark operator that arises from $i(\tilde{H}^\dagger D_\mu
H)\bar{u}_R\gamma^\mu d_R$. When the $W^\pm$ boson contained in $D_\mu$ is exchanged with the left-handed charge changing quark current, one obtains the following four-quark operator after EWSB:
\begin{eqnarray}
&&c_{4q}\bar{d}_L\gamma^\mu u_L\bar{u}_R\gamma_\mu d_R+h.c.\nonumber\\
&=&-\frac{2}{3}(c_{4q}\bar{Q}_R\frac{1+\tau_3}{2}Q_L\bar{Q}_L\frac{1-\tau_3}{2}Q_R+c_{4q}^*\bar{Q}_R\frac{1-\tau_3}{2}Q_L\bar{Q}_L\frac{1+\tau_3}{2}Q_R)\nonumber\\
&&-4(c_{4q}\bar{Q}_R\frac{1+\tau_3}{2}\frac{\lambda^a}{2}Q_L\bar{Q}_L\frac{1-\tau_3}{2}Q_R+c_{4q}^*\bar{Q}_R\frac{1-\tau_3}{2}\frac{\lambda^a}{2}Q_L\bar{Q}_L\frac{1+\tau_3}{2}Q_R)\label{eq:LR4Qspurion}
\end{eqnarray}
where the right hand side is obtained using a Fierz transformation.
We see that this operator breaks the chiral
symmetry as $c_{4q}X_{RL}\otimes
X_{LR}+c_{4q}^*X_{LR}^\dagger\otimes X_{RL}^\dagger$ where
$X_{RL}=(1+\tau_3)/2$, $X_{LR}=(1-\tau_3)/2$ and would be
chirally invariant if $X_{RL}\rightarrow V_RX_{RL}V_L^\dagger$ and
$X_{LR}\rightarrow V_LX_{LR}V_R^\dagger$ under a chiral rotation.
One observes that the part of the operator proportional to $\mathrm{Re}c_{4q}$ is has the structure $\bar{Q}_RQ_L\bar{Q}_LQ_R-\bar{Q}_R\tau_3Q_L\bar{Q}_L\tau_3Q_R$ (and terms with $\lambda_a$-insertions), so it is PCTC with isospin 0 or 2. Meanwhile, the part proportional to $\mathrm{Im}c_{4q}$ has the structure $\bar{Q}_RQ_L\bar{Q}_L\tau_3Q_R-\bar{Q}_R\tau_3Q_L\bar{Q}_LQ_R$ (and terms with $\lambda_a$-insertions). It is PVTV and with isospin 1. 

Up to this point we have discussed all the possible operators up to
dimension six that would break the QCD chiral symmetry. A complete
list of spurions induced by these operators can be found in Table
\ref{tab:spurion}. It is important to note that only the complex
quark mass term, the dipole-like operators, and the LR4Q operator are
chirally non-invariant and  contain both PCTC and
PVTV components. These three types of operators will be relevant
in the discussion of the matching formula for the $\bar{g}_\pi^{(i)}$ in
the upcoming sections.

\begin{table}

\begin{centering}
\begin{tabular}{|c|c|c|c|}
\hline Operators & Spurion & Constant value & ``Transformation
Rule"\tabularnewline \hline \hline Quark bilinears & $X$ &
$a+b\tau_{3}$ & $X\rightarrow V_RXV_L^\dagger$\tabularnewline \hline
Four Quark: & $a$ & $a$ & $a\rightarrow a$\tabularnewline
\cline{2-4} $(\bar{L}L)(\bar{L}L),$ $(\bar{R}R)(\bar{R}R),$ & $X_R$
& $\tau_{3}$ & $X_R\rightarrow V_RX_R V_R^\dagger$ \tabularnewline
\cline{2-3} $(\bar{L}L)(\bar{R}R),$ $(\bar{L}R)(\bar{L}R)$ &
$X_R\otimes X_R$ & $\tau_3\otimes \tau_3$ & \tabularnewline \hline
Induced LR4Q & $c_{4q}X_{RL}\otimes X_{LR}$ & $X_{RL}=(1+\tau_3)/2$,
& $X_{RL}\rightarrow V_RX_{RL}V_L^\dagger,$ \tabularnewline
 & $+c_{4q}^{*}X_{LR}^\dagger\otimes X_{RL}^\dagger$ & $X_{LR}=(1-\tau_3)/2$ & $X_{LR}\rightarrow V_LX_{LR}V_R^\dagger$\tabularnewline
\hline
\end{tabular}
\par\end{centering}

\caption{\totheleft\label{tab:spurion}Complete list of spurions
that enter the chiral Lagrangian. For each spurion, we show the constant value it takes during its implementation in the Lagrangian (third column), and how it would need to transform in order to leave the Lagrangian chirally invariant (fourth column). Among all the operators, only the quark
bilinears and the induced LR4Q operator are chirally non-invariant and
at the same time contain both PCTC and PVTV components. }

\end{table}

\section{\label{sec:linearOp}Chiral Symmetry Breaking Operators in a Linear
Representation}

Insertions of the spurion fields we discussed in Sec.
\ref{sec:spurion} into the chiral Lagrangian will give rise to CSB
operators consisting of baryons and pions. Among them, the leading PVTV $NN\pi$ operators and the hadron mass operators are of greatest importance because their Wilson coefficients will enter the matching formulae for the  $\bar{g}_\pi^{(i)}$ that is the focus this work. At the same time, the existence of such operators automatically implies the presence of a whole series of CSB operators with higher powers of pions whose operator coefficients are related  by chiral symmetry. The relation, however, depends on the explicit form of spurion. Consequently, it is not practical to write down a single CSB Lagrangian containing terms with an arbitrary number of pion fields without specifying the form of spurion. 

Nevertheless, a subset of CSB operators with higher powers of pion fields ({\em e.g.} $NN\pi\pi$, $NN\pi\pi\pi$ and $\pi\pi\pi\pi$ operators) must be included in this work because they contribute to $\bar{g}_\pi^{(i)}$ and hadron mass parameters at one-loop and will therefore modify the matching formulae from their tree-level expressions. For this purpose, it will be
convenient to express the Goldstone bosons ({\em i.e.} pions in our case) in the CSB operators in linear, in stead of
non-linear, representation. By doing so we pay the price of
losing the manifest  chiral structure of each term. On the other hand
results of the loop corrections will be completely
general and independent of any particular choice of spurion.
Eventually, when we need to apply the general result to specific effective operators (spurions) we simply refer back to the non-linear
representation, expand each term in powers of pion fields, and match the
coefficients with those in the general linear representation. 

We will also include the $\Delta$-baryons as explicit DOFs since the nucleon-$\Delta$ mass splitting vanishes in the large-$N_c$ limit \cite{Jenkins:1993zu} and since inclusion of $\Delta$s is generally required in order to  respect  $1/N_c$ power counting. As far as this work is concerned, the $\Delta$-baryons only appear as virtual particles in loop corrections to the $\bar{g}_\pi^{(i)}$ and nucleon masses.

\subsection{\label{sec:PVTV}PVTV operators}

Following the foregoing discussion , we  proceed to write down all possible forms of lowest-order PVTV operators involving nucleons, pions and $\Delta$-baryons in the linear representation of Goldstone bosons that are relevant to this work.
For the coefficient of these operators we adopt the following
unified notation, namely: the coefficient $\bar{g}_K^{(I,j)}$ is the
real coefficient of the $j$-th PVTV operator of type $K$ with
isospin $I$ (the superscript $j$ will however be suppressed if there
is only one operator with isospin $I$). Because the 
$\bar{g}_\pi^{(i)}$ can only have isospin $I=0,1,2$,  for the
renormalization of these operators at leading order we only need to
consider all PVTV operators with $I=0,1,2$. Furthermore, we
choose to parameterize $\bar{g}_K^{(I,j)}$ in such a way that all of
them are dimensionless by the inclusion of appropriate powers of
$F_\pi$ in front of each operator.

\subsubsection{$NN\pi$ operators} The PVTV
$NN\pi$ operators are defined as \cite{Engel:2013lsa}: 
\begin{equation}\mathcal{L}=\bar{g}_\pi^{(0)}\bar{N}\vec{\tau}\cdot\vec{\pi}N+\bar{g}_\pi^{(1)}\pi_0\bar{N}N-3\bar{g}_\pi^{(2)}\mathcal{I}^{ab}\pi_a\bar{N}\tau_bN
\end{equation} where $\mathcal{I}=(1/3)\mathrm{diag}(1\:\:1\:-2)$ is needed to combine two isospin triplets into
an $I=2$ quantity.

\subsubsection{$\Delta\Delta\pi$ operators}

The PVTV $\Delta\Delta\pi$ operators have the general form
$\bar{T}^a_\mu T^{b\mu}\pi$, where $T^a_\mu$ is the field representation of the $\Delta$-baryon as explained in Appendix \ref{sec:blocks}. They can be chosen as
\begin{eqnarray}
&&\mathcal{L}=\bar{g}_{\Delta\Delta\pi}^{(0)}\bar{T}^a_\mu\vec{\tau}\cdot\vec{\pi}T^{a\mu}
+i\bar{g}_{\Delta\Delta\pi}^{(1,1)}\epsilon^{abc}\pi_{b}[\bar{T}_{\mu}^{a}\tau_{c}T^{3\mu}-\bar{T}_{\mu}^{3}\tau_{c}T^{a\mu}]
+i\bar{g}_{\Delta\Delta\pi}^{(1,2)}\epsilon^{ab3}\bar{T}_{\mu}^{a}\vec{\tau}\cdot\vec{\pi}T^{b\mu}\nonumber\\
&&+\bar{g}_{\Delta\Delta\pi}^{(2,1)}\mathcal{I}^{ab}[\pi_{c}\bar{T}_{\mu}^{c}\tau_{a}T^{b\mu}+\pi_{c}\bar{T}_{\mu}^{b}\tau_{a}T^{c\mu}-\pi_{a}\bar{T}_{\mu}^{c}\tau_{c}T^{b\mu}-\pi_{a}\bar{T}_{\mu}^{b}\tau_{c}T^{c\mu}]
+\bar{g}_{\Delta\Delta\pi}^{(2,2)}\mathcal{I}^{ab}\pi_{b}\bar{T}_{\mu}^{c}\tau_{a}T^{c\mu}.
\end{eqnarray}

Note that the $T_\mu^a$ $(a=1,2,3)$ denote a set of three two-component vectors in isospin space, satisfying $\tau^a T_\mu^a=0$. This representation allows us to write down all expressions in terms of quantities such as $\tau^a$ and $\mathcal{I}^{ab}$ where the indices run from 1 to 3.

\subsubsection{$\pi\pi\pi$ operators}

The PVTV 3-pion operators should look like
$\pi\pi\pi$ as operators with derivatives are of higher order. In particular, the only operator relevant to us is the
$I=1$ operator (the others have $I=3$):
\begin{equation}
\mathcal{L}=\bar{g}_{\pi\pi\pi}^{(1)}F_\pi\vec{\pi}^2\pi_0.\end{equation}
It is T-odd because the neutral pion field changes sign under T under our conventions for the pion-nucleon interactions.

\subsubsection{$NN\pi\pi\pi$ operators}

The PVTV $NN\pi\pi\pi$ operators can be chosen as
\begin{equation}
\mathcal{L}=\frac{\bar{g}_{NN3\pi}^{(0)}}{F_\pi^2}\bar{N}\vec{\tau}\cdot\vec{\pi}N\vec{\pi}^2
+\frac{\bar{g}_{NN3\pi}^{(1)}}{F_\pi^2}\bar{N}N\pi_0\vec{\pi}^2
+\frac{\bar{g}_{NN3\pi}^{(2,1)}}{F_\pi^2}\mathcal{I}^{ab}\pi_a\pi_b\bar{N}\vec{\tau}\cdot\vec{\pi}N
+\frac{\bar{g}_{NN3\pi}^{(2,2)}}{F_\pi^2}\mathcal{I}^{ab}\pi_b\bar{N}\tau_aN\vec{\pi}^2.
\end{equation}





\subsection{\label{sec:CSB}PCTC operators}

Following the same line of thought as in the previous subsection, we
shall construct all relevant PCTC CSB operators that contribute
to the loop correction to hadron mass shifts. Again these operators
are defined using a linear representation of the Goldstone bosons.

\subsubsection{$\pi\pi$ operators} There are only two kinds of
CSB $\pi\pi$ operators that are the isospin invariant ($I=0$)
and isospin-breaking ($I=2$) mass terms respectively:
\begin{equation}\label{eq:defmpishift}\mathcal{L}=-\frac{1}{2}(\Delta
m_\pi^2)\vec{\pi}^2-\frac{3}{2}(\delta
m_\pi^2)\mathcal{I}^{ab}\pi^a\pi^b.\end{equation} Here we define
$(\Delta m_\pi^2)$ such that it does not include the LO-contribution
from the quark mass (i.e. the well-known $(m_\pi^2)_0=2B_0 \bar{m}$ contribution in ChPT, as we shall also discuss in Sec. \ref{sec:theta}). That is, we shall include only $(m_\pi^2)_0$ in the pion propagator while the $(\Delta m_\pi^2)$ and $(\delta m_\pi^2)$ defined above appear only in the form of two-pion vertex in Feynman diagrams, as depicted in Fig. \ref{fig:CBSloop}. Similar argument applies for the quantities $(\Delta m_\Delta)$, $(\delta m_\Delta)$ and $(\delta \tilde{m}_\Delta)$ which we shall define below: they appear only in the form of $\Delta-\Delta$ vertex, while the $\Delta$-propagator contains only $\delta_\Delta$, namely the nucleon-delta mass splitting in the chiral limit, as defined in Eq. \eqref{eq:invariantLag}.

\subsubsection{$\pi\pi\pi\pi$ operators} There are two four-pion
operators up to $I=2$. They can be written as:
\begin{equation}
\mathcal{L}=g_{4\pi}^{(0)}(\vec{\pi}^2)^2
+g_{4\pi}^{(2)}\vec{\pi}^2\mathcal{I}^{ab}\pi_a\pi_b.
\end{equation}
Again, we define $g_{4\pi}^{(0)}$ such that it does not include the
LO-contribution from the quark mass.

\subsubsection{$NN$ operators} Again there are only two kinds of
CSB $NN$ operators, corresponding to the nucleon $\sigma$-term
and the mass splitting term. We write them as
\begin{equation}\mathcal{L}=\label{eq:sigmaterm}(\Delta m_N)\bar{N}N
+\frac{(\delta m_N)}{2}\bar{N}\tau_3N.\end{equation} Even though the
operator $\bar{N}N$ is chirally invariant, it can still be obtained
through an insertion of a spurion (e.g. from the isospin-invariant
part of the quark mass matrix) so it must included for completeness.

\subsubsection{$NN\pi\pi$ operators}

We are only interested in the $I=0,1,2$ operators that are
\begin{equation}\mathcal{L}=\frac{g_{NN\pi\pi}^{(0)}}{F_\pi}\bar{N}N\vec{\pi}^2
+\frac{g_{NN\pi\pi}^{(1,1)}}{F_\pi}\bar{N}\tau_3N\vec{\pi}^2
+\frac{g_{NN\pi\pi}^{(1,2)}}{F_\pi}\bar{N}\vec{\tau}\cdot\vec{\pi}N\pi_0
+\frac{g_{NN\pi\pi}^{(2)}}{F_\pi}\mathcal{I}^{ab}\pi_a\pi_b\bar{N}N.
\end{equation}
There is another $I=2$ operator but it is T-odd.

\subsubsection{$\Delta\Delta$ operators}

There are four kinds of $\Delta\Delta$ operators corresponding to
four mass terms. However here we are only interested in the
$I=0,1,2$ operators. They are:
\begin{equation}\mathcal{L}=(\Delta m_\Delta)\bar{T}^a_\mu
T^{a\mu} +\frac{(\delta m_\Delta)}{2}\bar{T}^a_\mu\tau_3 T^{a\mu}
+3(\delta\tilde{m}_\Delta)\mathcal{I}^{ab}\bar{T}^a_\mu T^{b\mu}
\end{equation}
Again we define $(\Delta m_\Delta)$ such that it does not include
the original residual mass $\delta_\Delta$ in the chiral limit. Similar to $\bar{N}N$, the
operator $\bar{T}^a_\mu T^{a\mu}$ is chirally invariant yet it can
still be induced by a spurion-insertion so we need to include this
term.

\section{\label{sec:loopcorrection}One-loop correction to $\bar{g}_\pi^{(i)}$ and hadron mass shifts}

With all the relevant operators defined in Sec. \ref{sec:linearOp}
it is now straightforward to compute the most general one-loop
corrections to both $\bar{g}_\pi^{(i)}$ and the hadron mass shifts.
To obtain the total result one needs to compute both the
one-particle irreducible (1PI) diagrams and the wavefunction
renormalization graphs. The latter are quite standard and are summarized
in Appendix \ref{sec:ordloop}.

\begin{figure}
\centering
\begin{subfigure}[b]{0.25\textwidth}
\includegraphics[width=\textwidth]{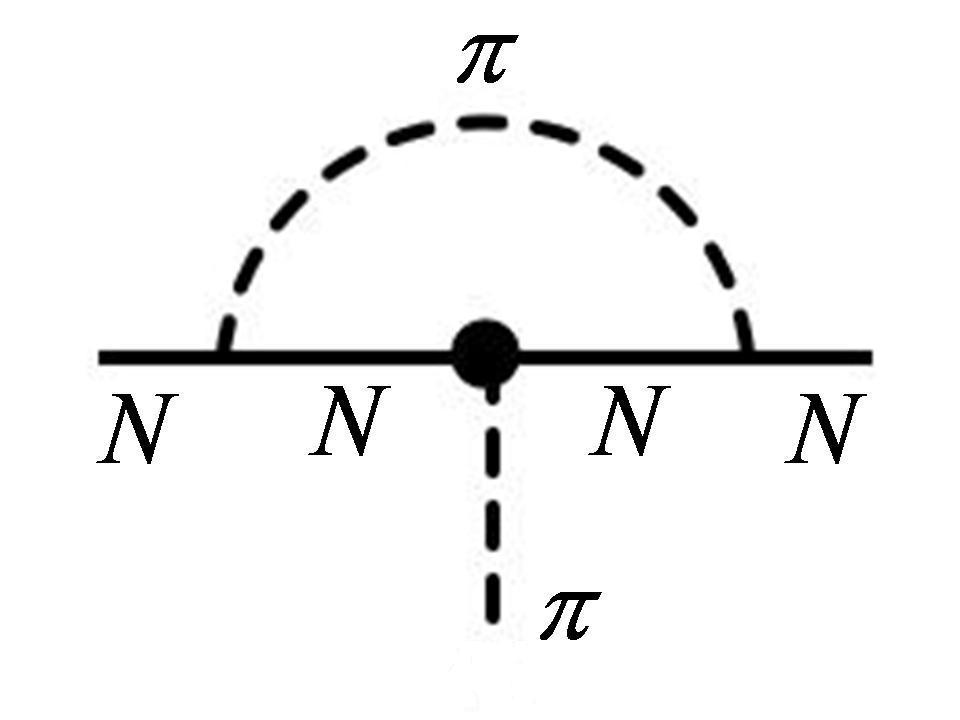}
\caption{\label{fig:1a}(a)}
\end{subfigure}
\begin{subfigure}[b]{0.25\textwidth}
\includegraphics[width=\textwidth]{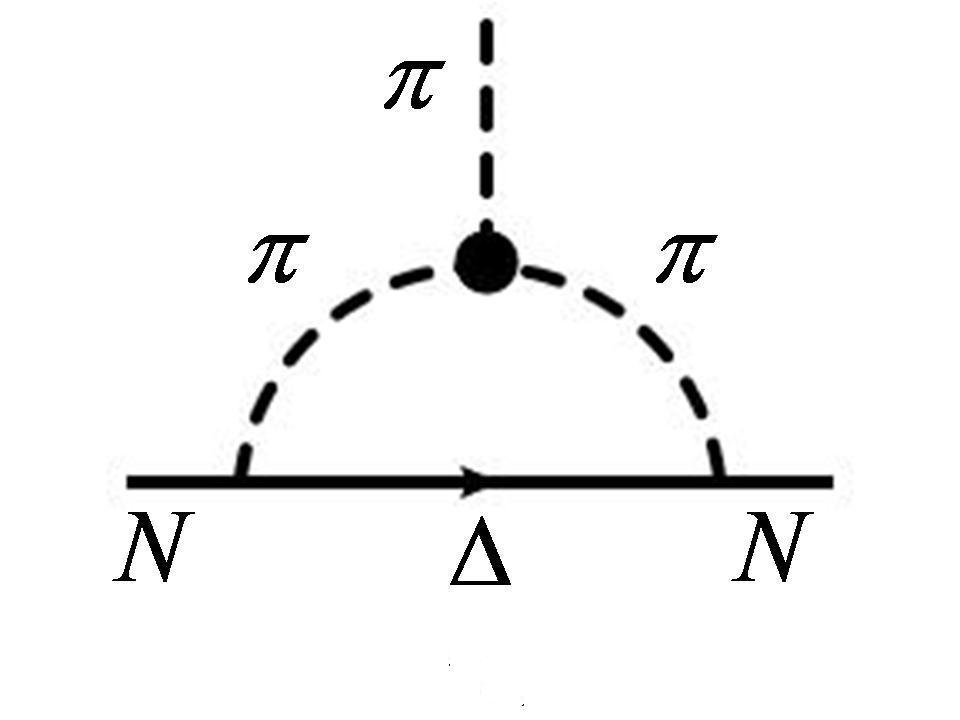}
\caption{(b)}
\end{subfigure}
\begin{subfigure}[b]{0.25\textwidth}
\includegraphics[width=\textwidth]{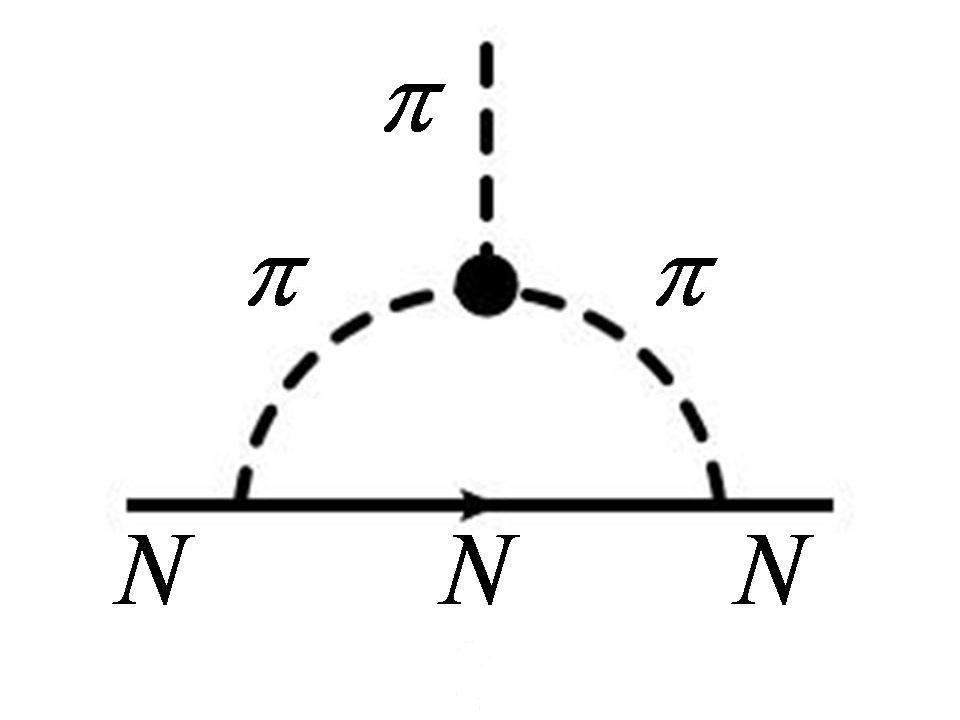}
\caption{(c)}
\end{subfigure}
\begin{subfigure}[b]{0.25\textwidth}
\includegraphics[width=\textwidth]{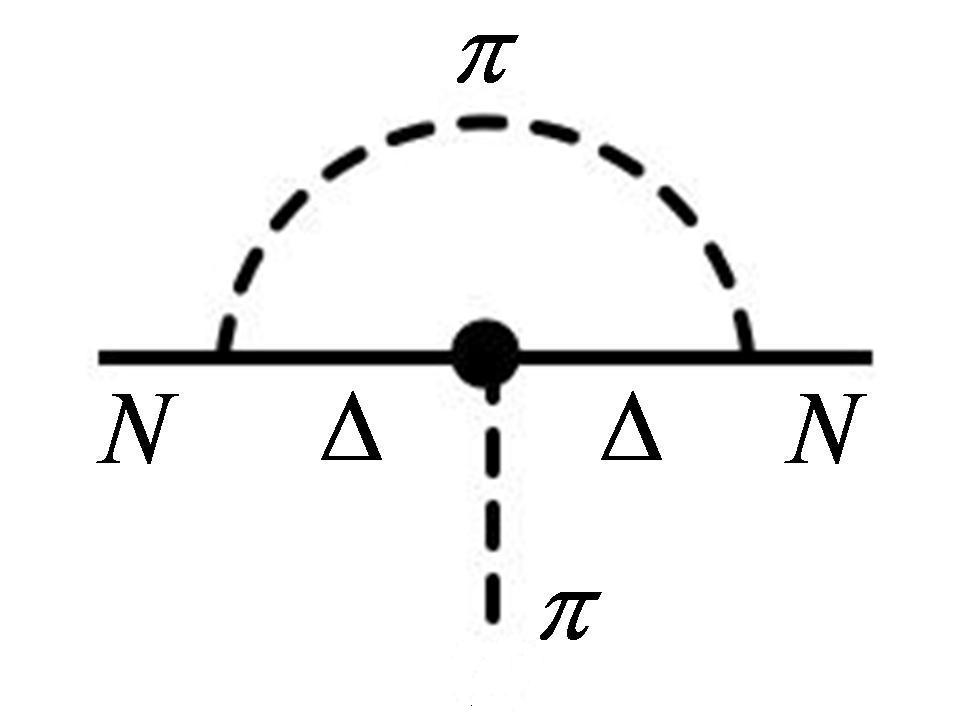}
\caption{(d)}
\end{subfigure}
\begin{subfigure}[b]{0.25\textwidth}
\includegraphics[width=\textwidth]{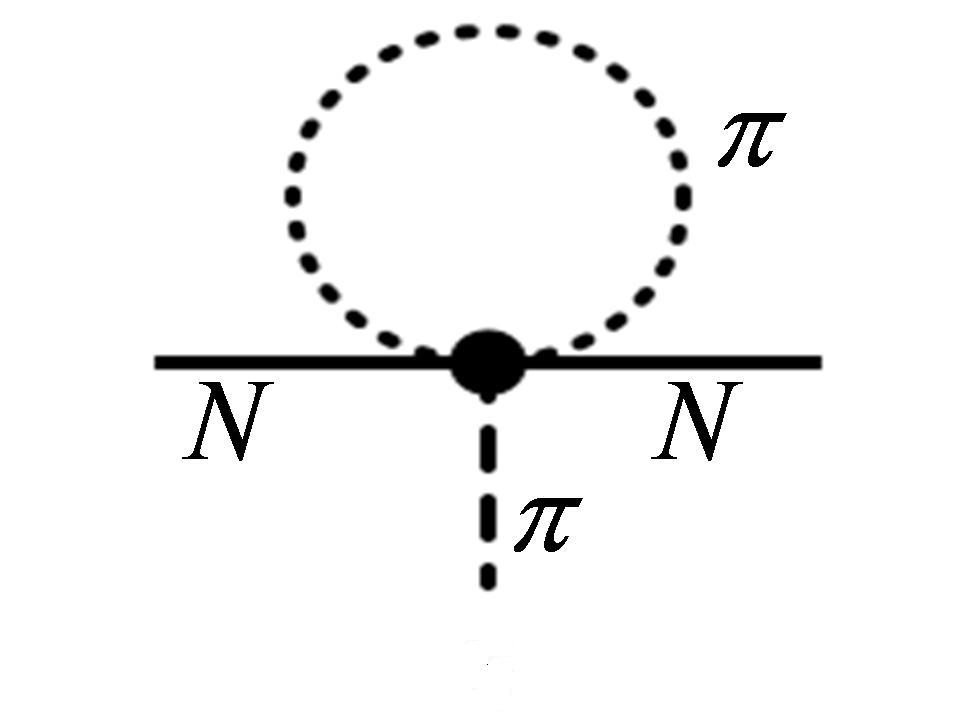}
\caption{(e)}
\end{subfigure}
\begin{subfigure}[b]{0.25\textwidth}
\includegraphics[width=\textwidth]{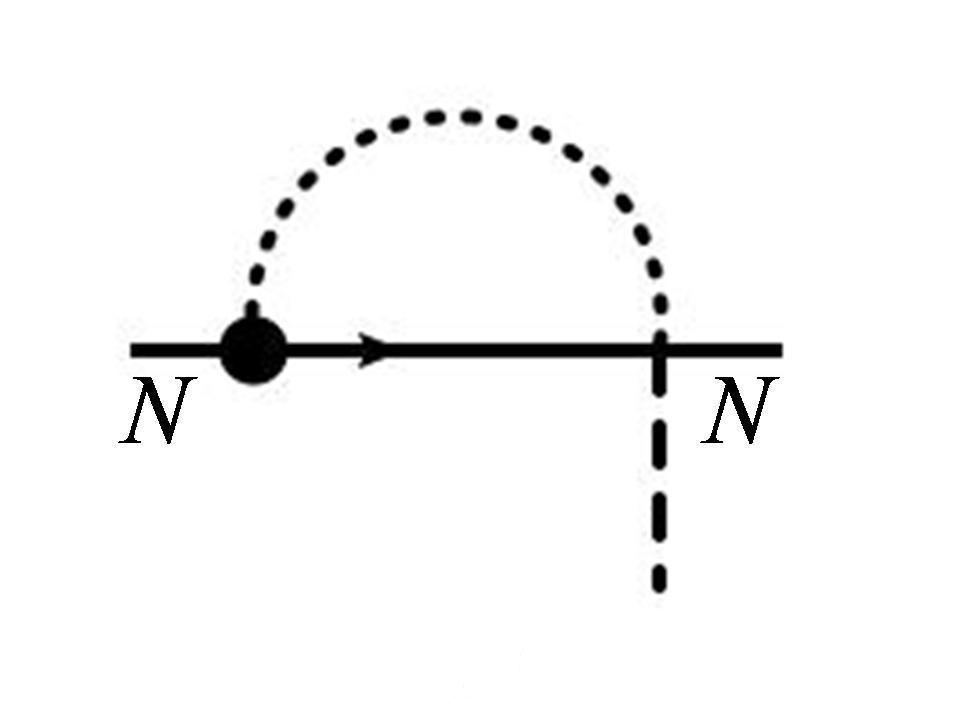}
\caption{(f)}
\end{subfigure}
\begin{subfigure}[b]{0.25\textwidth}
\includegraphics[width=\textwidth]{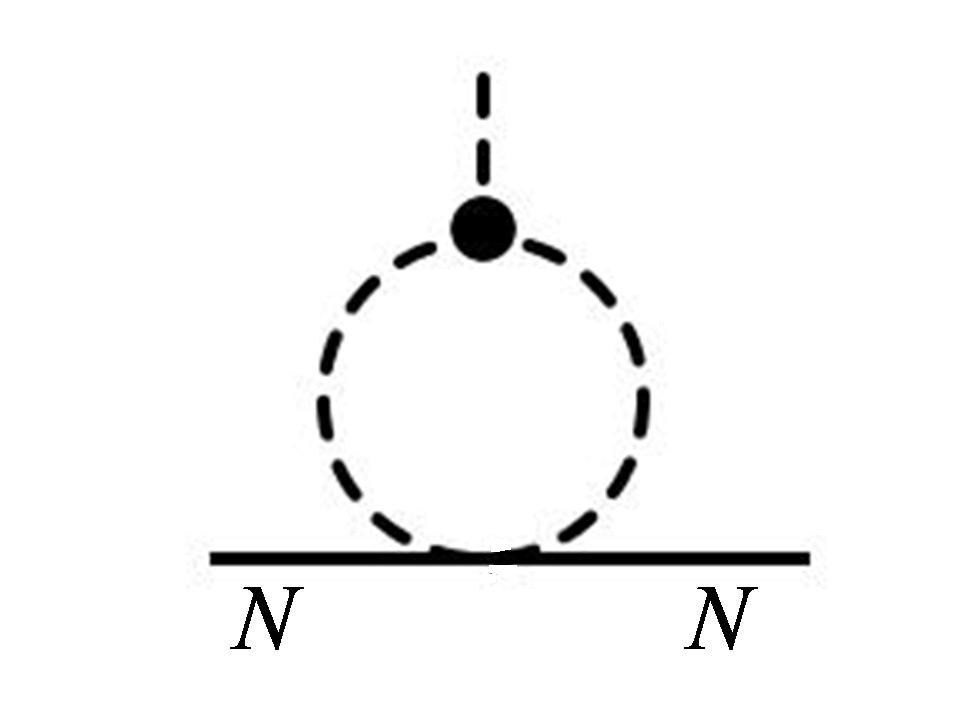}
\caption{(g)}
\end{subfigure}
\begin{subfigure}[b]{0.25\textwidth}
\includegraphics[width=\textwidth]{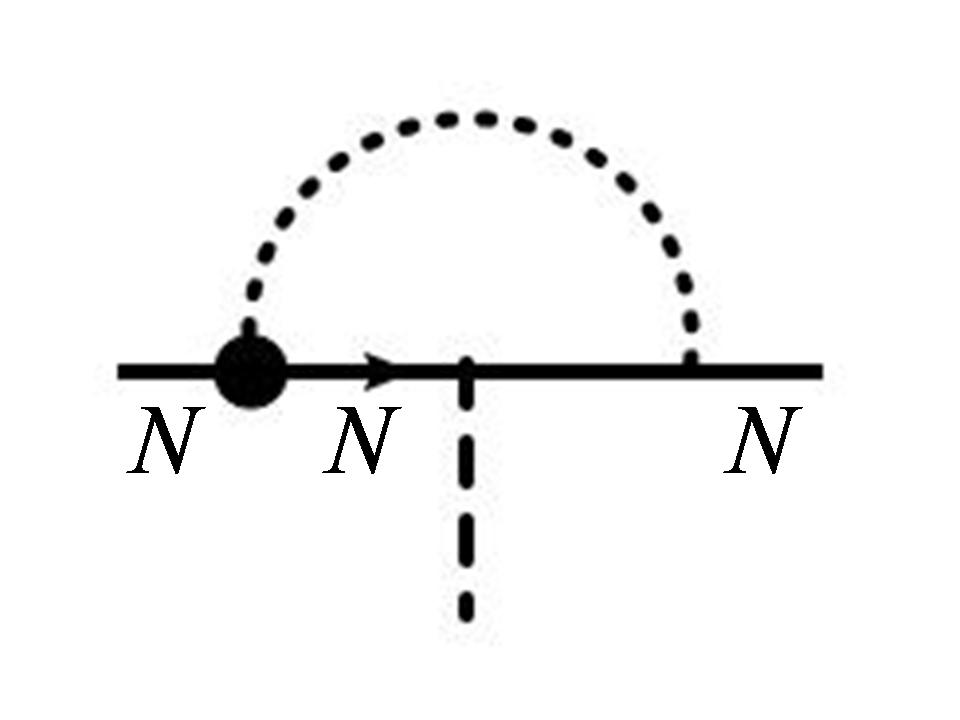}
\caption{(h)}
\end{subfigure}
\caption{\totheleft\label{fig:amputated}Loop diagrams that give rise to PVTV pion-nucleon interactions. Each circular vertex denotes a PVTV interaction vertex defined in Sec. \ref{sec:PVTV}. Diagram
$(g)$ involves $O(E^2)$ $NN\pi\pi$ coupling. The last diagram does
not contribute to $\bar{g}_\pi^{(i)}$ due to the derivative nature of the chiral-invariant pion-nucleon coupling.}
\end{figure}

Terms in the chiral effective Lagrangian at low energy are arranged according to increasing powers of $E$, a typical small energy scale in the theory. A valid power expansion in HBChPT requires $E/(2\pi F_\pi), E/m_N\ll 1$. Following usual conventions \cite{Hemmert:1997ye}, forms such as $\partial_\mu$, $m_\pi$ and the $\Delta-N$ mass splitting $\delta_\Delta$ count as $O(E^1)$ while the light quark mass $m_q$ and other quantities linearly proportional to $m_q$ count as $O(E^2)$ because we shall see later that $m_\pi^2\sim m_q$. Based on such power counting, there are seven types of 1PI diagrams that contribute to the
correction of $\bar{g}_\pi^{(i)}$ up to NNLO and they are summarized in the first seven
diagrams in Fig.\ref{fig:amputated}. Together with the wavefunction
renormalization, they give
\begin{eqnarray}\label{eq:total}
\delta(\bar{g}_\pi^{(0)})_\mathrm{loop}&=&\frac{4g_A^2\bar{g}_{\pi}^{(0)}}{F_\pi^2}I_a-\frac{40g_{\pi
N\Delta}^2\bar{g}_{\Delta\Delta\pi}^{(0)}}{9F_\pi^2}I_d+(\frac{4\bar{g}_{\pi}^{(0)}}{F_\pi^2}+\frac{5\bar{g}_{NN3\pi}^{(0)}}{F_\pi^2})I_e+(Z_N-1)\bar{g}_\pi^{(0)}\nonumber\\
&&+(\sqrt{Z_\pi}-1)_{\mathrm{loop}}\bar{g}_\pi^{(0)}\nonumber\\
\delta(\bar{g}_\pi^{(1)})_\mathrm{loop}&=&-\frac{12g_A^2\bar{g}_{\pi}^{(1)}}{F_\pi^2}I_a+(\frac{16g_{\pi
N\Delta}^2\bar{g}_{\Delta\Delta\pi}^{(1,1)}}{3F_\pi^2}+\frac{8g_{\pi
N\Delta}^2\bar{g}_{\Delta\Delta\pi}^{(1,2)}}{3F_\pi^2})I_d-\frac{40g_A^2\bar{g}_{\pi\pi\pi}^{(1)}}{F_\pi}I_c-\frac{80g_{\pi
N\Delta}^2\bar{g}_{\pi\pi\pi}^{(1)}}{3F_\pi}I_b\nonumber\\
&&+\frac{5\bar{g}_{NN3\pi}^{(1)}}{F_\pi^2}I_e+\frac{5m_\pi^2\bar{g}_{\pi\pi\pi}^{(1)}}{4\pi^2F_\pi^2}((\gamma_1+4\gamma_2)(L'+1)-2\gamma_2)+(Z_N-1)\bar{g}_\pi^{(1)}\nonumber\\
&&+(\sqrt{Z_\pi}-1)_{\mathrm{loop}}\bar{g}_\pi^{(1)}\nonumber\\
\delta(\bar{g}_\pi^{(2)})_\mathrm{loop}&=&\frac{4g_A^2\bar{g}_{\pi}^{(2)}}{F_\pi^2}I_a+(\frac{8g_{\pi
N\Delta}^2\bar{g}_{\Delta\Delta\pi}^{(2,1)}}{9F_\pi^2}+\frac{40g_{\pi
N\Delta}^2\bar{g}_{\Delta\Delta\pi}^{(2,2)}}{27F_\pi^2})I_d-(\frac{2\bar{g}_{\pi}^{(2)}}{F_\pi^2}+\frac{2\bar{g}_{NN3\pi}^{(2,1)}+5\bar{g}_{NN3\pi}^{(2,2)}}{3F_\pi^2})I_e\nonumber\\
&&+(Z_N-1)\bar{g}_\pi^{(2)}+(\sqrt{Z_\pi}-1)_{\mathrm{loop}}\bar{g}_\pi^{(2)}
\end{eqnarray}
with the loop integral functions $\{I_a\}$ defined in Appendix
\ref{sec:loop_function}. Throughout this paper, the UV-divergence of
the loop integral expressed in terms of the quantities $L$ and $L'$
that are defined as
\begin{equation}\label{eq:L}
L'\equiv L+\ln(\frac{\mu}{m_\pi})^2\equiv
\frac{2}{4-d}-\gamma+\ln4\pi+\ln(\frac{\mu}{m_\pi})^2.
\end{equation}

\begin{figure}
\centering
\begin{subfigure}[b]{0.25\textwidth}
\includegraphics[width=\textwidth]{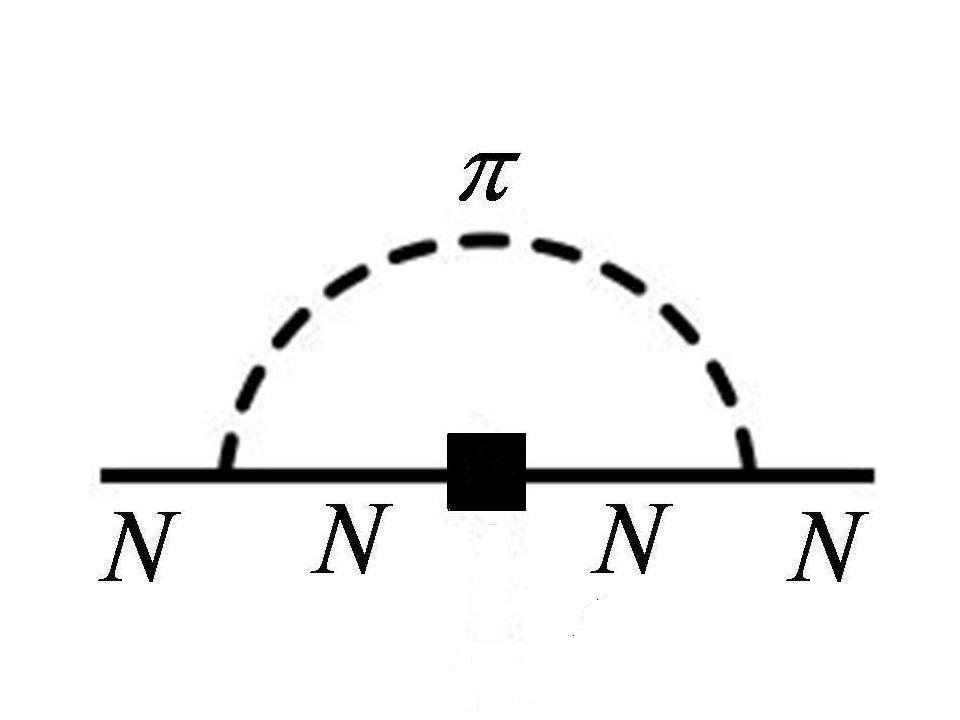}
\caption{\label{fig:2a}(a)}
\end{subfigure}
\begin{subfigure}[b]{0.25\textwidth}
\includegraphics[width=\textwidth]{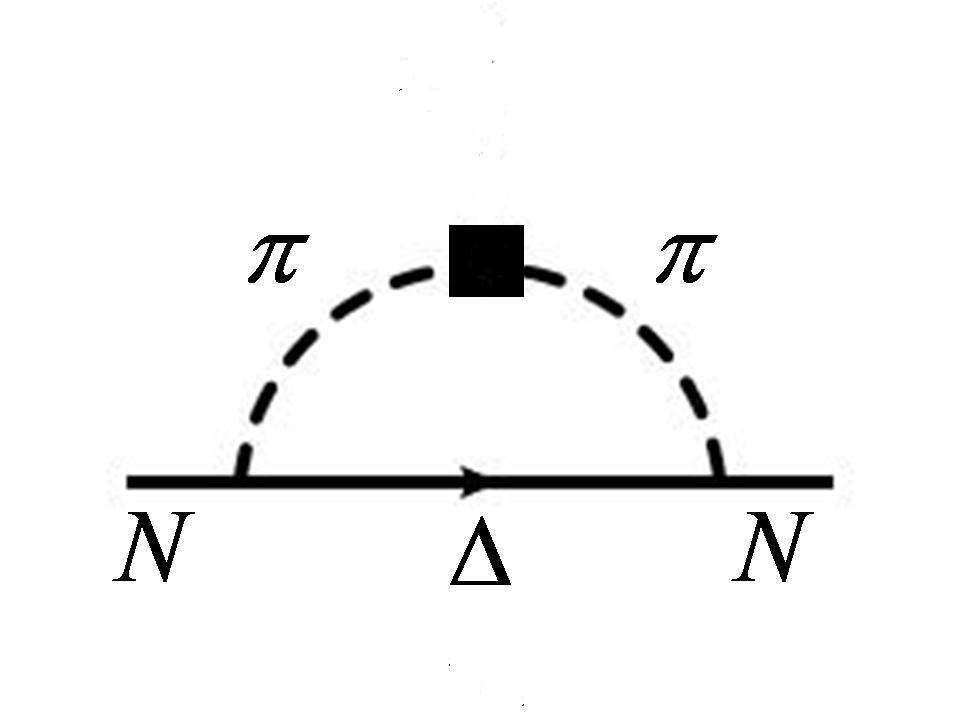}
\caption{(b)}
\end{subfigure}
\begin{subfigure}[b]{0.25\textwidth}
\includegraphics[width=\textwidth]{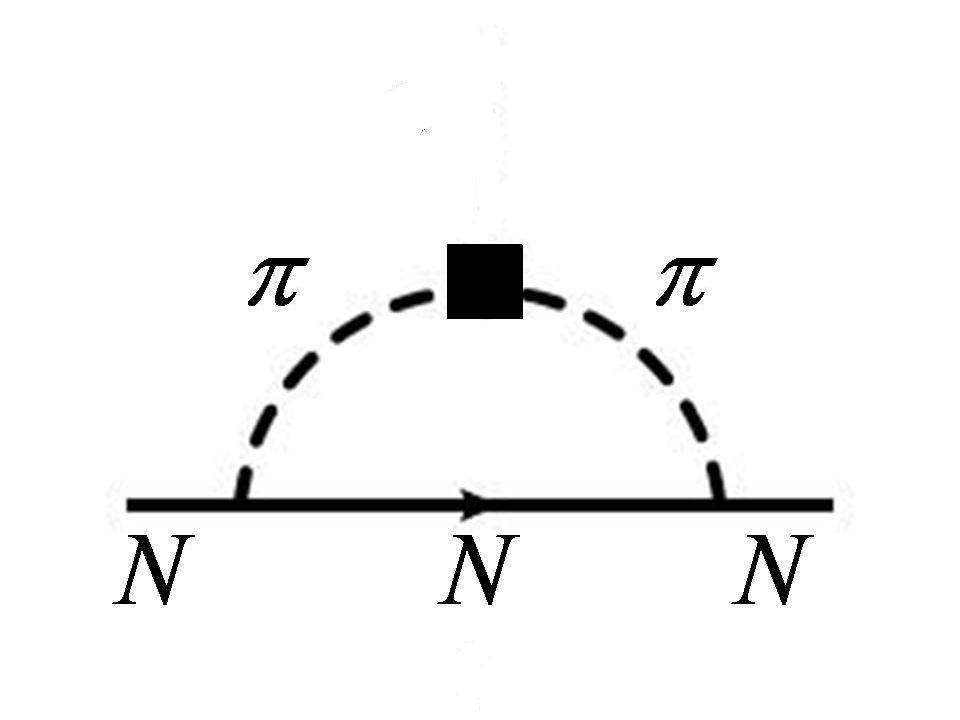}
\caption{(c)}
\end{subfigure}
\begin{subfigure}[b]{0.25\textwidth}
\includegraphics[width=\textwidth]{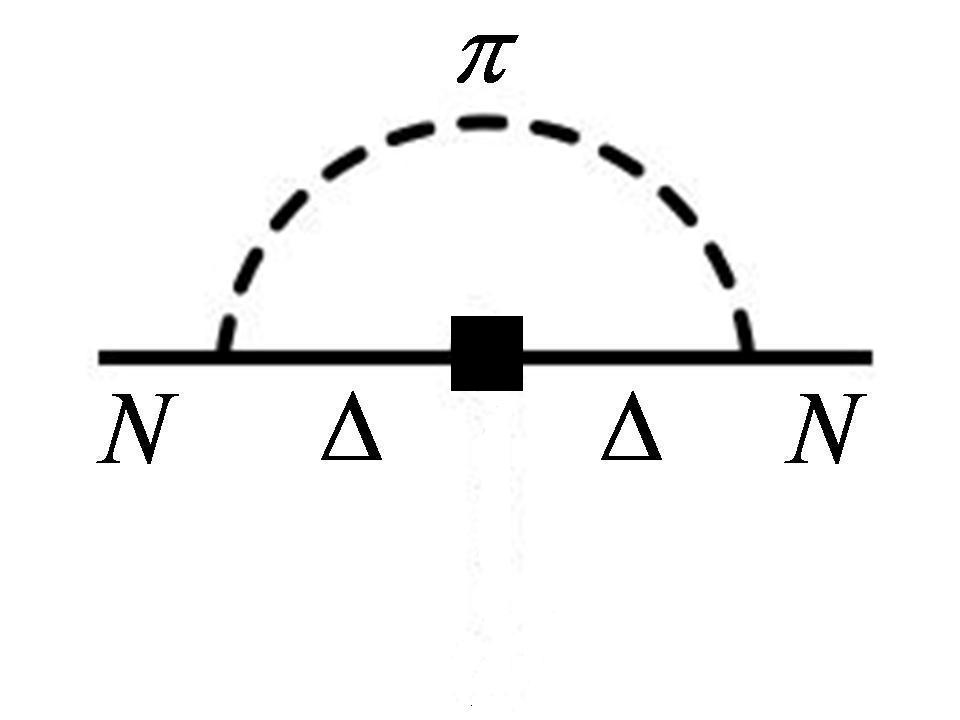}
\caption{(d)}
\end{subfigure}
\begin{subfigure}[b]{0.25\textwidth}
\includegraphics[width=\textwidth]{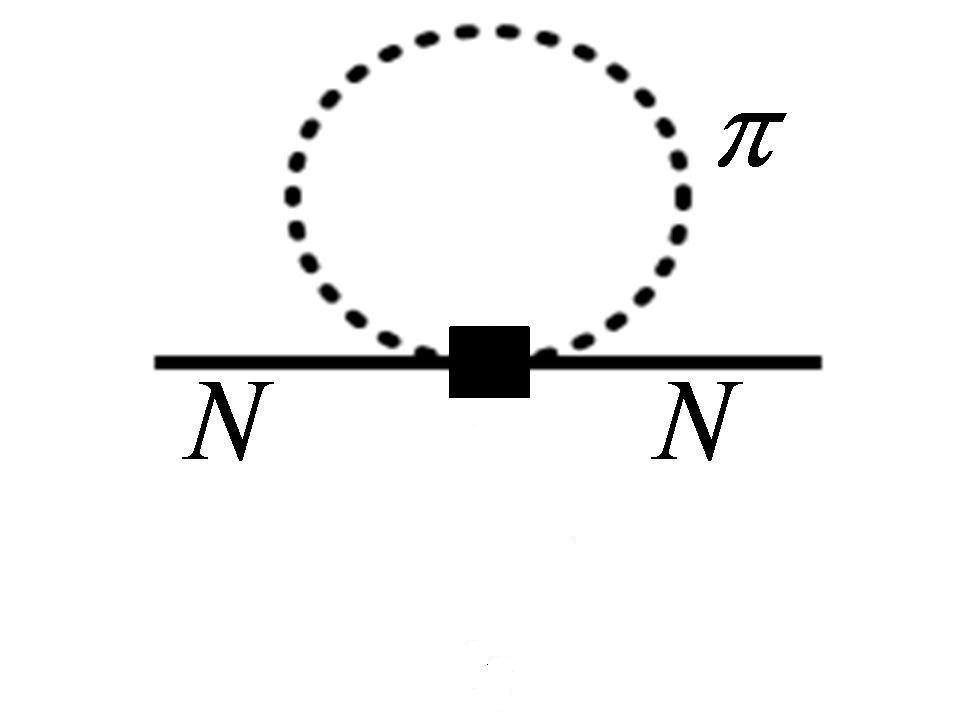}
\caption{(e)}
\end{subfigure}
\begin{subfigure}[b]{0.2\textwidth}
\includegraphics[width=\textwidth]{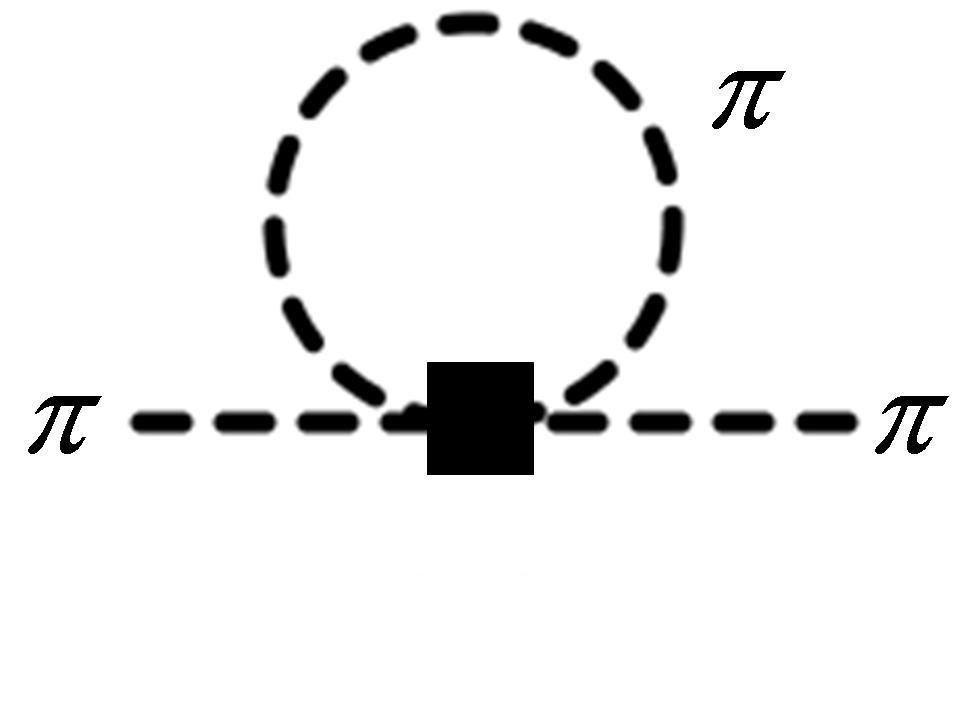}
\caption{(f)}
\end{subfigure}
\begin{subfigure}[b]{0.2\textwidth}
\includegraphics[width=\textwidth]{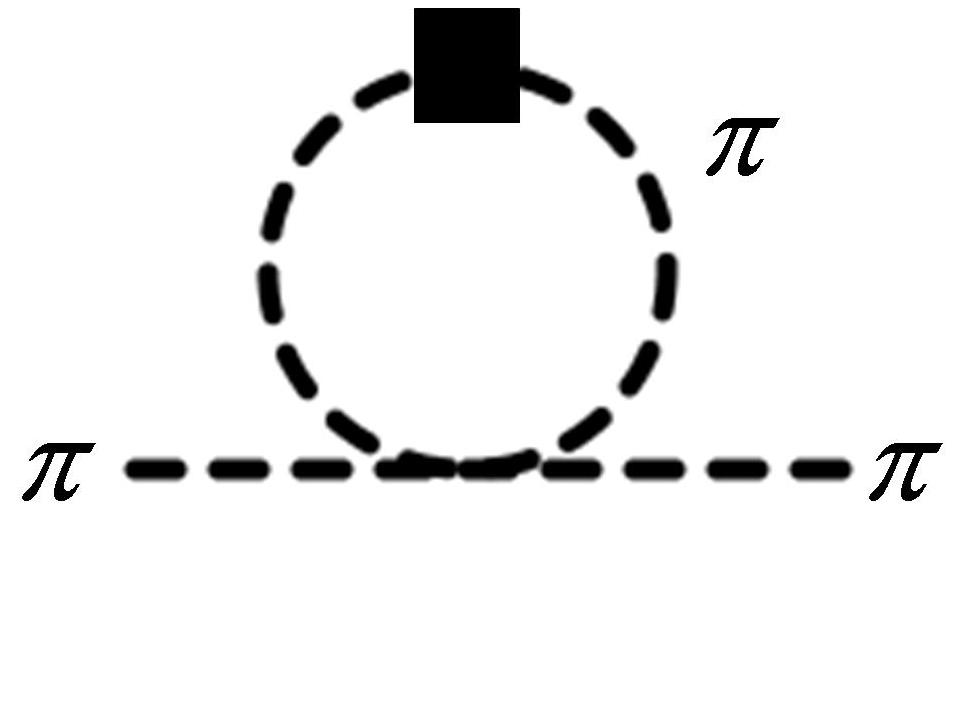}
\caption{(g)}
\end{subfigure}
\begin{subfigure}[b]{0.25\textwidth}
\includegraphics[width=\textwidth]{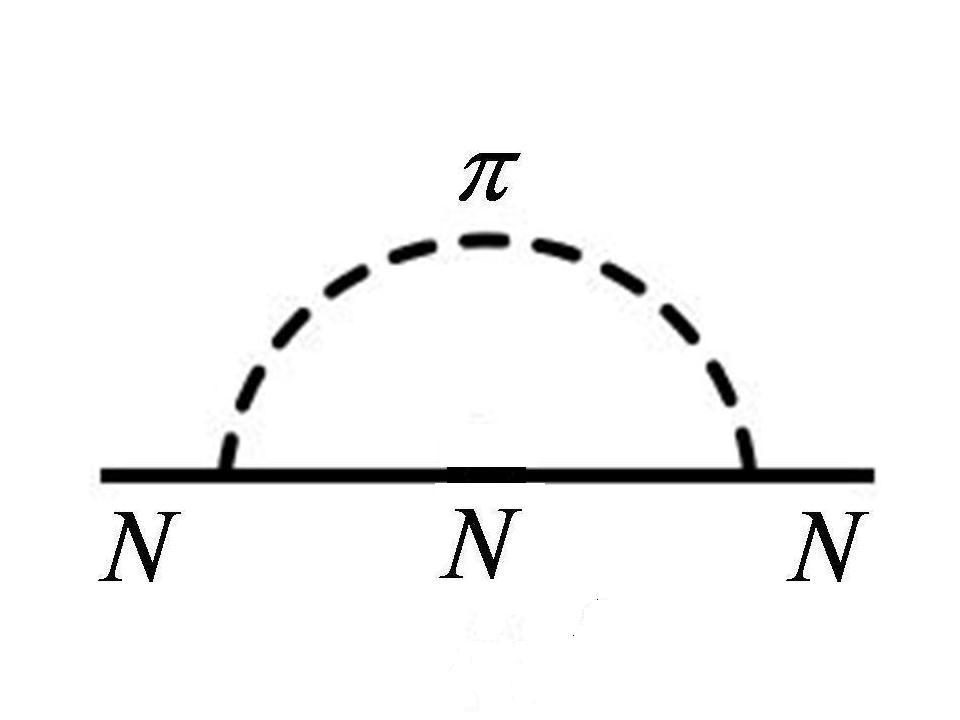}
\caption{\label{fig:2h}(h)}
\end{subfigure}
\begin{subfigure}[b]{0.25\textwidth}
\includegraphics[width=\textwidth]{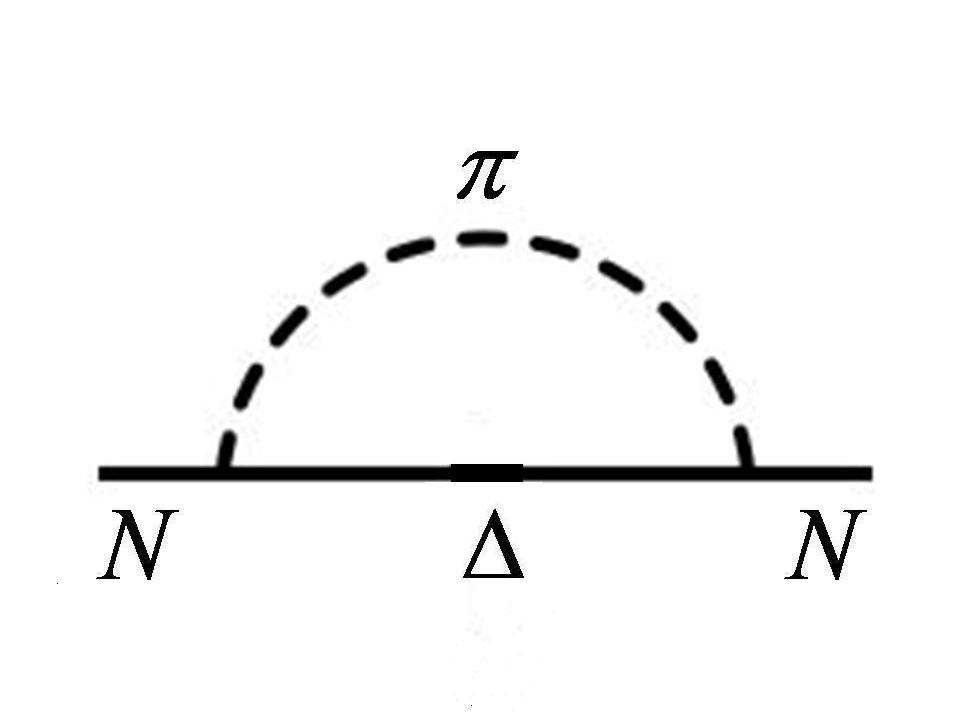}
\caption{\label{fig:2i}(i)}
\end{subfigure}
\begin{subfigure}[b]{0.25\textwidth}
\includegraphics[width=\textwidth]{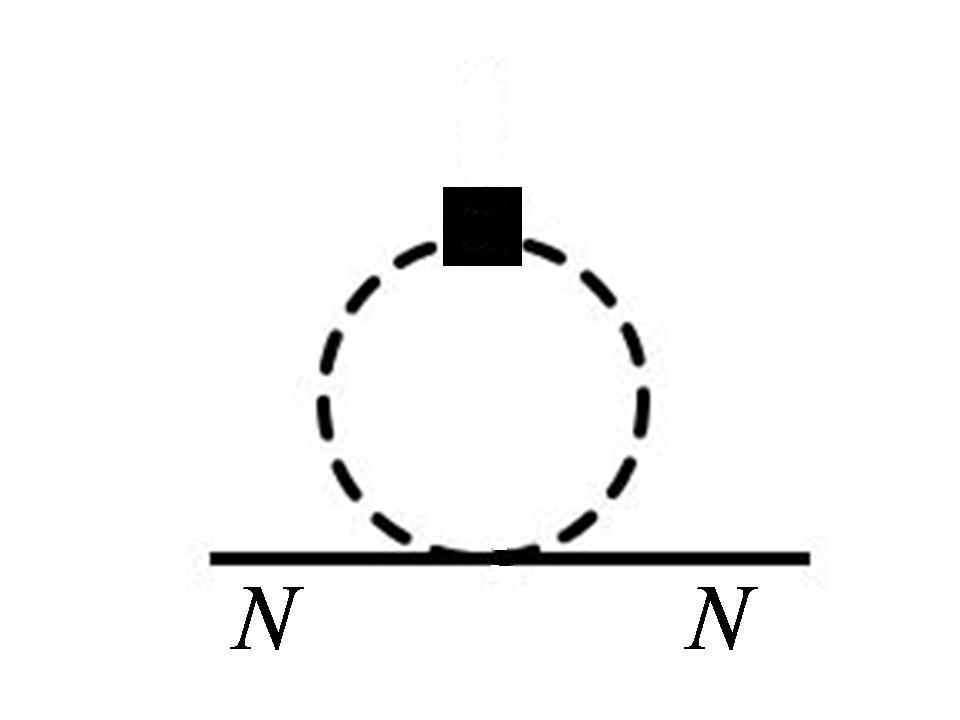}
\caption{(j)}
\end{subfigure}
\begin{subfigure}[b]{0.25\textwidth}
\includegraphics[width=\textwidth]{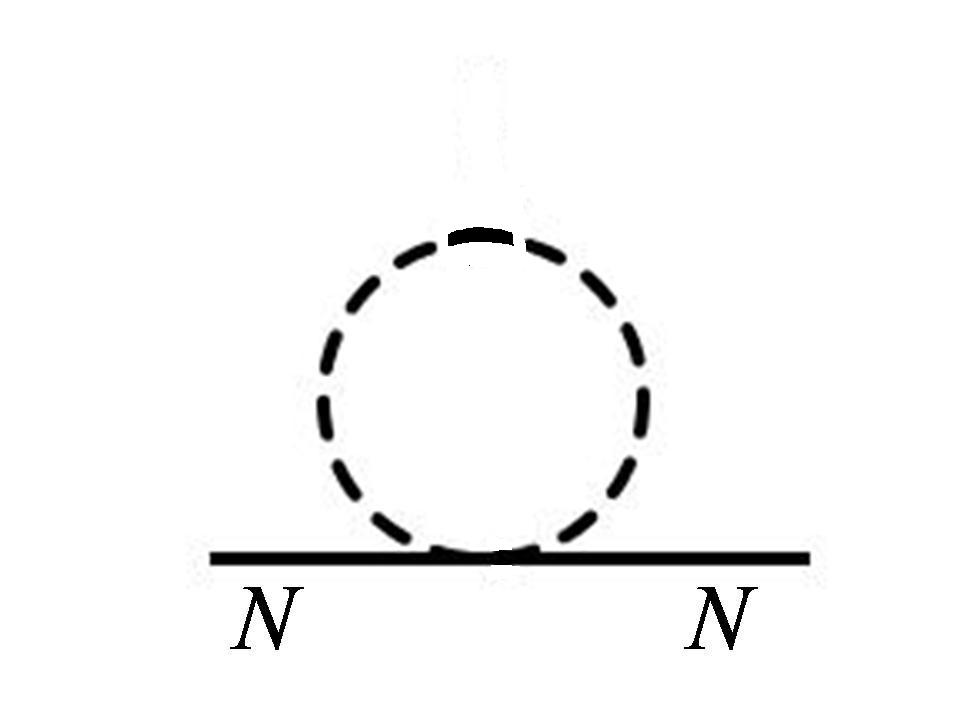}
\caption{\label{fig:2k}(k)}
\end{subfigure}
\caption{\totheleft\label{fig:CBSloop}The non-vanishing one-loop amputated
diagrams contribution to the hadron mass shifts. Each square denotes
a PCTC CSB-interaction defined in Sec. \ref{sec:CSB}. Diagram
$(j)$ and $(k)$ involve $O(E^2)$ $NN\pi\pi$ interaction vertex.}
\end{figure}

Similarly, we shall study the one-loop corrections to the hadron
mass shifts, i.e. $(\Delta m_\pi^2)$, $(\Delta m_N)$ and $(\delta
m_N)$ defined in Eqs. \eqref{eq:defmpishift} and \eqref{eq:sigmaterm}. The relevant 1PI-diagrams are given in Fig.
\ref{fig:CBSloop}. In the nucleon sector, the most general one-loop
corrections to the nucleon sigma term and mass splitting (defined in Eq. \eqref{eq:sigmaterm}) are given
by
\begin{eqnarray}\delta(\Delta m_N)_{i,\mathrm{loop}}&=&(Z_N-1)(\Delta m_N)_i-\frac{12(\Delta
m_N)_ig_A^2}{F_\pi^2}I_a-\frac{8g_{\pi N\Delta}^2(\Delta
m_\Delta)_i}{F_\pi^2}I_d+\frac{3(g_{NN\pi\pi}^{(0)})_i}{F_\pi}I_e\nonumber\\
&&+\frac{12(\Delta m_\pi^2)_ig_A^2}{F_\pi^2}I_c+\frac{8g_{\pi
N\Delta}^2(\Delta
m_\pi^2)_i}{F_\pi^2}I_b-\frac{3m_\pi^2(\Delta m_\pi^2)_i}{8\pi^2F_\pi^3}((\gamma_1+4\gamma_2)(L'+1)-2\gamma_2)\nonumber\\
&&+\delta_{iq}\{\frac{12g_A^2}{F_\pi^2}I_f+\frac{8g_{\pi N\Delta}^2}{F_\pi^2}I_g-\frac{3m_\pi^4}{16\pi^2F_\pi^3}((\gamma_1+4\gamma_2)(L'+1)+\frac{1}{2}\gamma_1)\}\nonumber\\
\delta(\delta m_N)_{i,\mathrm{loop}}&=&(Z_N-1)(\delta
m_N)_i+\frac{4g_A^2(\delta m_N)_i}{F_\pi^2}I_a-\frac{40g_{\pi
N\Delta}^2(\delta
m_\Delta)_i}{9F_\pi^2}I_d\nonumber\\
&&+\frac{6(g_{NN\pi\pi}^{(1,1))})_i+2(g_{NN\pi\pi}^{(1,2)})_i}{F_\pi}I_e.\end{eqnarray} Here the subscript $i$ denotes the specific choice
of effective operator that induces the spurion field, e.g. $i=q$
(quark mass), $c$ (quark cMDM/cEDM) and $4q$ (LR4Q). In the pion
sector, we will concentrate on the isospin-singlet pion mass shift. For the case of $\theta$-term and dipole operators, this is the only pion mass shift that comes into play. For the case of LR4Q, although $I=2$ pion mass shift is also generated, but it is not independent from its isosinglet counterpart. Therefore, we are allowed to choose only the $I=0$ pion mass shift to enter the matching relations. Its loop correction reads:
\begin{equation}\label{eq:mpiSU2}
\delta(\Delta m_\pi^2)_{i,\mathrm{loop}}=-\frac{11(\Delta
m_\pi^2)_im_\pi^2}{24\pi^2
F_\pi^2}(L'+\frac{8}{11})+\frac{5m_\pi^2(g_{4\pi}^{(0)})_i}{4\pi^2}(L'+1).
\end{equation}

\section{\label{sec:main}Discussion of the matching relations of $\bar{g}_\pi^{(i)}$}

With all the preparations in Sec.
\ref{sec:spurion}-\ref{sec:loopcorrection} we are now in a position
to discuss the matching relations between the $\bar{g}_\pi^{(i)}$ and the
hadron mass shifts induced by various effective operators. It is
obvious that a necessary condition for these matching relations to exist is
that the underlying operator should possess both PCTC and PVTV components simultaneously. This simple observation greatly
reduces the amount of relevant operators to four types, namely (1)
the complex quark mass operator, (2) the dipole-like operators, (3)
the LR4Q operator and (4) the two chirally invariant
$(\bar{L}R)(\bar{L}R)$-type operators. We shall apply our general
formalism in Sec. \ref{sec:spurion}-\ref{sec:loopcorrection} to
study the influences of these four types of operators separately.

\subsection{\label{sec:theta}Quark mass and QCD $\theta$-term}

We start by reviewing previous studies of the P and T-violation
generated by the QCD $\theta$-term (see, e.g. Ref. \cite{Mereghetti:2010tp,Ottnad:2009jw,deVries:2015una} and references therein). The QCD Lagrangian
with a non-zero $\theta$-term takes the following form:
\begin{equation}
\mathcal{L}(\{q_{iR},q_{iL}\})=\sum_i[\bar{q}_{i}i\slashed{D}q_{i}-\bar{q}_{iR}M_0q_{iL}-\bar{q}_{iL}M_0q_{iR}]-\frac{1}{4}G^a_{\mu\nu}G^{a\mu\nu}-\bar{\theta}\frac{g_s^2}{32\pi^2}G^a_{\mu\nu}\tilde{G}^{a\mu\nu}
\end{equation}
where $M_0=\mathrm{diag}(m_1\:m_2\:...)$ is the real quark mass matrix.
Due to the axial anomaly, if we perform an axial rotation
$q_i\rightarrow e^{i\theta_i\gamma_5}q_i$ to
the quark field $q_i$, the Lagrangian will change
as
\begin{equation}
\mathcal{L}(\{q_{iR},q_{iL}\})\rightarrow\mathcal{L}(\{e^{i\theta_i}q_{iR},e^{-i\theta_i}q_{iL}\})+\sum_i\theta_i\frac{g_s^2}{16\pi^2}G^a_{\mu\nu}\tilde{G}^{a\mu\nu}.
\end{equation}
Therefore, for a two-flavor QCD with $Q\equiv(u\:d)^T$, we may
perform the following rotation to eliminate the $\theta$-term:
\begin{equation}Q\rightarrow e^{\frac{i}{2}(\frac{\bar{\theta}}{2}-\alpha\tau_3)\gamma_5}Q
\end{equation}
Here $\alpha$ is so far a free parameter that will be fixed later
by the requirement of vacuum stability. The resulting Lagrangian
looks like
\begin{equation}\mathcal{L}=\bar{Q}i\slashed{D}Q-\bar{Q}_RX_qQ_L-\bar{Q}_LX_q^\dagger
Q_R-\frac{1}{4}G^a_{\mu\nu}G^{a\mu\nu}\end{equation} where now $X_q$
is the complex quark mass matrix that acts as a spurion as described
in Sec. \ref{sec:spurion}:
\begin{equation}\label{eq:xq}X_q=\bar{m}e^{-i\frac{\bar{\theta}}{2}}\{\cos\alpha-i\varepsilon\sin\alpha+(-\varepsilon\cos\alpha+i\sin\alpha)\tau_3\}\end{equation}
with $\bar{m}=(m_u+m_d)/2$ and $\varepsilon=(m_d-m_u)/(m_u+m_d)$. Throughout this paper we shall take $\bar{m}\approx 3.6$MeV and $\varepsilon\approx0.33$ from lattice calculation \cite{Aoki:2013ldr} whenever their values are needed.

\subsubsection{\label{sec:treetheta}Tree-level matching} Now we would like to generate PVTV
operators in the chiral Lagrangian by appropriate insertions of
$X_q$. In the pure pionic sector, the leading operator at $O(E^2)$ with an
$X_q$-insertion is:
\begin{eqnarray}\label{eq:thetapion}\frac{F_0^2 B_0}{8}\mathrm{Tr}[X_qU^\dagger+UX_q^\dagger]&=&\frac{F_0^2B_0\bar{m}}{2}(\cos\alpha\cos\frac{\bar{\theta}}{2}-\varepsilon\sin\alpha\sin\frac{\bar{\theta}}{2})(1-\frac{2\vec{\pi}^2}{F_0^2}+...)\nonumber\\
&&+F_0^2B_0\bar{m}(\sin\alpha\cos\frac{\bar{\theta}}{2}+\varepsilon\cos\alpha\sin\frac{\bar{\theta}}{2})(1-\frac{2\vec{\pi}^2}{3F_0^2}+...)\frac{\pi_0}{F_0}.\end{eqnarray}
where $F_0$ is just $F_\pi$ in the chiral limit.

Note that the existence of the $\pi_0$ term makes the vacuum
unstable as one may lower the energy of the system indefinitely by keep creating neutral pions from the vacuum. To avoid that, we simply impose the ``vacuum alignment"
condition that says the value of $\alpha$ should be chosen
such that the $\pi_0$ term vanishes
\cite{Crewther:1979pi,Baluni:1978rf}. For the case that the
$\theta$-term is the only source of T-violation, the vacuum
alignment condition is simply
$\alpha\approx-\varepsilon\bar{\theta}/2$ assuming $\bar{\theta}$ is
small. The complex quark mass matrix $X_q$ then turns into:
\begin{equation}
X_q=M_0-i\frac{\bar{m}}{2}(1-\varepsilon^2)\bar{\theta}\label{eq:Xqafteralignment}.
\end{equation}
After imposing the alignment condition we obtain
$m_\pi^2=2B_0\bar{m}$ where $m_\pi^2$ is defined as the squared mass
of charged pions $\pi_\pm$ (which leads to $B_0\approx2.7$GeV if we take the lattice value for $\bar{m}$ \cite{Aoki:2013ldr}). The mass splitting between charged and
neutral pion occurs at $O(E^4)$ level. Also, note that the
requirement of vacuum alignment kills the $\pi_0$ term as well as
all other terms that have an odd number of pions. Therefore the term in
Eq. \eqref{eq:thetapion} does not give any T-violating operator.
Instead, we obtain an isospin-invariant mass term for the pion
triplet. Also, an important feature one observes is that, after
vacuum alignment the PVTV term in $\tilde{X}_{q+}$ is an
isoscalar (recall the definition of $\tilde{X}_{\pm}$ in Eq. \eqref{eq:Xtildepm} and the subscript $q$ denotes the contribution from the complex quark mass matrix). That implies, at the leading order of $m_\pi^2$
expansion, the PVTV interactions induced by the $\theta$-term
are all isoscalars.

In the nucleon sector, the leading term that gives a non-zero
T-violating effect is\footnote{The operator coefficients are related to those in Bernard, Kaiser and Mei\ss{}ner \cite{Bernard:1995dp} by $c_1=2B_0c_5^{\mathrm{BKM}}$, $c_1'=2B_0c_1^{\mathrm{BKM}}$.}:
\begin{eqnarray}c_1\bar{N}\tilde{X}_{q+}N+c_1'\mathrm{Tr}[\tilde{X}_{q+}]\bar{N}N&=&2\bar{m}(c_1+2c_1')(\bar{N}N+...)-2\bar{m}\varepsilon c_1(\bar{N}\tau_3N+...)\nonumber\\
&&-\frac{2\bar{m}(1-\varepsilon^2)\bar{\theta}c_1}{F_0}(1-\frac{2\vec{\pi}^2}{3F_0^2}+...)\bar{N}\vec{\tau}\cdot\vec{\pi}N.
\label{eq:OE2nucleon}\end{eqnarray}
 where $+\cdots$ denotes terms additional  pion fields. This simply corresponds to the pion-nucleon Lagrangian with chiral index $\Delta=1$ by Mereghetti et al \cite{Mereghetti:2010tp}.
The first term contributes to the nucleon sigma term $(\Delta m_N)_q=2\bar{m}(c_1+2c_1')$ while the
second term contributes to the nucleon mass splitting $(\delta m_N)_q=-4\bar{m}\varepsilon c_1$. The third
term contributes to $\bar{g}_\pi^{(0)}$ and
$\bar{g}_{NN3\pi}^{(0)}$ with $\bar{g}_\pi^{(0)}=-2\bar{m}(1-\varepsilon^2)\bar{\theta}c_1/F_0$ and $\bar{g}_{NN3\pi}^{(0)}=-\frac{2}{3}\bar{g}_\pi^{(0)}$ (it is interesting to note that, in the SO(4)-representation of ChPT, e.g. in Ref. \cite{Mereghetti:2010tp}, the relation between $\bar{g}_{NN3\pi}^{(0)}$ and $\bar{g}_\pi^{(0)}$ is apparently different: $\bar{g}_{NN3\pi}^{(0)}=-\bar{g}_\pi^{(0)}$. However one is able to show their equivalence using the equation of motion (EOM) 
. Now since both $(\delta m_N)_q$ and $\bar{g}_\pi^{(0)}$ depend linearly on $c_1$, one may relate them as:
\begin{equation}\label{eq:thetamatching}
F_\pi\bar{g}_\pi^{(0)}=\frac{1-\varepsilon^2}{2\varepsilon}(\delta m_N)_q\bar{\theta}.
\end{equation}
Notice that we have made use of the fact that $F_\pi=F_0$ at leading order. This replacement is crucial so that the same equation holds even when higher-order corrections to the pion decay constant are included, as we shall discuss later. Eq. \eqref{eq:thetamatching} is exactly the
tree-level matching relation between $\bar{g}_\pi^{(0)}$ and $(\delta
m_N)_q$. The same procedure is used to determine all other matching relations at tree level so we may skip the intermediate steps when we introduce them later.

In the $\Delta$-sector we have an analogous leading term that gives
T-violation:
\begin{equation}c_2\bar{T}^i_\mu\tilde{X}_{q+}T^{i\mu}+c_2'\mathrm{Tr}[\tilde{X}_{q+}]\bar{T}^i_\mu T^{i\mu}\end{equation} and we have an analogous coefficient
matching:
\begin{eqnarray}
F_\pi\bar{g}_{\Delta\Delta\pi}^{(0)}
&=&\frac{1-\varepsilon^2}{2\varepsilon}(\delta
m_\Delta)_q\bar{\theta}\label{eq:thetamatchingdelta}
\end{eqnarray}

\subsubsection{Loop correction}

Next we shall consider the one-loop correction to the left-hand-side (LHS) and right-hand-side (RHS) of the tree-level matching relation \eqref{eq:thetamatching}. The loop correction to $F_\pi\bar{g}_\pi^{(0)}$ and $(\delta m_N)_q$ are given in Eq. \eqref{eq:loopgpibarq} and \eqref{eq:loopdmNq} in Appendix \ref{sec:detail} respectively. The former is expressed in terms of $\bar{g}_\pi^{(0)}$ and $\bar{g}_{\Delta\Delta\pi}^{(0)}$ that can be related to $(\delta m_N)_q$ and $(\delta m_\Delta)_q$ by Eq. \eqref{eq:thetamatching} and \eqref{eq:thetamatchingdelta} respectively because any correction is of higher order in power counting. With these we obtain
\begin{equation}\delta(F_\pi\bar{g}_\pi^{(0)})_{\mathrm{loop}}=\frac{1-\varepsilon^2}{2\varepsilon}\bar{\theta}\delta(\delta
m_N)_{q,\mathrm{loop}},\end{equation} 
{\em i.e.}, the tree-level matching
formula \eqref{eq:thetamatching} for $\bar{g}_\pi^{(0)}$ is
preserved at one-loop.

\subsubsection{\label{sec:thetacounter}LECs and the higher-order matching formula}

Higher-order terms in the chiral Lagrangian must be introduced to cancel the UV-divergence in the
loop corrections and make the full expression $\mu$-independent.
Apart from the baryon wavefunction and $F_\pi$ renormalization that
are well-known, for the case of the $\theta$-term CPV source we only need the
$O(E^4)$ terms that involve two insertions of
$\tilde{X}_{q\pm}$. Such terms in the pure pionic sector are
introduced in Appendix \ref{sec:ordloop}. In the nucleon sector we
have:
\begin{eqnarray}
\mathcal{L}_N^{O(E^4)}&=&F_\pi^{-3}B_0^2\{f_1\mathrm{Tr}[\tilde{X}_{q+}^2]\bar{N}N+f_2\mathrm{Tr}[\tilde{X}_{q+}]\bar{N}\tilde{X}_{q+}N+f_3\bar{N}\tilde{X}_{q+}^2N\nonumber\\
&&+f_4\mathrm{Tr}[\tilde{X}_{q-}^2]\bar{N}N+f_5\mathrm{Tr}[\tilde{X}_{q-}]\bar{N}\tilde{X}_{q-}N+f_6\bar{N}\tilde{X}_{q-}^2N\}+...\label{eq:OE4nucleon}\end{eqnarray}

Details of the $O(E^4)$ contribution to $F_\pi\bar{g}_\pi^{(i)}$ and the hadron masses are summarized in Appendix \ref{sec:thetadetail}. After some rearrangement we are able to match the final result with Eq. \eqref{eq:match_deviate} as:
\begin{equation}
F_\pi\bar{g}_\pi^{(0)}=\frac{1-\varepsilon^2}{2\varepsilon}\bar{\theta}(\delta
m_N)_q(1+\delta_{\mathrm{LEC}}^{(0)})
\end{equation} 
where the relative deviation from the LECs is given by:
\begin{equation}
\delta_{\mathrm{LEC}}^{(0)}=-\frac{4m_\pi^4\varepsilon}{(\delta m_N)_qF_\pi^3}(f_5^r+f_6^r)-\frac{64m_\pi^2\varepsilon^2}{F_\pi^2}(2L_7^r+L_8^r)
\end{equation}
Throughout this paper, we use the superscript ``$r$" to represent a
renormalized quantity of which the infinite value $L+1$ is
subtracted from the corresponding bare quantity following the
Gasser-Leutwyler subtraction scheme \cite{Gasser:1983yg}, i.e. a
bare quantity $A$ and its renormalized value $A^r$ are related by
$A=A^r+B(L+1)$ where $B$ is a finite number. The absence of $\delta_\mathrm{loop}^{(0)}$
shows that the LO-matching formula is modified at
higher order but the modification is analytic in the quark masses ({\em i.e.}, not logarithmic with respect to pion masses). This relation
has already been studied under the $\mathrm{SU}(3)$ version of ChPT in Ref.
\cite{deVries:2015una}.

One may perform a quick estimation of the size of $\delta_{\mathrm{LEC}}^{(0)}$ using lattice results and dimensional analysis arguments. First, for the contribution from $L_i^r$, we note that it contains a large prefactor 64 but is also suppressed by the square of the isospin breaking parameter $\varepsilon$. Furthermore, we have $L_{7,8}^r\sim 10^{-3}$ from meson data fits \cite{Bijnens:2014lea}. That gives a contribution of order $10^{-3}$ to $\delta_{\mathrm{LEC}}^{(0)}$ that is very small. On the other hand, the impact of $f_i^r$ is less transparent because the sizes of $f_{5,6}^r$ are not well-determined. Here we shall estimate their order of magnitude based on chiral power counting. For instance, one may compare the contribution of $O(p^2)$ (i.e. linear to $\bar{m}$) and $O(p^4)$ (i.e. quadratic to $\bar{m}$) contribution to the nucleon mass; they are proportional to $\bar{m}c_i$ and $F_\pi^{-3} B_0^2\bar{m}^2 f_i$ respectively, as one is able to read off from Eqs. \eqref{eq:OE2nucleon} and \eqref{eq:OE4nucleon}\footnote{The factor $F_\pi^{-3}B_0^2$ is just due to the definition of the coefficients of the $O(p^4)$ counterterms $f_i$ so that they are dimensionless.}. Chiral power-counting suggests that the latter should be suppressed with respect to the former by a factor of order  $(m_\pi/2\pi F_\pi)^2$. This implies
\begin{equation}F_\pi^{-3}B_0^2\bar{m}^2f_i\sim (m_\pi/2\pi F_\pi)^2c_i\bar{m}
\end{equation} 
which leads to $f_i\sim 3\times10^{-3}$ for $c_i\sim1$. This gives $\delta_{\mathrm{LEC}}^{(0)}\sim 0.1$ which is a 10\% correction to the tree-level matching relation.

\subsection{\label{sec:cEDM}The dipole-like operators}

Next we shall study the effect of the leading flavor-diagonal
dipole-like operator namely the quark cMDM/cEDM.  As discussed in
Sec. \ref{sec:spurion}, the spurion for this operator is simply
identical to the one for the complex quark mass. The Lagrangian in
the quark-gluon level reads
\begin{eqnarray}
\mathcal{L}&=&\sum_{q=u,d}g_s\tilde{d}^M_q\bar{q}\sigma^{\mu\nu}\frac{\lambda^a}{2}G^a_{\mu\nu}q-i\sum_{q=u,d}g_s\tilde{d}_q\bar{q}\sigma^{\mu\nu}\gamma_5\frac{\lambda^a}{2}G^a_{\mu\nu}q\nonumber\\
&=&g_s\bar{Q}_R\sigma^{\mu\nu}\frac{\lambda^a}{2}G^a_{\mu\nu}X_cQ_L+h.c\end{eqnarray}
where $\tilde{d}^M_q$ and $\tilde{d}_q$ are the cMDM and cEDM of the
quark $q$ respectively. The matrix $X_c$ acts as the spurion for
cMDM/cEDM as described in Sec. \ref{sec:spurion}, and is given by
\begin{equation}X_c=\frac{1}{2}(\tilde{d}^M_0+i\tilde{d}_0)+\frac{1}{2}(\tilde{d}^M_1+i\tilde{d}_1)\tau_3\end{equation}
where
$\tilde{d}_0(\tilde{d}_0^M)=\tilde{d}_u(\tilde{d}_u^M)+\tilde{d}_d(\tilde{d}_d^M)$
and
$\tilde{d}_1(\tilde{d}_1^M)=\tilde{d}_u(\tilde{d}_u^M)-\tilde{d}_d(\tilde{d}_d^M)$
are the isoscalar and isovector cEDM (cMDM) respectively.

\subsubsection{Tree-level matching} The implementation of the spurion $X_c$ into the chiral Lagrangian
works exactly in exactly the same way as the complex quark mass. In
the pionic sector the only operator at lowest order is:
\begin{eqnarray}\label{eq:cEDMpion}\beta F_0^5 \mathrm{Tr}[X_cU^\dagger+UX_c^\dagger]&=&2\beta F_0^5\tilde{d}_0^M(1-\frac{2\vec{\pi}^2}{F_0^2}+...)+4\beta F_0^5\tilde{d}_1(1-\frac{2\vec{\pi}^2}{3F_0^2}+...)\frac{\pi_0}{F_0}\end{eqnarray}
where $\beta$ is a dimensionless constant. This operator will again
induce a pion tadpole term that makes the vacuum unstable. In order
to cancel this term, we have to include the term with an
$X_q$-insertion given in Eq. \eqref{eq:thetapion} and choose an
appropriate value for the free parameter $\alpha$ to eliminate the
$\pi_0$ term induced by Eq. \eqref{eq:cEDMpion}. Assuming no
$\theta$-term, we obtain a pion mass shift $(\Delta m_\pi^2)_{c}=8\beta
F_0^3\tilde{d}_0^M$   as well as the vacuum alignment condition
\begin{equation}
\alpha\approx(-4\beta F_0^3\tilde{d}_1)/(B_0\bar{m})=-(\Delta m_\pi^2)_c\tilde{d}_1/(m_\pi^2\tilde{d}_0^M)\ \ \ 
\end{equation}. 
The non-zero value of $\alpha$ leads to an
interesting consequence, namely: in order to study the effect of
T-violation induced by the cEDM, it is not enough to consider only
terms with $X_c$ insertions. One needs to include all the terms with
$X_q$ insertions as given in Sec. \ref{sec:theta} because the quark
mass picks up a complex phase $\alpha$ even without the existence of
a $\theta$-term. Also, $\alpha$ is related but should not be confused with the so-called ``induced $\theta$-term" introduced by Pospelov and Ritz \cite{Pospelov:2005pr}. To see their relation, we take the complex quark mass matrix $X_q$ defined in Eq. \eqref{eq:xq} (without $\bar{\theta}$ just for simplicity) and expand it to the first power in $\alpha$. After plugging in the explicit expression of $\alpha$ one immediately sees that $X_q$ takes the following form 
\begin{equation}\label{eq:Xqinduced}X_q=M_0+i\frac{\bar{m}}{2}(1-\varepsilon^2)\bar{\theta}_{\mathrm{ind}}-i\frac{2(\Delta m_\pi^2)_c}{\tilde{d}_0^M}\frac{\bar{m}}{m_\pi^2}\tilde{\mathrm{D}}
\end{equation}
where $M_0$ is the real quark mass matrix and $\tilde{\mathrm{D}}=\mathrm{diag}(\tilde{d}_u\:\tilde{d}_d)$ is the cEDM matrix. The second term in Eq. \eqref{eq:Xqinduced} has the same form as the second term of Eq. \eqref{eq:Xqafteralignment} and defines an \lq\lq induced $\theta$ angle" whose value is given by
\begin{equation}
\bar{\theta}_{\mathrm{ind}}=2(\Delta m_\pi^2)_c(\tilde{d}_0+\varepsilon\tilde{d}_1)/(m_\pi^2(1-\varepsilon^2)\tilde{d}_0^M)\ \ \ .
\end{equation}
In the presence of a $\theta$-term one needs only to replace $\bar{\theta}_{\mathrm{ind}}$ by $ \bar{\theta}_{\mathrm{ind}}-\bar{\theta}$ (see Eq. \eqref{eq:Xqafteralignment}). Furthermore, if we assume Peccei-Quinn mechanism \cite{Peccei:1977hh} then $\bar{\theta}$ simply relaxes to $\bar{\theta}_{\mathrm{ind}}$. 

We may now construct other PVTV chiral operators induced by
the quark cEDM remembering that they can be generated by either the
$X_c$ or the $X_q$-insertion. It should be pointed out that the
tree-level matching relations we present below are already well-studied
previously using the chiral SO(4) formalism \cite{deVries:2012ab}.
Here we recast the analysis using the $\mathrm{SU}(2)_\mathrm{L}\times \mathrm{SU}(2)_\mathrm{R}$
formalism and also generalize it to include the $\Delta$-resonances
to show how the same physics works under different
representations. Also, our method has the advantage that it can be generalized more easily to the three-flavor case in order to study the role of the strange quark in the matching relations. In the nucleon sector the leading CSB
operators are
\begin{equation}c_1\bar{N}\tilde{X}_{q+}N+c_1'\mathrm{Tr}[\tilde{X}_{q+}]\bar{N}N+\tilde{c}_1F_0^2\bar{N}\tilde{X}_{c+}N+\tilde{c}_1'F_0^2\mathrm{Tr}[\tilde{X}_{c+}]\bar{N}N.\label{eq:cEDMnucleon}
\end{equation}
Following the same logic as in Sec. \ref{sec:treetheta}, one finds the following tree-level matching relations:
\begin{eqnarray}\label{eq:cEDMmatching}
F_\pi\bar{g}_\pi^{(0)}
&=&-(\delta m_N)_q\frac{(\Delta
m_\pi^2)_{c}}{m_{\pi}^2}\frac{\tilde{d}_1}{\tilde{d}^M_0}+(\delta
m_N)_{c}\frac{\tilde{d}_0}{\tilde{d}^M_1}\nonumber\\
F_\pi\bar{g}_\pi^{(1)}
&=&2\left[-(\Delta m_N)_q\frac{(\Delta
m_\pi^2)_{c}}{m_{\pi}^2}+(\Delta
m_N)_{c}\right]\frac{\tilde{d}_1}{\tilde{d}^M_0}
\end{eqnarray}
as well as $\bar{g}_{NN3\pi}^{(0,1)}=-2\bar{g}_\pi^{(0,1)}/3$. One observes that $\bar{g}_\pi^{(0)}$ depends on both $\tilde{d}_0$ and $\tilde{d}_1$ while $\bar{g}_\pi^{(1)}$ depends only on
$\tilde{d}_1$. However, if we take lattice calculations \cite{Bali:2012qs,Beane:2006fk} that give $(\Delta m_N)_q\approx-37\mathrm{MeV}$ and $(\delta m_N)_q\approx2.26\mathrm{MeV}$
then we find that $\bar{g}_\pi^{(1)}$ is about 30 times more sensitive to $\tilde{d}_1$ than $\bar{g}_\pi^{(0)}$ provided that there is no accidental
cancelation between the two terms in $\bar{g}_\pi^{(1)}$.

In the $\Delta$ sector, the most general terms at leading order are:
\begin{equation}c_2\bar{T}^i_\mu\tilde{X}_{q+}T^{i\mu}+c_2'\mathrm{Tr}[\tilde{X}_{q+}]\bar{T}^i_\mu
T^{i\mu}+\tilde{c}_2F_0^2\bar{T}^i_\mu\tilde{X}_{c+}T^{i\mu}+\tilde{c}_2'F_0^2\mathrm{Tr}[\tilde{X}_{c+}]\bar{T}^i_\mu
T^{i\mu}
\end{equation}
that lead to analogous tree-level matching relations\footnote{One
can easily show that
$i\epsilon^{abc}\pi_b[\bar{T}^{a}_\mu\tau_cT^{3\mu}-\bar{T}^{3}_\mu\tau_cT^{a\mu}]+i\epsilon^{ab3}\bar{T}^a_\mu\vec{\tau}\cdot\vec{\pi}T^{b\mu}=-\pi_0\bar{T}^i_\mu
T^{i\mu}$.}:
\begin{eqnarray}
F_\pi\bar{g}_{\Delta\Delta\pi}^{(0)}&=&-(\delta
m_\Delta)_q\frac{(\Delta
m_\pi^2)_{c}}{m_{\pi}^2}\frac{\tilde{d}_1}{\tilde{d}^M_0}+(\delta
m_\Delta)_{c}\frac{\tilde{d}_0}{\tilde{d}^M_1}\nonumber\\
F_\pi\bar{g}_{\Delta\Delta\pi}^{(1,1)}&=&-2\left[-(\Delta
m_\Delta)_q\frac{(\Delta m_\pi^2)_{c}}{m_{\pi}^2}+(\Delta
m_\Delta)_{c}\right]\frac{\tilde{d}_1}{\tilde{d}^M_0}\nonumber\\
&=&F_\pi\bar{g}_{\Delta\Delta\pi}^{(1,2)}
\end{eqnarray}

\subsubsection{Loop correction}
The one-loop corrections to the LHS and RHS of the tree-level matching relations \eqref{eq:cEDMmatching} can be inferred from Eq. \eqref{eq:loopgpibarcEDM} and \eqref{eq:loopmasscMDM} in Appendix \ref{sec:detail}. After some straightforward rearrangement, one obtains:
\begin{eqnarray}
\delta(F_\pi\bar{g}_\pi^{(0)})_{\mathrm{loop}}&=&\delta\left[-(\delta m_N)_q\frac{(\Delta
m_\pi^2)_c}{m_{\pi}^2}\frac{\tilde{d}_1}{\tilde{d}_0^M}+(\delta
m_N)_c\frac{\tilde{d}_0}{\tilde{d}_1^M}\right]_{\mathrm{loop}}-(\delta
m_N)_q\frac{\tilde{d}_1}{\tilde{d}_0^M}\frac{(\Delta m_\pi^2)_c}{8\pi^2F_\pi^2}L'\nonumber\\
\delta(F_\pi\bar{g}_\pi^{(1)})_{\mathrm{loop}}&=&2\delta\left[-(\Delta m_N)_q\frac{(\Delta m_\pi^2)_c}{m_{\pi}^2}+(\Delta
m_N)_c\right]_{\mathrm{loop}}\frac{\tilde{d}_1}{\tilde{d}_0^M}\nonumber\\
&&+2(\Delta m_\pi^2)_c\frac{\tilde{d}_1}{\tilde{d}_0^M}\left[-\frac{(\Delta
	m_N)_q}{8\pi^2F_\pi^2}L'-4\{\frac{3g_A^2}{F_\pi^2}(I_c-\frac{I_f}{m_\pi^2})+\frac{2g_{\pi
		N\Delta}^2}{F_\pi^2}(I_b-\frac{I_g}{m_\pi^2})\}\right]\nonumber\\
&&+\frac{3m_\pi^2(\Delta
	m_\pi^2)_c}{8\pi^2F_\pi^3}\frac{\tilde{d}_1}{\tilde{d}_0^M}\left((\gamma_1+4\gamma_2)(L'+1)-\frac{1}{2}\gamma_1-4\gamma_2\right)
\end{eqnarray}
{\em i.e.}, the one-loop corrections do not obey the
tree-level matching relations and the induced mismatch between the LHS and RHS of the relations are proportional to
$(\Delta m_\pi^2)_c$. For the case of $I=0$, the tree-level matching is preserved by the loop
correction only in the $\tilde{d}_1\rightarrow 0$ limit.

\subsubsection{\label{sec:cEDMcounter}LECs and the higher-order matching formula}

The relevant LECs that are needed to cancel the
UV-divergences in the loop diagrams are in the $O(E^4)$ Lagrangian and the
$O(E^2\tilde{d})$ terms in the chiral Lagrangian. The former have
already been discussed in Sec. \ref{sec:thetacounter} so we shall
concentrate on the latter. In the pionic sector the relevant
$O(E^2\tilde{d})$ Lagrangian is
\begin{eqnarray}\label{eq:cEDMpioncounter}
\mathcal{L}^{O(E^2\tilde{d})}_\pi&=&2B_0F_\pi^3\{G_1\mathrm{Tr}[X_qU^\dagger+UX_q^\dagger]\mathrm{Tr}[X_cU^\dagger+UX_c^\dagger]
+G_2\mathrm{Tr}[X_qU^\dagger-UX_q^\dagger]\mathrm{Tr}[X_cU^\dagger\nonumber\\
&&-UX_c^\dagger]
+G_3\mathrm{Tr}[U^\dagger X_cU^\dagger X_q+UX_c^\dagger
UX_q^\dagger]\}+F_\pi^3\{G_4\mathrm{Tr}[\partial_\mu U\partial^\mu
U^\dagger]\mathrm{Tr}[X_c U^\dagger+UX_c^\dagger]\nonumber\\
&&+G_5\mathrm{Tr}[\partial_\mu U^\dagger\partial^\mu U (X_c^\dagger
U+U^\dagger X_c)]\}.
\end{eqnarray}
In the equation above, the $E^2$ factor comes either from a factor
of $X_q$ or two derivatives. In particular, the $G_4$ and $G_5$
terms are required as they cancel divergences of both $F_\pi$ and
$Z_\pi$ that receive extra loop corrections due to the generation
of $(\Delta m_\pi^2)_c$. In the nucleon sector, the $O(E^2\tilde{d})$ Lagrangian can be chosen as
\begin{eqnarray}
\mathcal{L}_N^{O(E^2\tilde{d})}&=&2B_0\{g_1\mathrm{Tr}[\tilde{X}_{q+}\tilde{X}_{c+}]\bar{N}N+g_2\mathrm{Tr}[\tilde{X}_{q+}]\bar{N}\tilde{X}_{c+}N+g_3\mathrm{Tr}[\tilde{X}_{c+}]\bar{N}\tilde{X}_{q+}N\nonumber\\
&&+g_4\bar{N}\{\tilde{X}_{q+},\tilde{X}_{c+}\}N
+g_5\mathrm{Tr}[\tilde{X}_{q-}\tilde{X}_{c-}]\bar{N}N+g_6\mathrm{Tr}[\tilde{X}_{q-}]\bar{N}\tilde{X}_{c-}N\nonumber\\
&&+g_7\mathrm{Tr}[\tilde{X}_{c-}]\bar{N}\tilde{X}_{q-}N+g_8\bar{N}\{\tilde{X}_{q-},\tilde{X}_{c-}\}N\}.
\end{eqnarray}

With all these, one can straightforwardly deduce the modified
matching formula for $\bar{g}_\pi^{(i)}$. While all details are given in Appendix \ref{sec:dipoledetail}, the final outcome is:
\begin{eqnarray}
F_\pi\bar{g}_\pi^{(0)}&=&\left(-(\delta m_N)_q\frac{(\Delta m_\pi^2)_c}{m_\pi^2}\frac{\tilde{d}_1}{\tilde{d}_0^M}+(\delta m_N)_c\frac{\tilde{d}_0}{\tilde{d}_1^M}\right)
(1+\delta_\mathrm{loop}^{(0)}+\delta_\mathrm{LEC}^{(0)})\nonumber\\
F_\pi\bar{g}_\pi^{(1)}&=&2\left(-(\Delta m_N)_q\frac{(\Delta m_\pi^2)_c}{m_\pi^2}+(\Delta m_N)_c\right)\frac{\tilde{d}_1}{\tilde{d}_0^M}(1+\delta_\mathrm{loop}^{(1)}+\delta_\mathrm{LEC}^{(1)})
\end{eqnarray}
where the relative corrections $\{\delta^{(i)}\}$ due to loop and higher order LECs are given by
\begin{eqnarray}
\delta_{\mathrm{loop}}^{(0)}&=&\frac{m_\pi^2}{8\pi^2F_\pi^2}\left(-1+\ln(\frac{\mu}{m_\pi})^2\right)\left(1-\frac{m_\pi^2}{(\delta m_N)_q}\frac{(\delta m_N)_c}{(\Delta m_\pi^2)_c}\frac{\tilde{d}_0}{\tilde{d}_1^M}\frac{\tilde{d}_0^M}{\tilde{d}_1}\right)^{-1}
\nonumber\\
\delta_{\mathrm{loop}}^{(1)}&=&\left(
\frac{3g_A^2m_\pi^3}{16\pi(\Delta m_N)_qF_\pi^2}+\frac{8g_{\pi N\Delta}^2m_\pi^2}{(\Delta m_N)_qF_\pi^2}(I_b^r-\frac{I_g^r}{m_\pi^2})-\frac{3m_\pi^4}{16\pi^2(\Delta m_N)_qF_\pi^3}((\gamma_1+4\gamma_2)\ln(\frac{\mu}{m_\pi})^2\right.\nonumber\\
&&\left.-\frac{1}{2}\gamma_1-4\gamma_2)+\frac{m_\pi^2}{8\pi^2F_\pi^2}(-1+\ln(\frac{\mu}{m_\pi})^2)\right)\left(1-\frac{m_\pi^2}{(\Delta m_N)_q}\frac{(\Delta m_N)_c}{(\Delta m_\pi^2)_c}\right)^{-1}
\end{eqnarray}
and
\begin{eqnarray}
\delta_{\mathrm{LEC}}^{(0)}&=&\left(-\frac{4 m_\pi^4\varepsilon}{3(\delta m_N)_qF_\pi^3}[f_2^r+f_3^r+4(f_5^r+f_6^r)]+\frac{8m_\pi^4\tilde{d}_0^M}{(\delta m_N)_q(\Delta m_\pi^2)_c}[\varepsilon(\frac{1}{3}-\frac{\tilde{d}_0^M}{\tilde{d}_1^M}\frac{\tilde{d}_0}{\tilde{d}_1})(g_3^r+g_4^r)\right.\nonumber\\
&&-\frac{\tilde{d}_0}{\tilde{d}_1}(g_7^r+g_8^r)+\frac{\varepsilon}{3}(g_6^r+g_8^r)]-\frac{16F_\pi m_\pi^2\tilde{d}_0^M}{(\Delta m_\pi^2)_c}[(2G_1^r+G_3^r-2G_4^r-G_5^r)\nonumber\\
&&\left.-\frac{2}{3}\varepsilon(\frac{\tilde{d}_1^M}{\tilde{d}_0^M}-\frac{3}{2}\frac{\tilde{d}_0}{\tilde{d}_1})(2G_2^r+G_3^r)]-\frac{64m_\pi^2}{F_\pi^2}\varepsilon^2(2L_7^r+L_8^r)\right)\times\nonumber\\
&&\left(1-\frac{m_\pi^2}{(\delta m_N)_q}\frac{(\delta m_N)_c}{(\Delta m_\pi^2)_c}\frac{\tilde{d}_0}{\tilde{d}_1^M}\frac{\tilde{d}_0^M}{\tilde{d}_1}\right)^{-1}\nonumber\\
\delta_{\mathrm{LEC}}^{(1)}&=&\left(\frac{m_\pi^4}{(\Delta m_N)_qF_\pi^3}[2(1+\varepsilon^2)f_1^r+2f_2^r+(1+\varepsilon^2)f_3^r+4(1+\varepsilon^2)f_4^r+2(\varepsilon^2-1)f_5^r\right.\nonumber\\
&&+2(1+\varepsilon^2)(f_5^r+f_6^r)]+\frac{4m_\pi^4\tilde{d}_0^M}{(\Delta m_N)_q(\Delta m_\pi^2)_c}[\varepsilon\frac{\tilde{d}_0}{\tilde{d}_1}(g_1^r+g_4^r+g_5^r+g_6^r+g_7^r+g_8^r)\nonumber\\
&&-(g_5^r+g_8^r)-\varepsilon\frac{\tilde{d}_1^M}{\tilde{d}_0^M}(g_1^r+g_4^r)]-\frac{16F_\pi m_\pi^2\tilde{d}_0^M}{(\Delta m_\pi^2)_c}[(2G_1^r+G_3^r-2G_4^r-G_5^r)\nonumber\\
&&\left.-\frac{2}{3}\varepsilon(\frac{\tilde{d}_1^M}{\tilde{d}_0^M}-\frac{3}{2}\frac{\tilde{d}_0}{\tilde{d}_1})(2G_2^r+G_3^r)]-\frac{64m_\pi^2}{F_\pi^2}\varepsilon^2(2L_7^r+L_8^r)\right)\times\nonumber\\
&&\left(1-\frac{m_\pi^2}{(\Delta m_N)_q}\frac{(\Delta m_N)_c}{(\Delta m_\pi^2)_c}\right)^{-1}
\label{eq:deltacEDM}\end{eqnarray}
respectively. The functions $\{I_i^r\}$ are just the renormalized version of
the loop functions $\{I_i\}$ defined in Appendix
\ref{sec:loop_function} following the Gasser-Leutwyler subtraction
scheme \cite{Gasser:1983yg}. 

One may numerically estimate the loop corrections to
the tree-level matching relations upon neglecting the unknown matrix elements $(\Delta m_N)_c$ and $(\delta m_N)_c$. In the isoscalar channel, we find that
$\delta^{(0)}_\mathrm{loop}\approx0.021$ (taking
$\mu=1$GeV for the renormalization scale) therefore one has good convergence. On the hand, in the isovector channel we have
$\delta^{(0)}_\mathrm{loop}\approx-3.1$ that does not show any sign of
convergence . The reason is that $(\Delta m_N)_q\approx-37$MeV is much smaller than $\delta_\Delta$ and $m_\pi$, so terms in Eq. \eqref{eq:deltacEDM} such as  
\begin{equation}
\nonumber
(\delta_\Delta/(\Delta
m_N)_q)\ln(\frac{\mu}{m_\pi})^2,\quad (\delta_\Delta/(\Delta
m_N)_q)(\delta_\Delta/m_\pi)^2\ln(\frac{\mu}{m_\pi})^2,\quad
(m_\pi/(\Delta m_N)_q)\ln(\frac{\mu}{m_\pi})^2
\end{equation}
 may overcome the usual
chiral suppression. This implies that the matching formula for $I=1$ cEDM
has very limited practical use. Fortunately, there is a recent study by de Vries et al suggesting that the effect of $\delta_{\mathrm{loop}}^{(i)}$ can be completely eliminated by re-expressing the tree-level matching relations \eqref{eq:cEDMmatching} in terms of derivative operators \cite{deVries:2016jox}.

The impact of higher-order LECs encoded in $\delta_{\mathrm{LEC}}^{(i)}$ can also be studied following the power-counting argument in Sec. \ref{sec:thetacounter}. To make the discussion tractable, let us assume $\tilde{d}_0^M\sim \tilde{d}_1^M$, $\tilde{d}_0\sim\tilde{d}_1$ and take the denominators in $\delta_{\mathrm{LEC}}^{(i)}$ to be $O(1)$. The contribution from $L_i^r$ is negligible as we discussed before. The contribution from $f_i^r$ to $\delta_{\mathrm{LEC}}^{(0)}$ is around 0.1, similar to the case of $\theta$-term, while its contribution to $\delta_\mathrm{LEC}^{(1)}$ is expected to be much smaller because it is divided by $(\Delta m_N)_q$ instead of $(\delta m_N)_q$. New LECs appeared in Eq. \eqref{eq:deltacEDM} are $\{G_i^r\}$ and $\{g_i^r\}$. The estimation of their sizes involves two steps: first, the contribution from the $O(\tilde{d})$ Lagrangian \eqref{eq:cEDMpion} and \eqref{eq:cEDMnucleon} to the pion and nucleon masses can be estimated using Weinberg's counting rule \cite{Weinberg:1989dx}. Then, the effects from $\{G_i\}$ and $\{g_i\}$ are expected to receive a further $(m_\pi/2\pi F_\pi)^2$ suppression due to chiral power-counting. This implies, using Eq. \eqref{eq:cMDMpioncounterresult} and \eqref{eq:cMDMnucleoncounterresult}:
\begin{eqnarray}
(\Delta m_\pi^2)_{c,ct}^r&\sim&16m_\pi^2F_\pi \tilde{d}_i^MG_i^r\sim(\frac{m_\pi}{2\pi F_\pi})^2(\Delta m_\pi^2)_c\sim(\frac{m_\pi}{2\pi F_\pi})^2\frac{(2\pi F_\pi)^3\tilde{d}_i^M}{4\pi}\nonumber\\
(\Delta m_N)_{c,ct}^r&\sim&4m_\pi^2\tilde{d}_i^M g_i^r\sim(\frac{m_\pi}{2\pi F_\pi})^2(\Delta m_N)_c\sim(\frac{m_\pi}{2\pi F_\pi})^2\frac{(2\pi F_\pi)^2\tilde{d}_i^M}{4\pi}
\end{eqnarray}
which gives $G_i^r\sim0.03$ and $g_i^r\sim 0.02$. Applying them to Eq. \eqref{eq:deltacEDM}, we find that the contributions from $g_i^r$ to $\delta_{\mathrm{LEC}}^{(0)}$ is around 0.2 and all other effects are of order $10^{-2}$. Hence the only potentially-large LEC corrections to the tree-level matching relations are the $f^r_i$ and $g_i^r$-correction to $\bar{g}_\pi^{(0)}$.

Finally we shall mention briefly about the quark MDM/EDM operator
$\bar{q}_R\sigma^{\mu\nu}q_LF_{\mu\nu}$. Although its form is
analogous to that of the quark cMDM/cEDM, it involves an interaction
with photon that in turn introduces another CSB quantity, namely
the quark charge matrix. As a consequence, the chiral structure of
the resulting hadronic operators is much more complicated.
Interested readers are referred to Ref. \cite{deVries:2012ab} for
more discussion.

\subsection{\label{sec:LR4Q}LR4Q} The last type of chirally non-invariant operator
that contains both PCTC and PVTV components simultaneously
is the LR4Q operator. The form of its corresponding spurion is
already explained in Sec. \ref{sec:spurion} so we shall go straight
its application.

\subsubsection{Tree-level matching}

In the pionic sector, the only LO-operator we can write down
is\begin{equation}\label{eq:LR4Qpion} \mathcal{L}=\rho
F_0^6(c_{4q}\mathrm{Tr}[U^\dagger
X_{RL}]\mathrm{Tr}[UX_{LR}]+c_{4q}^*\mathrm{Tr}[U^\dagger
X_{LR}^\dagger]\mathrm{Tr}[UX_{RL}^\dagger])\end{equation} where
$X_{RL}=(1+\tau_3)/2$, $X_{LR}=(1-\tau_3)/2$ are the LR4Q spurion matrices defined after Eq. \eqref{eq:LR4Qspurion}, $c_{4q}$ is the complex Wilson coefficient of the LR4Q operator and $\rho$ is a real dimensionless number. Again, this Lagrangian
induces a pion tadpole term that must be removed by including the
term with an $X_q$-insertion and choosing the appropriate value of the free
parameter $\alpha$. This imposes the vacuum alignment
condition $\alpha\approx -8\rho
F_0^4\mathrm{Im}c_{4q}/(B_0\bar{m})=-16\rho
F_0^4\mathrm{Im}c_{4q}/m_\pi^2$. Therefore, to study the T-odd
effect from the LR4Q operator, we need to also include the contribution from $X_q$ with the value of $\alpha$ chosen
above. With these, the relevant quantities one could extract from
Eq. \eqref{eq:LR4Qpion} are:
\begin{eqnarray}
(\Delta m_\pi^2)_{4q}&=&\frac{64\rho
F_0^4\mathrm{Re}c_{4q}}{3}\nonumber\\
(\delta m_\pi^2)_{4q}&=&-\frac{16\rho
F_0^4\mathrm{Re}c_{4q}}{3}=-\frac{1}{4}(\Delta m_\pi^2)_{4q}\nonumber\\
\bar{g}_{\pi\pi\pi}^{(1)}&=&-16\rho
F_0^2\mathrm{Im}c_{4q}=-\frac{3(\Delta
m_\pi^2)_{4q}}{4F_0^2}\frac{\mathrm{Im}c_{4q}}{\mathrm{Re}c_{4q}}.
\end{eqnarray}
There are several differences compared to the case of quark
bilinears. Firstly, there exists an $I=2$ pion mass term (recall its definition in Sec. \ref{sec:CSB}) because the
spurion for the LR4Q contains all $I=0,1,2$ components while the
spurion for quark bilinears only have $I=0$ and $I=1$ pieces.
Secondly, we find a non-vanishing PVTV three-pion coupling
$\bar{g}_{\pi\pi\pi}^{(1)}$.In the case of the quark bilinears, the dipole operators and complex quark mass have the same spurion structure. Consequently, if one chooses the parameter $\alpha$ such that the tadpole contributions from the complex quark mass and dipole operators cancel (vacuum alignment), the corresponding contributions to $\bar{g}_{\pi\pi\pi}^{(1)}$ also cancel. In contrast, the spurion structure for the LR4Q operator differs from that for the complex quark mass, so the three pion term is not eliminated together with the pion tadpole.

In the nucleon sector, the leading operators are \footnote{One can
show that, other structures such as $c_{4q}\mathrm{Tr}[U^\dagger
X_{RL}]\bar{N}uX_{LR}uN+c_{4q}^*\mathrm{Tr}[U^\dagger
X_{LR}^\dagger]\bar{N}uX_{RL}^\dagger uN+h.c.$ are not independent
from the $\tilde{\tilde{c}}_1$ structure we just wrote down.}:
\begin{equation}c_1\bar{N}\tilde{X}_{q+}N+c_1'\mathrm{Tr}[\tilde{X}_{q+}]\bar{N}N+\tilde{\tilde{c}}_1F_0^3\{c_{4q}\mathrm{Tr}[U^\dagger X_{RL}]\mathrm{Tr}[UX_{LR}]+c_{4q}^*\mathrm{Tr}[U^\dagger X_{LR}^\dagger]\mathrm{Tr}[UX_{RL}^\dagger]\}\bar{N}N.\end{equation}
Again, following the same logic as in Sec. \ref{sec:treetheta}, they lead to the following tree-level matching relations \cite{Bsaisou:2014oka,Bsaisou:2014zwa,Dekens:2014jka,Cirigliano:2016yhc}:
\begin{eqnarray}\label{eq:LR4Qmatching}
F_\pi\bar{g}_\pi^{(0)}&=&-\frac{3(\Delta m_\pi^2)_{4q}}{4
m_{\pi}^2}(\delta
m_N)_q\frac{\mathrm{Im}c_{4q}}{\mathrm{Re}c_{4q}}\nonumber\\
F_\pi\bar{g}_\pi^{(1)}
&=&\left[-\frac{3(\Delta m_\pi^2)_{4q}}{2
m_{\pi}^2}(\Delta m_N)_q+4(\Delta
m_N)_{4q}\right]\frac{\mathrm{Im}c_{4q}}{\mathrm{Re}c_{4q}}
\end{eqnarray}
as well as $\bar{g}_{NN3\pi}^{(0)}=-\frac{2}{3}\bar{g}_\pi^{(0)}$ and $F_\pi\bar{g}_{NN3\pi}^{(1)}=(\frac{(\Delta m_\pi^2)_{4q}}{m_{\pi}^2}(\Delta
m_N)_q-\frac{32}{3}(\Delta
m_N)_{4q})\frac{\mathrm{Im}c_{4q}}{\mathrm{Re}c_{4q}}$. 
In particular, one observes that before considering the vacuum
alignment the only PVTV $NN\pi$ operator is $\bar{g}_\pi^{(1)}$. Including the vacuum alignment contribution gives
\begin{equation}
\left.\bar{g}_\pi^{(0)}/\bar{g}_\pi^{(1)}\right|_{\mathrm{vac}}=(\delta m_N)_q/2(
	\Delta m_N)_q \ \ \ .
\end{equation}
Taking the lattice inputs for $(\Delta m_N)_q$ and $(\delta m_N)_q$ gives $\bar{g}_\pi^{(0)}/\bar{g}_\pi^{(1)}\approx-0.03$, {\em i.e.}, the $I=1$ component is the dominant piece as long as there is no accidental cancellation between the direct and vacuum alignment contribution to $\bar{g}_\pi^{(1)}$. This is consistent with observations in Ref.~\cite{Dekens:2014jka,Bsaisou:2014oka}.
This is because the $I=0,2$ components of the LR4Q operator is PCTC while the $I=1$ component is PVTV (see the discussion in Sec. \ref{sec:LR4Qop}). Finally, in the $\Delta$-sector the leading operators are
\begin{eqnarray}
&&c_2\bar{T}^i_\mu\tilde{X}_{q+}T^{i\mu}+c_2'\mathrm{Tr}[\tilde{X}_{q+}]\bar{T}^i_\mu
T^{i\mu}+\tilde{\tilde{c}}_2 F_0^3\{c_{4q}\mathrm{Tr}[U^\dagger
X_{RL}]\mathrm{Tr}[UX_{LR}]\nonumber\\
&&+c_{4q}^*\mathrm{Tr}[U^\dagger
X_{LR}^\dagger]\mathrm{Tr}[UX_{RL}^\dagger]\}\bar{T}^i_\mu
T^{i\mu}\end{eqnarray} that lead to the following tree-level
matching:
\begin{eqnarray}
F_\pi\bar{g}_{\Delta\Delta\pi}^{(0)}&=&-\frac{3(\Delta
m_\pi^2)_{4q}}{4m_{\pi}^2}(\delta m_\Delta)_q\frac{\mathrm{Im}c_{4q}}{\mathrm{Re}c_{4q}}\nonumber\\
F_\pi\bar{g}_{\Delta\Delta\pi}^{(1,1)}&=&\left[\frac{3(\Delta
m_\pi^2)_{4q}}{2m_{\pi}^2}(\Delta m_\Delta)_q-4(\Delta
m_\Delta)_{4q}\right]\frac{\mathrm{Im}c_{4q}}{\mathrm{Re}c_{4q}}\nonumber\\
&&=F_\pi\bar{g}_{\Delta\Delta\pi}^{(1,2)}
\end{eqnarray}

\subsubsection{Loop correction}

The one-loop corrections to the LHS and RHS of the tree-level mataching relations \eqref{eq:LR4Qmatching} can be inferred from Eq. \eqref{eq:loopgpibarLR4Q} and \eqref{eq:loopmassLR4Q} in Appendix \ref{sec:detail}. After straightforward rearrangement on obtains:
\begin{eqnarray}
\delta(F_\pi\bar{g}_\pi^{(0)})_\mathrm{loop}&=&\delta\left[-\frac{3(\Delta m_\pi^2)_{4q}}{4m_\pi^2}(\delta
m_N)_q\right]_\mathrm{loop}
+(\delta m_N)_q\frac{3(\Delta m_\pi^2)_{4q}}{8\pi^2F_\pi^2}\frac{\mathrm{Im}c_{4q}}{\mathrm{Re}c_{4q}}(L'+\frac{5}{4})\nonumber\\
\delta(F_\pi\bar{g}_\pi^{(1)})_\mathrm{loop}&=&\frac{\mathrm{Im}c_{4q}}{\mathrm{Re}c_{4q}}\delta\left[-\frac{3}{2}(\Delta
m_N)_q\frac{(\Delta m_\pi^2)_{4q}}{m_\pi^2}+4(\Delta
m_N)_{4q}\right]_\mathrm{loop}\nonumber\\
&&+\frac{3}{2}(\Delta m_\pi^2)_{4q}\frac{\mathrm{Im}c_{4q}}{\mathrm{Re}c_{4q}}\left[\frac{(\Delta
	m_N)_q}{2\pi^2F_\pi^2}(L'+\frac{5}{4})-4\{\frac{3g_A^2}{F_\pi^2}(I_c-\frac{I_f}{m_\pi^2})\right.\nonumber\\
&&\left.+\frac{2g_{\pi
N\Delta}^2}{F_\pi^2}(I_b-\frac{I_g}{m_\pi^2})\}\right]+\frac{9m_\pi^2(\Delta
m_\pi^2)_{4q}}{32\pi^2F_\pi^3}\frac{\mathrm{Im}c_{4q}}{\mathrm{Re}c_{4q}}((\gamma_1+4\gamma_2)(L'+1)-\frac{1}{2}\gamma_1-4\gamma_2).\nonumber\\
\end{eqnarray}

Again, one clearly sees that the tree-level matching relations \eqref{eq:LR4Qmatching} are not obeyed by one-loop corrections. Analogous to the case of dipole operators, the mismatch between LHS and RHS of the matching relations is proportional to $(\Delta m_\pi^2)_{4q}$. 

\subsubsection{LECs and the higher-order matching formula}

The $O(E^2c_{4q})$ terms in pion and nucleon sector can be chosen as:
\begin{eqnarray}
\mathcal{L}_\pi^{O(E^2c_{4q})}&=&2B_0F_\pi^4\{K_1(c_4q\mathrm{Tr}[X_q^\dagger X_{RL}]\mathrm{Tr}[UX_{LR}]+c_{4q}^*\mathrm{Tr}[X_q^\dagger X_{LR}^\dagger]\mathrm{Tr}[UX_{RL}^\dagger])\nonumber\\
&&+K_2\mathrm{Tr}[U^\dagger X_q](c_{4q}\mathrm{Tr}[U^\dagger X_{RL}]\mathrm{Tr}[UX_{LR}]+c_{4q}^*\mathrm{Tr}[U^\dagger X_{LR}^\dagger]\mathrm{Tr}[UX_{RL}^\dagger])\}\nonumber\\
&&+F_\pi^4\{K_3c_{4q}\mathrm{Tr}[\partial_\mu U^\dagger X_{RL}]\mathrm{Tr}[\partial^\mu UX_{LR}]\nonumber\\
&&+K_4c_{4q}\mathrm{Tr}[\partial_\mu U\partial^\mu U^\dagger]\mathrm{Tr}[U^\dagger X_{RL}]\mathrm{Tr}[UX_{LR}]\}+h.c.\label{eq:LR4Qpioncounter}\\
\mathcal{L}_N^{O(E^2c_{4q})}&=&2B_0F_\pi\{h_1(c_{4q}\mathrm{Tr}[X_q^\dagger X_{RL}]\mathrm{Tr}[UX_{LR}]+c_{4q}^*\mathrm{Tr}[X_q^\dagger X_{LR}^\dagger]\mathrm{Tr}[UX_{RL}^\dagger])\bar{N}N\nonumber\\
&&+h_2\mathrm{Tr}[U^\dagger X_q](c_{4q}\mathrm{Tr}[U^\dagger X_{RL}]\mathrm{Tr}[UX_{LR}]+c_{4q}^*\mathrm{Tr}[U^\dagger X_{LR}^\dagger]\mathrm{Tr}[UX_{RL}^\dagger])\bar{N}N\nonumber\\
&&+h_3(c_{4q}\mathrm{Tr}[X_q^\dagger X_{RL}]\bar{N}uX_{LR}uN+c_{4q}^*\mathrm{Tr}[X_q^\dagger X_{LR}^\dagger]\bar{N}uX_{RL}^\dagger uN)\nonumber\\
&&+h_4(c_{4q}\mathrm{Tr}[U^\dagger X_{RL}U^\dagger X_q]\bar{N}uX_{LR}uN+c_{4q}^*\mathrm{Tr}[U^\dagger X_{LR}^\dagger U^\dagger X_q]\bar{N}uX_{RL}^\dagger uN)\nonumber\\
&&+h_5(c_{4q}\mathrm{Tr}[U^\dagger X_{RL}]\bar{N}\{u^\dagger X_qu^\dagger,uX_{LR}u\}N\nonumber\\
&&+c_{4q}^*\mathrm{Tr}[U^\dagger X_{LR}^\dagger]\bar{N}\{u^\dagger X_qu^\dagger,uX_{RL}^\dagger u\}N)\}+h.c.
\end{eqnarray}
respectively. One can thus straightforwardly deduce the matching formula
for $\bar{g}_\pi^{(i)}$ precise to $O(E^2c_{4q})$. With details provided in Appendix \ref{sec:LR4Qdetail}, the final result turns out to be
\begin{eqnarray}
F_\pi\bar{g}_\pi^{(0)}&=&-\frac{3}{4}\frac{(\Delta
m_\pi^2)_{4q}}{m_\pi^2}(\delta
m_N)_q\frac{\mathrm{Im}c_{4q}}{\mathrm{Re}c_{4q}}
(1+\delta_\mathrm{loop}^{(0)}+\delta_\mathrm{LEC}^{(0)})\nonumber\\
F_\pi\bar{g}_\pi^{(1)}&=&\left[-\frac{3(\Delta m_\pi^2)_{4q}}{2m_\pi^2}(\Delta m_N)_q+4(\Delta m_N)_{4q}\right]\frac{\mathrm{Im}c_{4q}}{\mathrm{Re}c_{4q}}(1+\delta_\mathrm{loop}^{(1)}+\delta_\mathrm{LEC}^{(1)})
\end{eqnarray}
with 
\begin{eqnarray}
\delta_{\mathrm{loop}}^{(0)}&=&-\frac{m_\pi^2}{2\pi^2F_\pi^2}\left[\frac{1}{4}+\ln(\frac{\mu}{m_\pi})^2\right]\nonumber\\
\delta_{\mathrm{loop}}^{(1)}&=&\left[\frac{3g_A^2m_\pi^3}{16\pi(\Delta m_N)_q F_\pi^2}+\frac{8g_{\pi N\Delta}^2m_\pi^2}{(\Delta m_N)_q F_\pi^2}(I_b^r-\frac{I_g^r}{m_\pi^2})-\frac{3m_\pi^4}{16\pi^2(\Delta m_N)_qF_\pi^3}((\gamma_1+4\gamma_2)\ln(\frac{\mu}{m_\pi})^2\right.\nonumber\\
&&\left.-\frac{1}{2}\gamma_1-4\gamma_2)-\frac{m_\pi^2}{2\pi^2F_\pi^2}(\frac{1}{4}+\ln(\frac{\mu}{m_\pi})^2)\right]\left[1-\frac{8}{3}\frac{m_\pi^2}{(\Delta m_N)_q}\frac{(\Delta m_N)_{4q}}{(\Delta m_\pi^2)_{4q}}\right]^{-1}
\end{eqnarray}
as well as
\begin{eqnarray}
\delta_{\mathrm{LEC}}^{(0)}&=&-\frac{4m_\pi^4\varepsilon}{3F_\pi^3(\delta
	m_N)_q}[f_2^r+f_3^r+4(f_5^r+f_6^r)]-\frac{16F_\pi m_\pi^4\mathrm{Re}c_{4q}\varepsilon}{9(\delta
	m_N)_q(\Delta
	m_\pi^2)_{4q}}[3h_3^r+5h_4^r-2h_5^r]\nonumber\\
&&-\frac{16F_\pi^2m_\pi^2\mathrm{Re}c_{4q}}{3(\Delta
	m_\pi^2)_{4q}}(-K_1^r+6K_2^r+K_3^r-6K_4^r)-\frac{64m_\pi^2}{F_\pi^2}\varepsilon^2(2L_7^r+L_8^r)\nonumber\\
\delta_{\mathrm{LEC}}^{(1)}&=&\left[\frac{m_\pi^4}{F_\pi^3(\Delta m_N)_q}[2(1+\varepsilon^2)f_1^r+2f_2^r+(1+\varepsilon^2)f_3^r+4(1+\varepsilon^2)f_4^r+2(\varepsilon^2-1)f_5^r\right.\nonumber\\
&&+2(1+\varepsilon^2)(f_5^r+f_6^r)]+\frac{8}{3}\frac{F_\pi m_\pi^4\mathrm{Re}c_{4q}}{(\Delta m_N)_q(\Delta m_\pi^2)_{4q}}[2h_1^r+h_3^r-h_4^r+2h_5^r]\nonumber\\
&&\left.-\frac{16}{3}\frac{F_\pi^2m_\pi^2\mathrm{Re}c_{4q}}{(\Delta m_\pi^2)_{4q}}(-K_1^r+6K_2^r+K_3^r-6K_4^r)-\frac{64m_\pi^2}{F_\pi^2}\varepsilon^2(2L_7^r+L_8^r)\right]\times\nonumber\\
&&\left[1-\frac{8}{3}\frac{m_\pi^2}{(\Delta m_N)_q}\frac{(\Delta m_N)_{4q}}{(\Delta m_\pi^2)_{4q}}\right]^{-1}.
\end{eqnarray}

Similarly, we may estimate the magnitude of the loop correction to the
tree-level matching relations. Upon neglecting the unknown matrix elements $(\Delta m_N)_{4q}$ and $(\delta m_N)_{4q}$, we have
$\delta^{(0)}_{\mathrm{loop}}=-0.12$ and $\delta^{(1)}_{\mathrm{loop}}=-3.2$ respectively, which implies a moderate convergence in the isoscalar channel and the non-convergence in the isovector channel; the $\bar{g}_\pi^{(1)}$ matching
formula for LR4Q is therefore not useful in practice. Meanwhile, the magnitudes of the LEC corrections 
$\delta_{\mathrm{LEC}}^{(i)}$ can be estimated following the procedure outlined at the end of Sec. \ref{sec:cEDMcounter}. The contributions from $f_i^r$ and $L_i^r$ are similar with the case of cEDM, while for the new LECs labeled as $\{K_i^r\}$ and $\{h_i^r\}$, we estimate their sizes again using the Weinberg counting rule. Eq. \eqref{eq:LR4Qpioncounterresult} and \eqref{eq:LR4Qnucleoncounterresult} then give:
\begin{eqnarray}
(\Delta m_\pi^2)_{4q,ct}^r&\sim&16m_\pi^2F_\pi^2\mathrm{Re}c_{4q}K_i^r\sim(\frac{m_\pi}{2\pi F_\pi})^2(\Delta m_\pi^2)_{4q}\sim(\frac{m_\pi}{2\pi F_\pi})^2\frac{(2\pi F_\pi)^4\mathrm{Re}c_{4q}}{(4\pi)^2}\nonumber\\
(\Delta m_N)_{4q,ct}^r&\sim&4m_\pi^2F_\pi\mathrm{Re}c_{4q}h_i^r\sim(\frac{m_\pi}{2\pi F_\pi})^2(\Delta m_N)_{4q}\sim(\frac{m_\pi}{2\pi F_\pi})^2\frac{(2\pi F_\pi)^3\mathrm{Re}c_{4q}}{(4\pi)^2}
\end{eqnarray}
leading to $K_i^r\sim0.02$ and $h_i^r\sim0.01$. With these, we find that the $h_i^r$-contribution to $\delta_{\mathrm{LEC}}^{(0)}$ is around 0.08 while all other contributions are of the order $10^{-2}$. Hence the conclusion we may draw here is again similar to the case of cEDM.

\subsection{\label{sec:Inv}The $(\bar{L}R)(\bar{L}R)$ operators}

Finally, let us discuss the last class of four-quark operators that
could be T-odd, namely the $(\bar{L}R)(\bar{L}R)$-type operators we
introduced in Sec. \ref{sec:LRLR}. Since they are chirally invariant
their ``spurion" is nothing but a complex number $a$.
Therefore, when their effects are implemented into the chiral
Lagrangian, terms proportional to $a$ and $a^*$ can in principle
both appear with independent coefficients (e.g. via terms like
$(\alpha_1 a+\alpha_2 a^*)\hat{O}+h.c.$ where $\alpha_1$ and
$\alpha_2$ are unrelated coefficients) so there is no definite
matching formula between the PCTC and PVTV observables.
Similar considerations apply for other chirally invariant operators, such as the
Weinberg three-gluon operator. Therefore, we focus only on the chirally non-invariant operators in this paper. 

\section{\label{sec:conclusion}Conclusion}

The computation of hadronic matrix elements induced by effective
quark-gluon operators that are relevant for tests of fundamental
symmetries is a non-trivial task. Among them, the PVTV
pion-nucleon couplings $\bar{g}_\pi^{(i)}$ that contribute to
nucleon and atomic EDMs are of particular interest in this paper.
These operators can be induced by PVTV effective operators that are either  chirally invariant nor
non-invariant . The latter class is 
interesting theoretically because the PCTC and PVTV
components of the CSB operator can be grouped into a single spurion
field that enters the effective chiral Lagrangian. Consequently,
there exist matching relations between the $\bar{g}_\pi^{(i)}$ induced
by the spurion field and various PCTC and CSB observables, such
as the pion mass and the nucleon mass shifts that are induced by
the same spurion field. These relationships are analogous to the relationships between
matrix elements of different components of a vector due to the
Wigner-Eckart theorem. The relations between PVTV and PCTC
hadronic matrix elements are extremely useful because one could
use studies of the PCTC hadronic observables (say, through
lattice) to obtain their PVTV counterparts.

A caveat to the use of this formalism is that the matching formulae
are derived at tree-level and may receive non-negligible
higher order corrections  from loop diagrams and/or higher order terms in the chiral Lagrangian. 
In order to study the higher order effects, we have performed a general
classifications of relevant operators that could generate loop
corrections to $\bar{g}_\pi^{(i)}$ and CSB observables, and
we have calculated the most general loop corrections to those quantities. We then applied this
general formalism  to study the loop corrections to
the matching formulae induced by all relevant effective operators (of
the lightest generation) up to dimension 6. In general, we found
that the matching relations for $\bar{g}_\pi^{(0)}$ are relatively
stable as the loop corrections lead to at most $\mathcal{O}(10\%)$
modifications. On the other hand, the robustness of the
$\bar{g}_\pi^{(1)}$ matching formulae is more complicated as the corrections
depend strongly on the ratio $F_\pi (\Delta m_N)_\mathcal{O}/(\Delta
m_\pi^2)_\mathcal{O}$. We also find that, the inclusion of $\Delta$-resonances in the loop diagrams does not spoil the matching relation of $\bar{g}_\pi^{(0)}$ but does affect the $\bar{g}_\pi^{(1)}$-matching significantly. For the impact of higher-order terms in the chiral Lagrangian, we find that the largest effects arise from the corresponding LECs in the nucleon sector, which may give rise to a (10-20)\% modification of the matching relation for $\bar{g}_\pi^{(0)}$. Contributions from the LECs in the pion sector are in general not much larger than 1\%.


\section{Acknowledgements}
We thank Barry Holstein, Emanuele Mereghetti and Jordy de
Vries for useful discussions. This work was supported in part under U.S. Department of Energy contract
number DE-SC0011095. CYS also acknowledges support by National Natural Science Foundation 
of China (NSFC)
under Grant Nos.11575110, 11655002, 11735010, Natural Science Foundation of 
Shanghai under Grant No.~15DZ2272100 and No.~15ZR1423100,  by Shanghai Key 
Laboratory for Particle Physics and Cosmology,  and by Key Laboratory
for Particle Physics, Astrophysics and Cosmology, Ministry of Education. He 
also appreciates the support through the Recruitment Program of Foreign Young 
Talents from the State Administration of Foreign Expert Affairs, China.

\appendix

\section{\label{sec:blocks}Chiral Building Blocks in $\mathrm{SU}(2)_\mathrm{L}\times \mathrm{SU}(2)_\mathrm{R}$}

Here we summarize the building blocks of $\mathrm{SU}(2)_\mathrm{L}\times \mathrm{SU}(2)_\mathrm{R}$
ChPT that are required to construct a chirally invariant Lagrangian and
implement the effect of chiral symmetry breaking. For most of the
notations and conventions we follow the pedagogical article by
Scherer \cite{Scherer:2002tk}. For the pion decay constant we take
$F_\pi=186\, \mathrm{MeV}$ following the convention 
\cite{Engel:2013lsa}. Although this is not the standard convention in the literature using $\mathrm{SU}(2)_\mathrm{L}\times \mathrm{SU}(2)_\mathrm{R}$ ChPT,  it allows us
to compare our results more easily with previous work on the study
of $\bar{g}_\pi^{(i)}$ that mostly adopt the SO(4)
representation.

\begin{enumerate}
\item $U=\exp\{i\frac{2\pi_a\tau_a}{F_{0}}\}$ transforms as $U\rightarrow
V_RUV_L^\dagger$ under $\mathrm{SU}(2)_\mathrm{L}\times \mathrm{SU}(2)_\mathrm{R}$.
\item $u=\sqrt{U}$ transforms as $u\rightarrow
V_RuK^\dagger=KuV_L^\dagger$ where $K=K(V_R,V_L,U)$ is a unitary
matrix. It reduces to isospin transformation matrix when $V_R=V_L$.
\item The SU(2) nucleon field: $N=(p\;\;n)^T$ transforms as $N\rightarrow
KN$.
\item The chiral axial vector $u_\mu=iu^\dagger(\partial_\mu
U)u^\dagger$ is a Hermitian and traceless operator. It transforms as
$u_\mu\rightarrow Ku_\mu K^\dagger$.
\item The $\Delta$-resonance field $T^i_\mu$ transforms as $T^i_\mu\rightarrow
\tilde{K}^{ij}KT_\mu^i$, where:
\begin{equation}\tilde{K}^{ij}=\delta^{ij}+\epsilon^{ijk}\theta^k_V+\frac{1}{F_0}(\pi^i\theta^j_A-\pi^j\theta^i_A)+O(\theta^2,\pi^{(2)})\end{equation}
satisfying $\tilde{K}^{ij}\tilde{K}^{ij'}=\delta^{jj'}$. For
$\mathrm{SU}(2)_\mathrm{V}$, the matrix $\tilde{K}^{ij}$ simply reduces to the
transformation matrix of an isospin triplet. Also, in order to
eliminate the spin-1/2 and isospin-1/2 components, $T^i_{\mu}$ is
subject to the following constraints:
\begin{eqnarray}\label{eq:constraints}\gamma^\mu T^i_\mu&=&0\nonumber\\
\tau^i T^i_\mu&=&0.\end{eqnarray} In particular, the first relation,
combining with the (relativistic) free-field EOM
$(i\partial\!\!\!/-m_\Delta)T^i_\mu=0$, gives $\partial_\mu
T^{i\mu}=0$ that reduces to $v^\mu T^i_\mu=0$ in the HBChPT
formalism. In terms of the physical $\Delta$-fields, $T_{\mu}^{i}$
can be expressed as:\begin{eqnarray} T_{\mu}^{1} & = &
\frac{1}{\sqrt{2}}\left(\begin{array}{c}
\Delta^{++}-\frac{1}{\sqrt{3}}\Delta^{0}\nonumber\\
\frac{1}{\sqrt{3}}\Delta^{+}-\Delta^{-}\end{array}\right)_{\mu}\\
T_{\mu}^{2} & = & \frac{i}{\sqrt{2}}\left(\begin{array}{c}
\Delta^{++}+\frac{1}{\sqrt{3}}\Delta^{0}\nonumber\\
\frac{1}{\sqrt{3}}\Delta^{+}+\Delta^{-}\end{array}\right)_{\mu}\\
T_{\mu}^{3} & = & -\sqrt{\frac{2}{3}}\left(\begin{array}{c}
\Delta^{+}\\
\Delta^{0}\end{array}\right)_{\mu}.\end{eqnarray} Details of the
inclusion of $\Delta$-resonance in ChPT can be found in Ref.
\cite{Hemmert:1997ye}.
\item $\omega_{\mu}^{i}=\frac{1}{2}\mathrm{Tr}[\tau^{i}u_{\mu}]$ is a Hermitian operator that transforms
as $\omega_{\mu}^{i}\rightarrow\tilde{K}^{ij}\omega_{\mu}^{j}$.
\end{enumerate}

\section{\label{sec:ChiralInvariantLag}Relevant Chirally Invariant Lagrangian}

Here we write down the PCTC, chirally invariant Lagrangian involving pion, nucleon and $\Delta$-resonance fields that is relevant to our work, expanded to
$O(E^2)$ in ChPT using the $\mathrm{SU}(2)_\mathrm{L}\times \mathrm{SU}(2)_\mathrm{R}$ formalism. It
is given by \cite{Scherer:2002tk,Hemmert:1997ye,Zhu:2000gn}:

\begin{eqnarray}
\mathcal{L}&=&\frac{F_0^2}{16}\mathrm{Tr}[\partial_\mu
U\partial^\mu U^\dagger]+\bar{N}iv\cdot
\mathcal{D} N+g_A\bar{N}u_\mu S^\mu N\nonumber\\
&&+F_0^{-1}\bar{N}[\gamma_1(v\cdot u)^2+\gamma_2 u\cdot u]N\nonumber\\
&&-\bar{T}^\mu_i[iv\cdot\mathcal{D}^{ij}-\delta_\Delta]T_{j\mu}+g_{\pi
N\Delta}[\bar{T}^\mu_i\omega^i_\mu N+\bar{N}\omega^i_\mu
T^\mu_i]+...
\label{eq:invariantLag}\end{eqnarray}
where $\mathcal{D}_\mu$ and $\mathcal{D}^{ij}_\mu$ are the chiral
covariant derivatives on the nucleon and $\Delta$ respectively while
the $\Delta-N$ mass-splitting is given by
$\delta_\Delta=m_\Delta-m_N$. In the absence of external fields, we
have $\mathcal{D}_\mu
=\partial_\mu+\frac{1}{2}\{u^\dagger,\partial_\mu u\}$ and
$\mathcal{D}^{ij}_\mu=\delta^{ij}\mathcal{D}_\mu-\frac{i}{2}\epsilon^{ijk}\mathrm{Tr}[\tau^k\{u^\dagger,\partial_\mu
u\}]$. The value of the $\Delta$-nucleon-pion coupling constant is given by
$g_{\pi N\Delta}\approx1.05$ according to \cite{Hemmert:1997ye}. Fitting to scattering observables yields \cite{Scherer:2002tk}:
\begin{equation}\gamma_1\approx
0.621,\gamma_2\approx-0,984
\end{equation}
Also, we have dropped the terms that are not needed in our work,
e.g. coupling term of the form $\bar{N}[S^\mu,S^\nu]u_\mu u_\nu N$ as well as the $\Delta\Delta\pi$ interaction terms.

The free propagators of pion, nucleon and $\Delta$-resonance are
\begin{eqnarray}iD_\pi(k)&=&\frac{i}{k^2-m_{\pi,0}^2+i\epsilon}\nonumber\\
iS_N(k)&=&\frac{i}{v\cdot
k+i\epsilon}\nonumber\\
iD_\Delta(k)^{ij}_{\mu\nu}&=&\frac{-i}{v\cdot
k-\delta_\Delta+i\epsilon}P^{3/2}_{\mu\nu}\xi^{ij}_{3/2}\end{eqnarray}
respectively, where
\begin{eqnarray}
P^{3/2}_{\mu\nu}&=&g_{\mu\nu}-v_\mu v_\nu+\frac{4}{d-1}S_\mu
S_\nu\nonumber\\
\xi^{ij}_{3/2}&=&\frac{2}{3}\delta^{ij}-\frac{i}{3}\epsilon^{ijk}\tau^k\end{eqnarray}
are the projection operators for spin-3/2 and isospin-3/2
respectively.

\section{\label{sec:loop_function} Loop Integral Functions}

Here we define several loop integral functions $\{I_i\}$ that
always appear when one calculate loop diagrams given in
Fig.\ref{fig:amputated} and \ref{fig:CBSloop}:

\begin{eqnarray}
I_a&\equiv&\mu^{4-d}\int\frac{d^dl}{(2\pi)^d}S\cdot
l\frac{i}{-v\cdot l+i\epsilon}\frac{i}{-v\cdot l+i\epsilon}S\cdot
l\frac{i}{l^2-m_\pi^2+i\epsilon}\nonumber\\
&=&\frac{3m_\pi^2}{64\pi^2}(L'+\frac{1}{3})\\
I_b&\equiv&\mu^{4-d}\int\frac{d^dl}{(2\pi)^d}l_\alpha\frac{-i}{-v\cdot
l-\delta_\Delta+i\epsilon}P^{\alpha\beta}_{3/2}l_\beta\frac{i}{l^2-m_\pi^2+i\epsilon}\frac{i}{l^2-m_\pi^2+i\epsilon}\nonumber\\
&=&\frac{\delta_\Delta}{8\pi^2}L'+\frac{1}{8\pi^2}\left[\delta_\Delta-2\sqrt{\delta_\Delta^2-m_\pi^2}\ln\frac{\delta_\Delta+\sqrt{\delta_\Delta^2-m_\pi^2}}{m_\pi}\right]\\
I_c&\equiv&\mu^{4-d}\int\frac{d^dl}{(2\pi)^d}S\cdot
l\frac{i}{-v\cdot l+i\epsilon}S\cdot l\frac{i}{l^2-m_\pi^2+i\epsilon}\frac{i}{l^2-m_\pi^2+i\epsilon}\nonumber\\
&=&\frac{3m_\pi}{64\pi}\\
I_d&\equiv&\mu^{4-d}\int\frac{d^dl}{(2\pi)^d}l_\alpha\frac{-i}{-v\cdot
l-\delta_\Delta+i\epsilon}P^{\alpha\beta}_{3/2}g_{\beta\delta}\frac{-i}{-v\cdot
l-\delta_\Delta+i\epsilon}P^{\delta\rho}_{3/2}l_\rho\frac{i}{l^2-m_\pi^2+i\epsilon}\nonumber\\
&=&\frac{1}{8\pi^2}(2\delta_\Delta^2-m_\pi^2)L'+\frac{\delta_\Delta^2}{4\pi^2}-\frac{\delta_\Delta}{2\pi^2}\sqrt{\delta_\Delta^2-m_\pi^2}\ln\frac{\delta_\Delta+\sqrt{\delta_\Delta^2-m_\pi^2}}{m_\pi}
\\
I_e&\equiv&\mu^{4-d}\int\frac{d^dl}{(2\pi)^d}\frac{i}{l^2-m_\pi^2}\nonumber\\
&=&-\frac{m_\pi^2}{16\pi^2}(L'+1)\\
I_f&\equiv&i\mu^{4-d}\int\frac{d^dl}{(2\pi)^d}S\cdot
l\frac{i}{-v\cdot l+i\epsilon} S\cdot
l\frac{i}{l^2-m_\pi^2+i\epsilon}\nonumber\\
&=&\frac{m_\pi^3}{32\pi}\\
I_g&\equiv&i\mu^{4-d}\int\frac{d^dl}{(2\pi)^d}l_\mu\frac{-i}{-v\cdot
l-\delta_\Delta+i\epsilon}P^{\mu\nu}_{3/2}l_\nu\frac{i}{l^2-m_\pi^2+i\epsilon}\nonumber\\
&=&\frac{\delta_\Delta}{12\pi^2}(-\delta_\Delta^2+\frac{3}{2}m_\pi^2)L'+\frac{1}{72\pi^2}\left[2\delta_\Delta(6m_\pi^2-5\delta_\Delta^2)+12(\delta_\Delta^2-m_\pi^2)^{3/2}\ln\frac{\delta_\Delta+\sqrt{\delta_\Delta^2-m_\pi^2}}{m_\pi}\right]\nonumber\\
\end{eqnarray}
with the divergent quantity $L'$ defined in Eq. \eqref{eq:L}.

In Table \ref{tab:loopfunction} we give the numerical values of the
renormalized loop functions $I^r_i$ where the divergent quantity
$L+1$ is subtracted and the renormalization scale $\mu$ is taken to
be 1 GeV.

\begin{table}
\begin{centering}
\begin{tabular}{|c|c|c|c|c|c|c|}
\hline $I_{a}^{r}$ & $I_{b}^{r}$ & $I_{c}^{r}$ & $I_{d}^{r}$ &
$I_{e}^{r}$ & $I_{f}^{r}$ & $I_{g}^{r}$\tabularnewline \hline \hline
303MeV$^{2}$ & 5.66MeV & 2.08MeV & 2590MeV$^{2}$ & -486MeV$^{2}$ &
27000MeV$^{3}$ & -273000MeV$^{3}$\tabularnewline \hline
\end{tabular}
\par\end{centering}

\caption{\totheleft\label{tab:loopfunction}Numerical values of the
renormalized loop functions with $\mu=1$ GeV.}

\end{table}

\section{\label{sec:ordloop}Relevant One-Loop Corrections in Ordinary ChPT}
Here we summarize some important results for ordinary ChPT at one
loop that are necessary in our work. First we introduce the relevant
$O(E^3)$ and $O(E^4)$ Lagrangian that are crucial in the cancellation of one-loop UV-divergence.
They are \cite{Bernard:1998gv,Gasser:1983yg}:
\begin{eqnarray}
\mathcal{L}_{O(E^3)}&=&\frac{B_{20}}{(2\pi F_\pi)^2}
B_0\mathrm{Tr}[\tilde{X}_{q+}]\bar{N}iv\cdot\mathcal{D}N+...\nonumber\\
\mathcal{L}_{O(E^4)}&=&2B_0 L_4\mathrm{Tr}[\partial_\mu
U\partial^\mu
U^\dagger]\mathrm{Tr}[X_qU^\dagger+UX_q^\dagger]+2B_0L_5\mathrm{Tr}[\partial_\mu
U^\dagger\partial^\mu U (U^\dagger X_q+X_q^\dagger U)]\nonumber\\
&&+4B_0^2L_6\mathrm{Tr}[X_qU^\dagger+UX_q^\dagger]^2+4B_0^2L_7\mathrm{Tr}[X_qU^\dagger-UX_q^\dagger]^2\nonumber\\
&&+4B_0^2L_8\mathrm{Tr}[UX_q^\dagger UX_q^\dagger+U^\dagger
X_qU^\dagger X_q]+...\label{eq:E4pion}\end{eqnarray}

The loop and LEC corrections to $Z_N$ and $Z_\pi$ are given
by:
\begin{eqnarray}
Z_N-1&=&\frac{9g_A^2m_\pi^2}{16\pi^2F_\pi^2}\left[-\frac{2}{3}+\ln(\frac{\mu}{m_\pi})^2\right]-\frac{g_{\pi N\Delta}^2}{\pi^2F_\pi^2}[(2\delta_\Delta^2-m_\pi^2)L'\nonumber\\
&&+2\delta_\Delta^2-4\delta_\Delta\sqrt{\delta_\Delta^2-m_\pi^2}\ln\frac{\delta_\Delta+\sqrt{\delta_\Delta^2-m_\pi^2}}{m_\pi}]-\frac{2B_{20}^rm_\pi^2}{\pi^2F_\pi^2}\nonumber\\
\sqrt{Z_\pi}-1&=&-\frac{m_\pi^2}{12\pi^2F_\pi^2}(L'+1)-\frac{16m_\pi^2}{F_\pi^2}(2L_4+L_5)\nonumber\\
&&=\frac{4}{3F_\pi^2}I_e-\frac{16m_\pi^2}{F_\pi^2}(2L_4+L_5).
\end{eqnarray}
It is worth pointing out that the wavefunction renormalization of the
nucleon field $Z_N$ is finite as its infinity is absorbed by the
LEC $B_{20}$. Meanwhile, the wavefunction renormalization of
the pion field $Z_\pi$ remains infinite as the LEC $2L_4+L_5$ is
not used to subtract the infinity in $Z_\pi$. This is not an issue
because $Z_\pi$ is not a physical observable. On the other hand, the
pion decay constant $F_\pi$ is a physical observable; therefore its
renormalization must be finite. The same combination of LECs
$2L_4+L_5$ is used to subtract the divergence entering $F_\pi$ instead of
$Z_\pi$. It gives:
\begin{equation}
F_\pi=F_0\left[1+\frac{m_\pi^2}{4\pi^2F_\pi^2}(L'+1)+\frac{16m_\pi^2}{F_\pi^2}(2L_4+L_5)\right]=F_0\left[1-\frac{4}{F_\pi^2}I_e^r+\frac{16m_\pi^2}{F_\pi^2}(2L_4^r+L_5^r)\right].\end{equation}
Also, the one-loop correction to the squared charged-pion mass is
useful:
\begin{equation}\label{eq:mpishift}
m_{\pi}^2=m_{\pi,0}^2\left[1+\frac{2}{F_\pi^2}I_e^r-\frac{32m_\pi^2}{F_\pi^2}(2L_4^r+L_5^r-4L_6^r-2L_8^r)\right].\end{equation}

Finally, since the LEC $B_{20}$ is only used for $Z_N$ and
not other quantities (as far as this work is concerned), we will not
 distinguish the loop and LEC contribution to $Z_N$
in the main text. Rather, $Z_N$ is simply taken as 
a finite quantity.

\section{\label{sec:detail}Some important details in Section \ref{sec:main}}

Here we collect some important intermediate results -- including loop corrections, LEC corrections and implications of  higher-order vacuum alignment -- that are crucial in order to derive the main conclusions in Sec. \ref{sec:main}  yet are too long to be put in the main text.

\subsection{$\theta$-term}\label{sec:thetadetail}
First we consider the one-loop renormalization to $F_\pi\bar{g}_\pi^{(0)}$. Eq.
\eqref{eq:total} together with Sec. \ref{sec:treetheta} gives
\begin{eqnarray}
\delta(F_\pi\bar{g}_\pi^{(0)})_{\mathrm{loop}}
&=&F_\pi\left[(Z_N-1)\bar{g}_\pi^{(0)}+(-\frac{2}{F_\pi^2}I_e+\frac{4g_A^2}{F_\pi^2}I_a)\bar{g}_\pi^{(0)}-\frac{40g_{\pi
		N\Delta}^2\bar{g}_{\Delta\Delta\pi}^{(0)}}{9F_\pi^2}I_d\right].\label{eq:loopgpibarq}
\end{eqnarray}
Similarly, the nucleon sigma term and mass splitting (recall their definitions in Sec. \ref{sec:CSB}) also receive
loop corrections from the CSB operators we wrote down in Sec.
\ref{sec:treetheta}. With the identification
$(g_{NN\pi\pi}^{(0)})_q=-2(\Delta m_N)_q/F_\pi$,
$(g_{NN\pi\pi}^{(1,1)})_q=0$ and $(g_{NN\pi\pi}^{(1,2)})_q=-(\delta
m_N)_q/F_\pi$, we obtain:
\begin{eqnarray}
\delta(\Delta m_N)_{q,\mathrm{loop}}&=&(Z_N-1)(\Delta
m_N)_q-(\frac{12g_A^2}{F_\pi^2}I_a+\frac{6}{F_\pi^2}I_e)(\Delta
m_N)_q-\frac{8g_{\pi N\Delta}^2(\Delta
	m_\Delta)_q}{F_\pi^2}I_d\nonumber\\
&&+\frac{12g_A^2}{F_\pi^2}I_f+\frac{8g_{\pi N\Delta}^2}{F_\pi^2}I_g-\frac{3m_\pi^4}{16\pi^2F_\pi^3}\left((\gamma_1+4\gamma_2)(L'+1)+\frac{1}{2}\gamma_1\right)\label{eq:loopDmNq}\\
\delta(\delta m_N)_{q,\mathrm{loop}}&=&(Z_N-1)(\delta
m_N)_q+(-\frac{2}{F_\pi^2}I_e+\frac{4g_A^2}{F_\pi^2}I_a)(\delta
m_N)_q-\frac{40g_{\pi N\Delta}^2(\delta
	m_\Delta)_q}{9F_\pi^2}I_d.\label{eq:loopdmNq}\end{eqnarray}
Here we include loop corrections for both $(\Delta m_N)_q$ and $(\delta m_N)_q$ even though the tree-level matching relation only involves the latter in the case of $\theta$-term, because the former will appear in the matching relations induced by cEDM and LR4Q. 

Next we consider consequences of the introduction of $O(E^4)$ Lagrangian. First, the $O(E^4)$ Lagrangian in pion sector defined in Eq. \eqref{eq:E4pion} leads to a modification of the  vacuum-alignment condition:
\begin{equation}
\alpha=-\frac{\varepsilon\bar{\theta}}{2}\left[1+\frac{64m_\pi^2}{F_\pi^2}(1-\varepsilon^2)(2L_7^r+L_8^r)\right]\end{equation}
where $L_i^r$ are the renormalized $O(E^4)$ LECs in the pion sector. This gives rise to an extra $O(E^4)$ contribution to $\bar{g}_\pi^{(i)}$:
\begin{eqnarray}
\delta(\bar{g}_\pi^{(0)})_v&=&-\frac{32(\delta m_N)_qm_\pi^2\varepsilon(1-\varepsilon^2)(2L_7^r+L_8^r)}{F_\pi^3}\bar{\theta}=\delta(\bar{g}_\pi^{(0)})_v^r\nonumber\\
\delta(\bar{g}_\pi^{(1)})_v&=&-\frac{64(\Delta
	m_N)_qm_\pi^2\varepsilon(1-\varepsilon^2)(2L_7^r+L_8^r)}{F_\pi^3}\bar{\theta}=\delta(\bar{g}_\pi^{(1)})_v^r
\end{eqnarray}
The subscript ``$v$" denotes the contribution from
higher-order vacuum alignment.

In the nucleon sector, the $O(E^4)$ Lagrangian \eqref{eq:OE4nucleon} provides LEC-contributions to the nucleon sigma term and
mass-splitting:
\begin{eqnarray}
\delta(\Delta
m_N)_{q,ct}&=&\frac{m_\pi^4}{F_\pi^3}\left[2(1+\varepsilon^2)f_1+2f_2+(1+\varepsilon^2)f_3\right]\nonumber\\
\delta(\delta m_N)_{q,ct}&=&-\frac{4\varepsilon
	m_\pi^4}{F_\pi^3}(f_2+f_3).\label{eq:OE4ctnucleon}
\end{eqnarray}
The subscript ``$ct$" denotes direct contributions from higher-order LECs (which also play the role of counterterms, hence the naming of the subscript).
Meanwhile, the LEC contributions for $\bar{g}_\pi^{(i)}$ are given by:
\begin{eqnarray}
\delta(F_\pi\bar{g}_\pi^{(0)})_{ct}&=&\frac{2(\varepsilon^2-1)(f_2+f_3+f_5^r+f_6^r)m_\pi^4}{F_\pi^3}\bar{\theta}\nonumber\\
\delta(F_\pi\bar{g}_\pi^{(1)})_{ct}&=&-\frac{2\varepsilon(\varepsilon^2-1)(2f_1^r+f_3^r+2f_4^r+2f_5^r+f_6^r)m_\pi^4}{F_\pi^3}\bar{\theta}.
\end{eqnarray}
Note that the LECs for $F_\pi$ and $\sqrt{Z_\pi}-1$ will always cancel each other so they never appear in $\delta(F_\pi\bar{g}_\pi^{(i)})_{ct}$ (see Appendix \ref{sec:ordloop}).

\subsection{dipole-operators}\label{sec:dipoledetail}

First, the one-loop renormalization to $F_\pi\bar{g}_\pi^{(i)}$ is:
\begin{eqnarray}
\delta(F_\pi\bar{g}_\pi^{(0)})_{\mathrm{loop}}&=&F_\pi\left[(\frac{4g_A^2}{F_\pi^2}I_a-\frac{2}{F_\pi^2}I_e)\bar{g}_{\pi}^{(0)}-\frac{40g_{\pi
		N\Delta}^2\bar{g}_{\Delta\Delta\pi}^{(0)}}{9F_\pi^2}I_d+(Z_N-1)\bar{g}_\pi^{(0)}\right]\nonumber\\
\delta(F_\pi\bar{g}_\pi^{(1)})_{\mathrm{loop}}&=&F_\pi\left[-(\frac{12g_A^2}{F_\pi^2}I_a+\frac{6}{F_\pi^2}I_e)\bar{g}_{\pi}^{(1)}+\frac{8g_{\pi
		N\Delta}^2\bar{g}_{\Delta\Delta\pi}^{(1,1)}}{F_\pi^2}I_d
+(Z_N-1)\bar{g}_\pi^{(1)}\right].\label{eq:loopgpibarcEDM}
\end{eqnarray}

Meanwhile, with $(g_{4\pi}^{(0)})_c=(\Delta m_\pi^2)_c/6F_\pi^2$,
$(g_{NN\pi\pi}^{(0)})_c=-2(\Delta m_N)_c/F_\pi$,
$(g_{NN\pi\pi}^{(1,1)})_c=0$ and $(g_{NN\pi\pi}^{(1,2)})_c=-(\delta
m_N)_c/F_\pi$ (recall their definitions in Sec. \ref{sec:CSB}), the one-loop renormalization to hadron mass parameters is:
\begin{eqnarray}
\delta(\Delta m_N)_{c,\mathrm{loop}}&=&(Z_N-1)(\Delta m_N)_c-(\Delta
m_N)_c(\frac{6}{F_\pi^2}I_e+\frac{12g_A^2}{F_\pi^2}I_a)-\frac{8g_{\pi
		N\Delta}^2(\Delta m_\Delta)_c}{F_\pi^2}I_d\nonumber\\
&&+\frac{12g_A^2(\Delta
	m_\pi^2)_c}{F_\pi^2}I_c+\frac{8g_{\pi N\Delta}^2(\Delta
	m_\pi^2)_c}{F_\pi^2}I_b-\frac{3m_\pi^2(\Delta m_\pi^2)_c}{8\pi^2F_\pi^3}((\gamma_1+4\gamma_2)(L'+1)\nonumber\\
&&-2\gamma_2)\nonumber\\
\delta(\delta m_N)_{c,\mathrm{loop}}&=&(Z_N-1)(\delta
m_N)_c+(-\frac{2}{F_\pi^2}I_e+\frac{4g_A^2}{F_\pi^2}I_a)(\delta
m_N)_c-\frac{40g_{\pi N\Delta}^2(\delta
	m_\Delta)_c}{9F_\pi^2}I_d\nonumber\\
\delta(\Delta m_\pi^2)_{c,\mathrm{loop}}&=&-\frac{(\Delta
	m_\pi^2)_cm_\pi^2}{4\pi^2F_\pi^2}(L'+\frac{1}{2}).\label{eq:loopmasscMDM}
\end{eqnarray}
 
 Next we consider consequences of the introduction of $O(E^2\tilde{d})$ LECs. First, $O(E^2\tilde{d})$ terms in the pion sector \eqref{eq:cEDMpioncounter}, together with the $O(E^4)$ pion
 Lagrangian, modify the vacuum-alignment formula as
 \begin{eqnarray}
 \alpha&=&-\frac{(\Delta
 	m_\pi^2)_{c,0}}{(m_\pi^2)_0}\frac{\tilde{d}_1}{\tilde{d}_0^M}+\frac{64(\Delta
 	m_\pi^2)_c}{F_\pi^2}[(2L_6^r+L_8^r)+\varepsilon^2(2L_7^r+L_8^r)]\frac{\tilde{d}_1}{\tilde{d}_0^M}\nonumber\\
 &&-16F_\pi[(2G_1^r+G_3^r)\tilde{d}_1-\varepsilon(2G_2^r+G_3^r)\tilde{d}_0].\end{eqnarray}
 Note that the divergent pieces of the LECs cancel each
 other, leaving $\alpha$ finite. This leads to extra vacuum-alignment contribution to $\bar{g}_\pi^{(i)}$:
 \begin{eqnarray}
 \delta(\bar{g}_\pi^{(0)})_v&=&\frac{64(\delta m_N)_q(\Delta
 	m_\pi^2)_c}{F_\pi^3}[(2L_6^r+L_8^r)+\varepsilon^2(2L_7^r+L_8^r)]\frac{\tilde{d}_1}{\tilde{d}_0^M}\nonumber\\
 &&-16(\delta
 m_N)_q[(2G_1^r+G_3^r)\tilde{d}_1-\varepsilon(2G_2^r+G_3^r)\tilde{d}_0]\nonumber\\
 &=&\delta(\bar{g}_\pi^{(0)})_v^r\nonumber\\
 \delta(\bar{g}_\pi^{(1)})_v&=&\frac{128(\Delta m_N)_q(\Delta
 	m_\pi^2)_c}{F_\pi^3}[(2L_6^r+L_8^r)+\varepsilon^2(2L_7^r+L_8^r)]\frac{\tilde{d}_1}{\tilde{d}_0^M}\nonumber\\
 &&-32(\Delta
 m_N)_q[(2G_1^r+G_3^r)\tilde{d}_1-\varepsilon(2G_2^r+G_3^r)\tilde{d}_0]\nonumber\\
 &=&\delta(\bar{g}_\pi^{(1)})_v^r
 \end{eqnarray}
 
 Eq. \eqref{eq:cEDMpioncounter} also provides LEC-contributions for the
 $I=0$ and $I=2$ pion mass shift:
 \begin{eqnarray}
 \delta(\Delta
 m_\pi^2)_{c,ct}&=&\frac{32}{3}m_\pi^2F_\pi[3(2G_1+G_3)\tilde{d}_0^M-\varepsilon(2G_2^r+G_3^r)\tilde{d}_1^M]\nonumber\\
 &&-16\tilde{d}_0^M m_\pi^2 F_\pi(2G_4+G_5)-\frac{32m_\pi^2}{F_\pi^2}(2L_4+L_5)(\Delta m_\pi^2)_c\nonumber\\
 \delta(\delta m_\pi^2)_{c,ct}&=&\frac{32}{3}m_\pi^2
 F_\pi\varepsilon(2G_2^r+G_3^r)\tilde{d}_1^M.\label{eq:cMDMpioncounterresult}
 \end{eqnarray}
 Note that there is no loop contribution to $(\delta m_\pi^2)_c$ so
 the corresponding LECs must be finite, therefore $2G_2+G_3=2G_2^r+G_3^r$. Also notice that terms like $2L_4+L_5$ and $2G_4+G_5$ come from the
 wavefunction renormalization.
 
 In the nucleon sector, the introduction of $O(E^2\tilde{d})$ Lagrangian gives the following consequences. First, it modifies the cMDM-induced nucleon mass
 shifts:
 \begin{eqnarray}
 \delta(\Delta
 m_N)_{c,ct}&=&4m_\pi^2[(\tilde{d}_0^M-\varepsilon\tilde{d}_1^M)(g_1^r+g_4^r)+\tilde{d}_0^M(g_2+g_3)]\nonumber\\
 \delta(\delta
 m_N)_{c,ct}&=&8m_\pi^2[(g_2+g_4)\tilde{d}_1^M-\varepsilon(g_3^r+g_4^r)\tilde{d}_0^M].\label{eq:cMDMnucleoncounterresult}
 \end{eqnarray}
 Also, combining with the $O(E^4)$ LECs and the leading-order
 vacuum alignment, we obtain the LEC-contributions for
 $\bar{g}_\pi^{(i)}$:
 \begin{eqnarray}
 \delta(F_\pi\bar{g}_\pi^{(0)})_{ct}&=&\frac{16(\Delta m_\pi^2)_c
 	m_\pi^2\varepsilon}{3F_\pi^3}\frac{\tilde{d}_1}{\tilde{d}_0^M}[f_2+f_3+f_5^r+f_6^r]\nonumber\\
 &&+\frac{8m_\pi^2}{3}[3\tilde{d}_0(g_2+g_4+g_7^r+g_8^r)-\varepsilon\tilde{d}_1(g_3^r+g_4^r+g_6+g_8)]\nonumber\\
 \delta(F_\pi\bar{g}_\pi^{(1)})_{ct}&=&-\frac{4(\Delta m_\pi^2)_c
 	m_\pi^2}{F_\pi^3}\frac{\tilde{d}_1}{\tilde{d}_0^M}[2(1+\varepsilon^2)f_1+2f_2+(1+\varepsilon^2)f_3+2(1+\varepsilon^2)f_4+(\varepsilon^2-1)f_5\nonumber\\
 &&+(1+\varepsilon^2)(f_5^r+f_6^r)]+8m_\pi^2[\tilde{d}_1(g_1^r+g_2+g_3+g_4^r+g_5+g_8)\nonumber\\
 &&-\varepsilon\tilde{d}_0(g_1^r+g_4^r+g_5^r+g_6^r+g_7^r+g_8^r)]\nonumber\\
 \delta(F_\pi\bar{g}_\pi^{(2)})_{ct}&=&\frac{4(\Delta m_\pi^2)_c
 	m_\pi^2\varepsilon}{3F_\pi^3}\frac{\tilde{d}_1}{\tilde{d}_0^M}[f_2^r+f_3^r+f_5^r+f_6^r]-\frac{8m_\pi^2\varepsilon}{3}\tilde{d}_1(g_3^r+g_4^r+g_6^r+g_8^r)
 \end{eqnarray}

\subsection{LR4Q}\label{sec:LR4Qdetail}

First, the one-loop renormalization to $F_\pi\bar{g}_\pi^{(i)}$ is:
\begin{eqnarray}
\delta(F_\pi\bar{g}_\pi^{(0)})_\mathrm{loop}&=&F_\pi\left[(\frac{4g_A^2}{F_\pi^2}I_a-\frac{2}{F_\pi^2}I_e)\bar{g}_{\pi}^{(0)}-\frac{40g_{\pi
		N\Delta}^2\bar{g}_{\Delta\Delta\pi}^{(0)}}{9F_\pi^2}I_d+(Z_N-1)\bar{g}_\pi^{(0)}\right]\nonumber\\
\delta(F_\pi\bar{g}_\pi^{(1)})_\mathrm{loop}&=&F_\pi\left[(-\frac{12g_A^2}{F_\pi^2}I_a-\frac{8}{3F_\pi^2}I_e)\bar{g}_{\pi}^{(1)}+\frac{8g_{\pi
		N\Delta}^2\bar{g}_{\Delta\Delta\pi}^{(1,1)}}{F_\pi^2}I_d-(\frac{40g_A^2}{F_\pi}I_c+\frac{80g_{\pi
		N\Delta}^2}{3F_\pi}I_b)\bar{g}_{\pi\pi\pi}^{(1)}\right.\nonumber\\
&&\left.+\frac{5\bar{g}_{NN3\pi}^{(1)}}{F_\pi^2}I_e-\frac{15m_\pi^2(\Delta m_\pi^2)_{4q}}{16\pi^2F_\pi^4}\frac{\mathrm{Im}c_{4q}}{\mathrm{Re}c_{4q}}((\gamma_1+4\gamma_2)(L'+1)-2\gamma_2)+(Z_N-1)\bar{g}_\pi^{(1)}\right].\nonumber\\
\label{eq:loopgpibarLR4Q}\end{eqnarray}

Meanwhile, for LR4Q-induced hadron mass parameters, with the identification that
$(g_{NN\pi\pi}^{(0)})_{4q}=-16(\Delta m_N)_{4q}/3F_\pi$,
$(g_{4\pi}^{(0)})_{4q}=2(\Delta m_\pi^2)_{4q}/3F_\pi^2$ and
$(g_{4\pi}^{(2)})_{4q}=2(\delta m_\pi^2)_{4q}/F_\pi^2$ at
tree-level, we obtain:
\begin{eqnarray}
\delta(\Delta m_N)_{4q,\mathrm{loop}}&=&(Z_N-1)(\Delta
m_N)_{4q}-(\frac{12g_A^2}{F_\pi^2}I_a+\frac{16}{F_\pi^2}I_e)(\Delta
m_N)_{4q}-\frac{8g_{\pi N\Delta}^2(\Delta
	m_\Delta)_{4q}}{F_\pi^2}I_d\nonumber\\
&&+(\frac{12g_A^2}{F_\pi^2}I_c+\frac{8g_{\pi N\Delta}^2}{F_\pi^2}I_b)(\Delta m_\pi^2)_{4q}-\frac{3m_\pi^2(\Delta m_\pi^2)_{4q}}{8\pi^2F_\pi^3}((\gamma_1+4\gamma_2)(L'+1)-2\gamma_2)\nonumber\\
\delta(\Delta m_\pi^2)_{4q,\mathrm{loop}}&=&\frac{3(\Delta
	m_\pi^2)_{4q}m_\pi^2}{8\pi^2
	F_\pi^2}(L'+\frac{4}{3})\nonumber\\
\delta(\delta m_\pi^2)_{4q,\mathrm{loop}}&=&\frac{3(\delta
	m_\pi^2)_{4q}m_\pi^2}{4\pi^2 F_\pi^2}(L'+\frac{2}{3}).\label{eq:loopmassLR4Q}
\end{eqnarray}

 Next we consider consequences of the introduction of $O(E^2c_{4q})$ LECs. First, $O(E^2c_{4q})$ terms in the pion sector \eqref{eq:LR4Qpioncounter}, together with the $O(E^4)$ pion Lagrangian, modify the vacuum-alignment condition as:
 \begin{eqnarray}
 \alpha
 &=&-\frac{3}{4}\frac{(\Delta m_\pi^2)_{4q,0}}{(m_\pi^2)_0}\frac{\mathrm{Im}c_{4q}}{\mathrm{Re}c_{4q}}+\frac{48(\Delta m_\pi^2)_{4q}}{F_\pi^2}[(2L_6^r+L_8^r)+\varepsilon^2(2L_7^r+L_8^r)]\frac{\mathrm{Im}c_{4q}}{\mathrm{Re}c_{4q}}\nonumber\\
 &&-16F_\pi^2\mathrm{Im}c_{4q}(K_1^r+4K_2^r)+\frac{15(\Delta
 	m_\pi^2)_{4q}}{32\pi^2F_\pi^2}\frac{\mathrm{Im}c_{4q}}{\mathrm{Re}c_{4q}}(L+1).
 \end{eqnarray}
 Unlike the case of the $\theta$-term and cEDM, the angle $\alpha$ here
 is UV-divergent. It does not cause any problem though since $\alpha$
 itself is not a physical observable. The modified vacuum alignment
 condition leads to extra contributions to $\bar{g}_\pi^{(0)}$ and
 $\bar{g}_\pi^{(1)}$:
 \begin{eqnarray}
 \delta(\bar{g}_\pi^{(0)})_v&=&\frac{48(\delta m_N)_q(\Delta m_\pi^2)_{4q}}{F_\pi^3}[(2L_6+L_8)+\varepsilon^2(2L_7^r+L_8^r)]\frac{\mathrm{Im}c_{4q}}{\mathrm{Re}c_{4q}}\nonumber\\
 &&-16(\delta m_N)_qF_\pi\mathrm{Im}c_{4q}(K_1+4K_2)\nonumber\\
 &=&\delta(\bar{g}_\pi^{(0)})_v^r+\frac{15(\Delta m_\pi^2)_{4q}(\delta m_N)_q}{32\pi^2F_\pi^3}\frac{\mathrm{Im}c_{4q}}{\mathrm{Re}c_{4q}}(L+1)\nonumber\\
 \delta(\bar{g}_\pi^{(1)})_v&=&\frac{96(\Delta m_N)_q(\Delta m_\pi^2)_{4q}}{F_\pi^3}[(2L_6+L_8)+\varepsilon^2(2L_7^r+L_8^r)]\frac{\mathrm{Im}c_{4q}}{\mathrm{Re}c_{4q}}\nonumber\\
 &&-32(\Delta m_N)_qF_\pi\mathrm{Im}c_{4q}(K_1+4K_2)\nonumber\\
 &=&\delta(\bar{g}_\pi^{(0)})_v^r+\frac{15(\Delta
 	m_\pi^2)_{4q}(\Delta
 	m_N)_q}{16\pi^2F_\pi^3}\frac{\mathrm{Im}c_{4q}}{\mathrm{Re}c_{4q}}(L+1).
 \end{eqnarray}
 that are also UV-divergent. This should not bother us because these divergences, together with the divergences from the one-loop corrections to $F_\pi\bar{g}_\pi^{(i)}$, will be canceled by the $O(E^2c_{4q})$ LECs as we shall discuss later.
 
 Finally, Eq. \eqref{eq:LR4Qpioncounter} also contributes to
 the pion mass shifts:
 \begin{eqnarray}
 \delta(\Delta m_\pi^2)_{4q,ct}&=&-\frac{32m_\pi^2}{F_\pi^2}(2L_4+L_5)(\Delta m_\pi^2)_{4q}-16F_\pi^2\mathrm{Re}c_{4q}(2K_4-\frac{1}{3}K_3)m_\pi^2\nonumber\\
 &&+\frac{16}{3}F_\pi^2m_\pi^2\mathrm{Re}c_{4q}(3K_1+22K_2)\nonumber\\
 \delta(\delta
 m_\pi^2)_{4q,ct}&=&-\frac{32m_\pi^2}{F_\pi^2}(2L_4+L_5)(\delta
 m_\pi^2)_{4q}-\frac{16}{3}F_\pi^2\mathrm{Re}c_{4q}K_3
 m_\pi^2-\frac{64}{3}F_\pi^2m_\pi^2\mathrm{Re}c_{4q}K_2.\nonumber\\
 &&\label{eq:LR4Qpioncounterresult}
 \end{eqnarray}
 
 The introduction of $O(E^2c_{4q})$ LECs in nucleon sector contributes to the LR4Q-induced nucleon mass
 shifts:
 \begin{eqnarray}
 \delta(\Delta m_N)_{4q,ct}&=&2F_\pi m_\pi^2\mathrm{Re}c_{4q}[2h_1+4h_2+h_3+h_4+2h_5]\nonumber\\
 \delta(\delta m_N)_{4q,ct}&=&4\varepsilon F_\pi
 m_\pi^2\mathrm{Re}c_{4q}[h_3+h_4-2h_5].\label{eq:LR4Qnucleoncounterresult}
 \end{eqnarray}
 Next, combining with the $O(E^4)$ LECs and the leading-order
 vacuum alignment, we obtain the LEC-contributions for
 $\bar{g}_\pi^{(i)}$:
 \begin{eqnarray}
 \delta(F_\pi\bar{g}_\pi^{(0)})_{ct}&=&\frac{4(\Delta
 	m_\pi^2)_{4q}m_\pi^2\varepsilon}{F_\pi^3}\frac{\mathrm{Im}c_{4q}}{\mathrm{Re}c_{4q}}[f_2+f_3+f_5^r+f_6^r]
 +\frac{4F_\pi\mathrm{Im}c_{4q}m_\pi^2\varepsilon}{3}[3h_3+5h_4-2h_5]\nonumber\\
 \delta(F_\pi\bar{g}_\pi^{(1)})_{ct}&=&-\frac{3(\Delta m_\pi^2)_{4q}m_\pi^2}{F_\pi^3}\frac{\mathrm{Im}c_{4q}}{\mathrm{Re}c_{4q}}[2(1+\varepsilon^2)f_1+2f_2+(1+\varepsilon^2)f_3+2(1+\varepsilon^2)f_4\nonumber\\
 &&+(\varepsilon^2-1)f_5+(1+\varepsilon^2)(f_5^r+f_6^r)]+4F_\pi\mathrm{Im}c_{4q}m_\pi^2[2h_1+8h_2+h_3+3h_4+2h_5]\nonumber\\
 \delta(F_\pi\bar{g}_\pi^{(2)})_{ct}&=&\frac{(\Delta
 	m_\pi^2)_{4q}m_\pi^2\varepsilon}{F_\pi^3}\frac{\mathrm{Im}c_{4q}}{\mathrm{Re}c_{4q}}[f_2+f_3+f_5^r+f_6^r]
 +\frac{8F_\pi\mathrm{Im}c_{4q}m_\pi^2\varepsilon}{3}[h_4-h_5]。
 \end{eqnarray}
 As we noticed before, one may choose the values of the combinations $3h_3+5h_4-2h_5$, $2h_1+8h_2+h_3+3h_4+2h_5$ and $h_4-h_5$ to subtract out the UV-divergences from $\delta(F_\pi\bar{g}_\pi^{(i)})_{\mathrm{loop}}+F_\pi\delta(\bar{g}_\pi^{(i)})_v$ together with the residual UV-divergences coming from the LECs ${f_i}$ in $\delta(F_\pi\bar{g}_\pi^{(i)})_{ct}$. Details of such subtractions are given in Table \ref{tab:subt1}.
 
\section{\label{sec:ct}Divergence Subtraction and Renormalized LECs}

In this section we summarize the divergence subtractions by the
counter terms. Following the Gasser-Leutwyler subtraction scheme, the
relation between the bare and renormalized LEC is given by
\begin{equation}
A=A^r+\frac{B}{\pi^2}(L+1)
\end{equation}
where $A$ is the bare LEC, $A^r$ is the renormalized
LEC and $B$ is a finite quantity. Also, since any physical result must be $\mu$-independent,
the renormalized LEC $A^r$ must be $\mu$-dependent in the following way:
\begin{equation}A^r(\mu')=A^r(\mu)+\frac{B}{\pi^2}\ln\left(\frac{\mu'}{\mu}\right)^2\end{equation}
in order to cancel the $\mu$-dependence in the divergent loop integral.

The values of the finite quantity $B$ for different combinations of
LECs are summarized in Table \ref{tab:subt1}.

\begin{table}[h]
\begin{center}
\begin{tabular}{|c|c|}
\hline $A$ & $B$ \tabularnewline \hline \hline $2L_{4}+L_{5}$&
$-\frac{1}{64}$ \tabularnewline \hline $2L_{6}+L_{8}$ &
$-\frac{3}{512}$ \tabularnewline \hline $2L_7+L_8$ &
$0$\tabularnewline\hline $B_{20}$ & $\frac{9\pi^2g_A^2}{32}+\frac{\pi^2g_{\pi
N\Delta}^2}{2}\frac{m_\pi^2-2\delta_\Delta^2}{m_\pi^2}$\tabularnewline \hline $2f_1+f_3,2f_4+f_5,f_5+f_6$ &0\tabularnewline\hline
$f_2$&$\frac{F_\pi
(c_1+2c_1')}{2B_0}(\frac{9g_A^2}{16}-\frac{3}{8})+\frac{F_\pi
g_{\pi N\Delta}^2(c_2+2c_2')}{2B_0}\frac{2\delta_\Delta^2-m_\pi^2}{m_\pi^2}+\frac{F_\pi
g_{\pi
N\Delta}^2\delta_\Delta}{6m_\pi^2}\frac{2\delta_\Delta^2-3m_\pi^2}{m_\pi^2}+\frac{3(\gamma_1+4\gamma_2)}{32}$\tabularnewline\hline
$f_2+f_3$ & $-\frac{F_\pi c_1}{2
B_0}(\frac{3g_A^2}{16}+\frac{1}{8})+\frac{5F_\pi
g_{\pi N\Delta}^2c_2}{18B_0}\frac{2\delta_\Delta^2-m_\pi^2}{
m_\pi^2}$\tabularnewline\hline
$2G_1+G_3$ & $-\frac{3\beta}{16}$\tabularnewline\hline
$2G_2+G_3$ & $0$ \tabularnewline\hline $2G_4+G_5$ & $-\frac{\beta}{4}$\tabularnewline\hline
$g_1+g_4,g_3+g_4$ & $0$\tabularnewline $g_5+g_6,g_7+g_8$ & \tabularnewline\hline
$g_2+g_3$ & $\frac{\tilde{c}_1+2\tilde{c}_1'}{4}(\frac{9g_A^2}{16}-\frac{3}{8})+\frac{g_{\pi
N\Delta}^2(\tilde{c}_2+2\tilde{c}_2')}{4}\frac{2\delta_\Delta^2-m_\pi^2}{m_\pi^2}-\frac{2 g_{\pi
N\Delta}^2\beta F_\pi\delta_\Delta}{m_\pi^2}+\frac{3\beta(\gamma_1+4\gamma_2)}{4}$\tabularnewline\hline
$g_2+g_4$ &
$-\frac{\tilde{c}_1}{4}(\frac{3g_A^2}{16}+\frac{1}{8})+\frac{5g_{\pi
N\Delta}^2\tilde{c}_2}{36}\frac{2\delta_\Delta^2-m_\pi^2}{m_\pi^2}$\tabularnewline\hline
$g_6+g_8$ & $-\frac{2F_\pi\beta c_1}{
B_0}(\frac{3g_A^2}{16}+\frac{1}{8})+\frac{10F_\pi
g_{\pi N\Delta}^2\beta c_2}{9B_0}\frac{2\delta_\Delta^2-m_\pi^2}{
m_\pi^2}$\tabularnewline\hline
$2f_4-f_5$ & $-\frac{F_\pi(c_1+2c_1')}{2B_0}(\frac{9g_A^2}{16}-\frac{3}{8})-\frac{F_\pi g_{\pi N\Delta}^2(c_2+2c_2')}{2B_0}\frac{2\delta_\Delta^2-m_\pi^2}{m_\pi^2}$ \tabularnewline
$-\frac{1}{4\beta}(g_5+g_8)$ & $-\frac{F_\pi g_{\pi N\Delta}^2\delta_\Delta}{6m_\pi^2}\frac{4\delta_\Delta^2-3m_\pi^2}{m_\pi^2}-\frac{3(\gamma_1+4\gamma_2)}{16}$\tabularnewline\hline
$2K_4-\frac{1}{3}K_3$ & $-\frac{2\rho}{3}$\tabularnewline\hline
$K_3$ & $-\frac{\rho}{4}$\tabularnewline\hline
$3K_1+22K_2$ & $-\frac{11\rho}{2}$\tabularnewline\hline
$K_2$ & $-\frac{\rho}{4}$\tabularnewline\hline
$2h_1+4h_2+h_3+h_4+2h_5$ & $\tilde{\tilde{c}}_1(\frac{9g_A^2}{16}-1)+g_{\pi N\Delta}^2\tilde{\tilde{c}}_2\frac{2\delta_\Delta^2-m_\pi^2}{m_\pi^2}
-\frac{32 g_{\pi N\Delta}^2\rho}{3}\frac{F_\pi\delta_\Delta}{m_\pi^2}+4\rho(\gamma_1+4\gamma_2)$\tabularnewline\hline
$h_3+h_4-2h_5$ & $0$\tabularnewline\hline
$3h_3+5h_4-2h_5$ & $\frac{32F_\pi\rho c_1}{B_0}(\frac{3g_A^2}{64}+\frac{1}{2})-\frac{40F_\pi g_{\pi N\Delta}^2\rho c_2}{9B_0}\frac{2\delta_\Delta^2-m_\pi^2}{m_\pi^2}$\tabularnewline\hline
$2f_4-f_5$ & $-\frac{1}{6}(\frac{9g_A^2}{16}-1)(\frac{3F_\pi(c_1+2c_1')}{B_0}+\frac{3\tilde{\tilde{c}}_1}{4\rho})$\tabularnewline
$-\frac{1}{16\rho}(2h_1+8h_2+h_3$ & $-\frac{ g_{\pi N\Delta}^2}{6}(\frac{2\delta_\Delta^2-m_\pi^2}{m_\pi^2})(\frac{3F_\pi(c_2+2c_2')}{B_0}+\frac{3\tilde{\tilde{c}}_2}{4\rho})$\tabularnewline
$+3h_4+2h_5)$ & $-\frac{F_\pi g_{\pi N\Delta}^2\delta_\Delta}{6 m_\pi^2}\frac{4\delta_\Delta^2-11m_\pi^2}{m_\pi^2}-\frac{\gamma_1+4\gamma_2}{2}$\tabularnewline\hline
$h_4-h_5$ & $\frac{4F_\pi\rho c_1}{
	B_0}(\frac{3g_A^2}{16}+\frac{1}{8})-\frac{20F_\pi
	g_{\pi N\Delta}^2\rho c_2}{9B_0}\frac{2\delta_\Delta^2-m_\pi^2}{
	m_\pi^2}$\tabularnewline\hline
\end{tabular}
\par\end{center}
\caption{\totheleft\label{tab:subt1}Infinity subtraction by the LECs.}
\end{table}

\bibliographystyle{prsty}
\bibliography{CYSbib}

\end{document}